\documentclass[12pt,aps,rmp,longbibliography,preprint,notitlepage]{revtex4-1}
\usepackage{amsmath}
\usepackage{textcomp}
\usepackage{graphicx}
\begin{document}
\title{
  Compatible Quantum Theory}

\author{R Friedberg} \affiliation{Department of Physics, Columbia University, New York, NY}

\author{P C Hohenberg}

\affiliation{Department of Physics, New York University, New York, NY}

\date{May 6, 2014)\\
\vspace{20mm}
Abstract:\\
Formulations of quantum mechanics can be characterized as realistic, operationalist, or a combination of the two. In this paper a
realistic theory is defined as describing a closed system entirely by means of entities and concepts pertaining to the system. An
operationalist theory, on the other hand, requires in addition entities external to the system. A realistic formulation comprises an
\emph{ontology}, the set of (mathematical) entities that describe the system, and \emph{assertions}, the set of correct statements
(predictions) the theory makes about the objects in the ontology. Classical mechanics is the prime example of a realistic physical
theory. A straightforward generalization of classical mechanics to quantum mechanics is hampered by the inconsistency of quantum
properties with classical logic, a circumstance that was noted many years ago by Birkhoff and von Neumann. The present realistic formulation of the histories approach originally introduced by Griffiths, which we call \lq Compatible Quantum Theory (CQT)', consists of a 'microscopic' part (MIQM), which applies to a closed quantum system of \emph{any} size, and a 'macroscopic' part (MAQM), which
requires the participation of a large (ideally, an infinite) system. The first (MIQM) can be fully formulated based solely on the assumption of a
Hilbert space ontology and the noncontextuality of probability values, relying in an essential way on Gleason's theorem and on an application to dynamics
due in large part  to Nistico. Thus, the present formulation, in contrast to earlier ones, \emph{derives} the Born probability formulas and the
consistency (decoherence) conditions for frameworks. The microscopic theory does not, however, possess a unique corpus of assertions, but rather a multiplicity of contextual truths (\lq c-truths'), each one
associated with a different framework. This circumstance leads us to consider the microscopic theory to be \emph{physically
indeterminate} and therefore \emph{incomplete}, though logically coherent. The completion of the theory requires a \emph{macroscopic} mechanism for selecting a physical framework, which is part of the macroscopic theory (MAQM). The selection of a physical framework involves the breaking of the microscopic \lq framework symmetry', which
can proceed either phenomenologically as in the standard quantum measurement theory, or more fundamentally by considering the quantum
system under study to be a subsystem of a macroscopic quantum system. The Decoherent Histories formulation of Gell-Mann and Hartle, as
well as that of Omn\`es, are theories  of this fundamental type, where the physical framework is selected by a coarse-graining
procedure in which the physical phenomenon of decoherence plays an essential role. Various well-known interpretations of quantum mechanics are
described from the perspective of CQT. Detailed definitions and proofs are presented in the appendices.}
\maketitle

\nopagebreak
\section{\label{1}Introduction and Motivation}

It is nearly 90 years since the mathematical formalism of nonrelativistic quantum mechanics (hereafter QM) was created and there have been essentially no revisions to the present day. Yet there remains significant controversy regarding the physical \lq interpretation' of the theory, specifically the meaning of the wave function or quantum state. There is even controversy as to whether QM requires an interpretation! \cite{van2008,fuchs-peres}. This situation, in which what is arguably the most successful theory in all of science is still a subject of debate as to its basic meaning, is unprecedented in modern science.

The present paper agrees with those who say that QM requires no interpretation. Instead it requires a proper \emph{formulation} that specifies the assumptions of the theory and its relationship to physical systems and their properties. There are, however, two types of foundational questions regarding QM:

 $\bullet$ Is QM the \lq whole truth', with respect to experimental consequences?

 $\bullet$ If so, what is the best formulation of QM?

These two questions are often confounded in discussions of quantum foundations, but we would like to distinguish them. If the answer to the first question is \lq no', then all formulations of QM are doomed, by definition. It has become customary, however, to include in the list of formulations some that are based on the answer \lq no,' i.e. revisions of the physical content of QM that are motivated by the difficulty of proper formulations of the existing theory. The most popular example is the class of \lq spontaneous collapse' formulations, see e.g. \textcite{ab5}.  Unless otherwise stated we shall assume the answer \lq yes' to the first question, always leaving open the possibility that experiment will rule otherwise, in which case a different theory will be sought. This paper is devoted almost exclusively to the second question.

Formulations of QM can be broadly divided into two classes, which we shall refer to as \lq realistic' and \lq operationalist'. We define a theory to be \emph{realistic} if for a system \textbf{S} it is formulated entirely in terms of entities and concepts referring to \textbf{S} itself. An \emph{operationalist} formulation, on the other hand, \emph{requires}, in addition to \textbf{S}, entities \emph{external} to \textbf{S} such as \lq observers', \lq measurement apparatus', or \lq agents', for a full specification. There are also theories that are partly realistic and partly operationalist, in that some properties are defined in terms of \textbf{S} and some in terms of external entities.

For a realistic formulation we shall distinguish two parts, the \lq ontology' and the \lq assertions'. The ontology, represented in the formalism by mathematical entities, comprises the list of objects or properties that characterize the system under study. They are what the theory is \lq about'. We shall sometimes refer to the objects in the ontology, following \textcite{b8}, as the \lq beables' (or existables) of the theory. The assertions are the set of correct mathematical or physical statements (predictions) that are made by the theory, regarding the beables. Note that the concept of truth is inextricably linked to the assertions.

An operationalist formulation, of which the Bohr (or \lq Copenhagen') version of QM is the prime example, requires a dividing line (a \lq cut') between the system \textbf{S} and the outside and a separate characterization of the outside. In a totally operationalist theory the term ontology is inappropriate for the properties of the unobserved system, since these properties are not what the predictions (assertions) are about.

Note that we use the terms realism and operationalism exclusively to characterize formulations of physical theories, not as descriptions of a philosophical point of view. In this restricted sense, the terms are well defined. As explained below, on the other hand, \lq reality' and \lq real', referring to beables or their properties, are terms whose meaning is much more obscure and we shall avoid their use altogether. We recommend this practice in all discussions of the foundations of QM.

The present paper focusses on the \emph{nonrelativistic} case with Euclidean space-time, since most of the traditional questions and paradoxes of QM already appear in that limit. As emphasized by earlier authors, however, the histories approach generalizes naturally to the relativistic case, including general relativity with more complicated space-time geometries. This is in contrast to some other formulations of QM (see below).

The aim of the present paper is to provide a realistic formulation of QM, paying careful attention to basic philosophical issues and to precise use of language. In the interest of clarity, we are led to redefine certain terms, as compared to their standard usage. We summarize our definitions in Appendix E.

Section II presents a formulation of classical mechanics, which serves as a model for a realistic theory, in a form designed to permit generalization to the quantum case. Section III, entitled  ``The Basic Conundrum of Quantum Mechanics'', explains why this generalization is not straightforward. Section IV, the core of the paper, is a realistic formulation of QM which we call \lq Compatible Quantum Theory (CQT)'. It is a reformulation of the Consistent and Decoherent Histories theories introduced by \textcite{grif2}, \textcite{gh3} and \textcite{omnes1992}. Although there is no contradiction with these earlier theories, the point of view and language differ somewhat, whence the (slightly) different name we give to our version. The two principal ways in which CQT differs from the earlier approaches are (i) an essential distinction is introduced between the microscopic (MIQM) and macroscopic (MAQM) theories, and (ii) in the microscopic theory the Born formulas and consistency (decoherence) conditions  are \emph{derived} rather than posited, assuming only the Hilbert space ontology and the noncontextuality of probability values. Section V discusses the main alternative formulations of QM, seen through the lens of CQT. Detailed mathematical definitions and proofs are presented in the appendixes.

\section{\label{2}	Classical Mechanics}
Classical mechanics provides the prime example of a realistic formulation of a physical theory.
\subsection{\label{A.}	Standard Formulation}
In its simplest but nevertheless general form, nonrelativistic classical mechanics begins with a phase space as a $6N$-dimensional Euclidean space representing the positions and momenta of a system of $N$ particles or $3N$ degrees of freedom. The system under study is represented by a point $x$ in the phase space, with \lq initial' value $x_0$ at $t=t_0$. Then according to classical mechanics there is a unique trajectory $x(t)$ going through $x_0$ at $t=t_0$, for all times $t<t_0$ or $t>t_0$. This trajectory can in principle be calculated from Newton's laws of motion once the forces are known, and any property of the system can be determined from $x(t)$.

The above is a complete formulation of classical mechanics for a \emph{deterministic} system, by which we mean a system for which a specific value of the initial system point $x_0$ has been assumed. In the \emph{stochastic} case (statistical mechanics) we replace the initial system point by an \emph{ensemble} of points $x_i$ with a probability distribution
at $t= t_0 $ such that
\begin {equation}
\int \rho(x,t_0) dx =1,
\end {equation}
where the integral extends over all the points in the ensemble. Then the time evolution of the probability distribution $\rho(x,t)$ is obtained from the Liouville equation, which is the generalization of Newton's laws to the stochastic case. Properties of the system can be obtained at any time by averaging over $x$ with weight $\rho(x,t)$. The Liouville equation ensures that (1) will remain true if any other time $t$ is substituted for $t_0$.

\subsection{\label{B.}	Reformulation in Terms of Set Theory and Logic}

For the purpose of generalization to QM it is useful to rephrase the previous formulation in the language of set theory and logic, see e.g. \textcite{bub}. A brief outline of the necessary mathematical background is presented in Appendix A.

\noindent
\underline{Ontology}

The ontology of the theory is represented by the elements of phase space, which consist of \emph{system points} and \emph{system properties}. We shall refer to the system point, somewhat loosely, as the \emph{state} of the system, or sometimes more carefully, as the \emph{representative point} of the state. A property is represented by any (Borel) subset of phase space.  Under the operations of union, intersection, and complementation, these subsets form a Boolean lattice (see Appendix A).

\noindent
\underline{Assertions}

As mentioned earlier, the assertions are the set of true statements (predictions) that the theory makes about the elements of the ontology. We shall say that a system \lq has' the property $A$ if the representative point for its state is contained in the subset of phase space specified by $A$.   Thus a state $x$ defines a \emph{truth function} $\mathcal{T}_x(A)$ on the properties of the system, which selects out of all possible properties (i.e., Borel subsets of phase space) of the system, those that contain the representative point of the state. This truth function can be written as
\begin{eqnarray}
\mathcal{T}_x(A)= \text{T(True)} \; \text{if} \;\;x\; \in A,\\
\mathcal{T}_x(A)= \text{F(False)} \; \text{if} \;\;x \;\notin A.
\end{eqnarray}

\noindent All classical properties are determinate with respect to any state: they are either true (contain the state) or false (do not contain the state).
We can thus say that the state is the \lq source of truth' for any system property. The truth values then satisfy all the requirements of classical logic (see Appendix A).

\noindent
\underline{Determinism}

Classical mechanics is deterministic, in that \emph{if} the state is known exactly at some time, any property is exactly determined at any later time
to be either true or false. There are, however, two important qualifications to this determinism. The first is the well-known phenomenon of chaos, namely that in order to determine a property with an accuracy $\alpha$ at time $t>0$, one would in general have to know the state at time $t=0$ with an accuracy $\alpha exp(-\beta t)$, where $\beta$ is a time
constant characteristic of the system. Thus in practice the predictions are almost always approximate.

The second qualification is less often discussed but no less important. Classical mechanics does not tell you \lq what happens': all predictions of properties are contingent upon (\lq relative to' or \lq \emph{contextual} to') an assumed state at some \lq initial time', but the theory does not tell us what that initial state is. By definition the state is declared to be \lq true' at some time (technically, the set containing only this state is \lq true')  and the assertions are the set of other true propositions that follow from this assumption.  We shall see below that QM in its CQT formulation retains the contextuality with respect to the assumed truth of the initial state and it adds the contextuality of frameworks.

\noindent
\underline{Statistical Mechanics}

In the stochastic case we assume an initial probability distribution $\rho( x, t_0)$, which governs the choice of the state $x$ and which generates a well-defined probability function on the properties,
\begin{equation}
\mathcal{P}_{\rho}(A,t) = \int_{x\in A} {\rho(x,t)} dx,
\end{equation}

\noindent where the quantity $\rho(x,t)$ evolves in time according to the classical laws governing $x (t)$. In Appendix A, Kolmogorov's set-theoretic formulation of probability theory is presented and shown to be closely related to classical logic. It follows that the probability function $\mathcal{P}_{\rho}$ can be thought of as a distributed truth function obtained by combining strict truth functions ${\cal T}_x$ according to the weights given by $\rho(x)$. In the probabilistic case it is statements about $\mathcal{P}_{\rho}$ that are the assertions of the theory.

In this connection we may say that the function $\rho$ represents the state; this requires us to say that in the deterministic case the state is represented not strictly by the system point $x$, but by the characteristic function of the set $\{x\}$, if the system points are discrete - otherwise, by a delta function.  This shift anticipates the shift, in quantum mechanics, from a pure state $|\psi\rangle$ to a mixed state $\rho$, in Section IVB. In the general case we can say that the state $\rho$ is the source of probability for the system.

\noindent
\underline{Measurements and state preparation}

It is not traditional to discuss measurements in formulations of classical mechanics, not because measurements are considered unimportant, but because they are thought to be unproblematic. In order to verify or test the predictions of the theory, the system in question is coupled to some measurement apparatus (if the original definition of the system does not contain any apparatus) and the \lq pointer readings' on the apparatus are correlated with the properties of the original system. One either assumes that the effects of the apparatus on the system are negligible, or that they can be calculated and subtracted out to determine the \lq true' behavior of the uncoupled system.

Similarly, the designation of a particular element of phase space as the \lq state of the system' can be realized by a physical action analogous to a measurement, whereby the state in question is \lq prepared' and selected. Carrying out and describing such a procedure in practice may not be easy but there are no problems of principle.

\noindent
\underline {Microscopic and macroscopic theories}

The formulation of classical mechanics described above represents what we refer to as a \emph{microscopic} theory, namely one that applies to
a system of any size.  Measurements, to be sure, involve macroscopic devices capable of recording the truth value of the property they are designed to measure, but in classical mechanics the necessary addition to the microscopic theory does not involve any new concepts in order to deal with such devices, so there is no need for a special auxiliary theory for this purpose.  The classical theory is thus physically complete.  In the quantum world, however, we will see that the \emph{choice} of a measurement has an influence on what statements can be correctly made about the system under study; this gives the measurement process a special character that differs from its role in classical mechanics. We shall thus exclude measurements from the quantum mechanical microscopic theory we develop, which then implies that the latter theory is physically incomplete, and it will need to be augmented by a \emph{macroscopic} theory that can deal with the special character of measurement.

Classical mechanics also illustrates the point made earlier about the terms \lq reality' and \lq real'. One could simply say that any element of the ontology is real, but this includes elements that are true as well as elements that are false, or elements that are indeterminate in case no initial state has been designated as true. We prefer to avoid the use of \lq real' and \lq reality' altogether in this context. The terms \lq realism' and \lq realistic', on the other hand, have a well-defined meaning
(see Section I)
 when referring to the formulation of a theory.

The above microscopic formulation of classical mechanics is illustrated schematically in Fig. 1.

\begin{figure}
\begin{center}
 \includegraphics[width=6in]{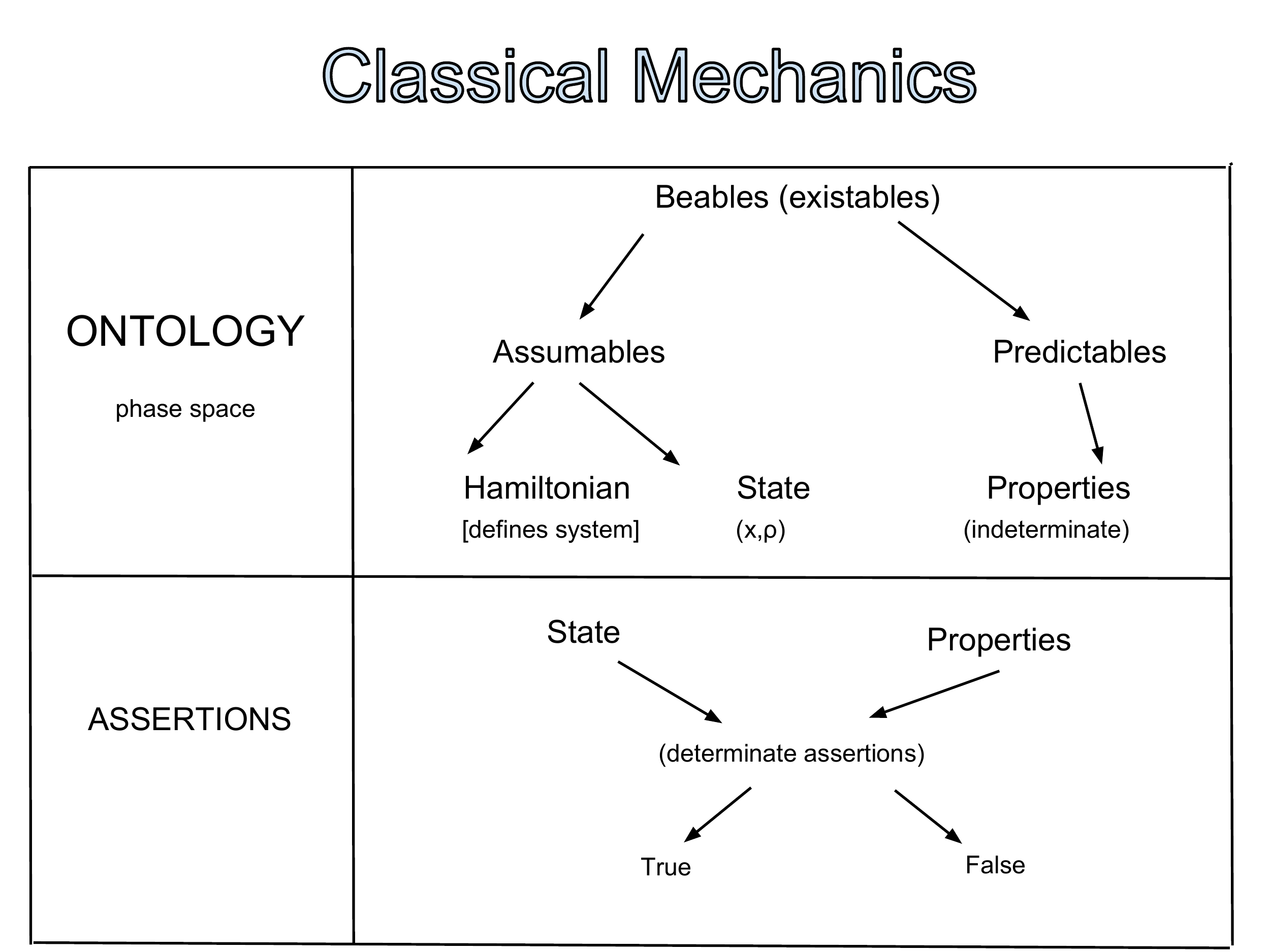}
\end{center}
\caption{Schematic representation of classical mechanics. The top row describes the ontology (or beables), which consists of a Hamiltonian as well as states and properties. A state can be either a point $x$ in phase space (pure) or a probability function $\rho$ (mixed). Properties are subspaces of phase space and their truth values remain indeterminate as long as a state has not been assumed. The bottom row describes the assertions, which consist of truth values that the state (the source of truth) confers on properties.}

\end{figure}

\section{\label{3}	The Basic Conundrum of Quantum Mechanics}

 Before presenting our formulation of QM, we highlight what we consider the essential difficulty in passing from classical to quantum mechanics. This is what we call the \lq fundamental conundrum of quantum mechanics' and it can be illustrated by considering an isolated spin-\mbox{\textonehalf} degree of freedom. Let us consider the statement ``The value of $S_x           \; \text{is}\; + \mbox{\textonehalf}$'' and write it as
\begin{equation}
[S_{x+}]= \text{T},
\end{equation}
 \noindent i.e. the property $S_{x+}$ is true. Similarly, we can write $[S_{i+}]$= T or $[S_{i+}]$= F for any spin component $S_i$.  Such statements about the truth of a property, just like the corresponding ones in classical mechanics, are meant to refer to an intrinsic attribute of the quantum system, and they are thus part of the microscopic theory.

Now Nature tells us (or so it appears) that an essential experimental fact about quantum spins is that $S_i$ is either $+ \mbox{\textonehalf}$ or $- \mbox{\textonehalf}$ but not both, and it can have no other value. We can therefore write, in particular,
\begin{eqnarray}\label{parspins}
[S_{i+}]  \;\text{OR}\;  [S_{i-}] \equiv \text{T}.
\end{eqnarray}

\noindent To be concrete, we mean that a Stern-Gerlach machine oriented along the $i$-axis will (apart from experimental imperfection) send the particle either in the $i+$ or in the $i-$ direction.

We claim that Nature also tells us

\begin{equation}\label{nonparspins}
[S_{i\pm}] \; \text{AND} \; [S_{j\pm}] \equiv \text{F},
\end{equation}

\noindent for any $i, j$ such that $S_i$ and $S_j$ are not collinear.  This does not seem obvious, as a pair of Stern-Gerlach measurements, first along $i$ and then along $j$, may well give affirmative results for both $S_{i+}$ and $S_{j+}$.  Nevertheless, we may interpret $[S_{i+}] \; \text{AND} \; [S_{j+}]$ concretely as meaning that the spin-\mbox{\textonehalf} system can be \emph {prepared} so that a single Stern-Gerlach measurement  of one of $i+$ and $j+$ will give an affirmative result no matter which measurement is chosen.  This is certainly not true according to experience.  Thus (repeating the argument for $[S_{i+}] \; \text{AND} \; [S_{j-}]$, etc.) we have \eqref{nonparspins}.

Let us assume, then, that $i$, $j$ are not collinear. According to the distributive law of logic, Eq. \eqref{clogdist} of Appendix A, written with AND/OR instead of $\wedge,\vee$], we would have
\begin{eqnarray}\label{distrib}
 ([S_{i+}]  \text{OR}  [S_{i-}])\; & \text{AND}\; & ([S_{j+}]  \text{OR} [S_{j-}]) =\\  \nonumber
 ([S_{i+}]\text{AND} [S_{j+}]) \; \text{OR} \; ([S_{i-}]\text{AND} [S_{j+}])  & \text{OR} & ([S_{i+}]\text{AND} [S_{j-}])\;  \text{OR}\; ([S_{i-}]\text{AND} [S_{j-}]).
 \end{eqnarray}

\noindent The above equation states that if either $[S_{i+}]$ or  $[S_{i-}]$ is true and either $[S_{j+}]$ or  $[S_{j-}]$ is true, then one of the pairs $([S_{i\pm}],[S_{j\pm}])$ must be true. But the left-hand side of  \eqref{distrib} is true since  the propositions in parentheses are true by \eqref{parspins},  whereas the right-hand side is false since each proposition in the parentheses is false by \eqref{nonparspins}.  Having \lq proved' that T = F, we must therefore conclude that the \lq essential  experimental fact about quantum mechanics' mentioned above defies logic! \emph{Every realistic formulation of quantum mechanics is some way of coming to terms with this fundamental conundrum.}

 An operationalist does not encounter the conundrum, since he does not consider that closed quantum systems \lq have' intrinsic properties, without specifying an external measurement or agent. Thus both \eqref{parspins} and \eqref{nonparspins} are rejected in an operationalist formulation. Among the most common \emph{realistic} formulations of quantum mechanics there are three  possible choices.
 \newline
 (a) What we will call the \lq classical' choice is to consider states corresponding to the different directions of $S$ to be the separate  members of a ray-ontology considered as a classical set. This choice denies the \lq eigenstate-eigenvalue link' for quantum variables and thus invalidates \eqref{parspins}.
 \newline
 (b) The Consistent Histories formulation of \textcite{grif1} declares that propositions such as those  appearing in \eqref{nonparspins}, involving the combination of \lq incompatible propositions', are meaningless; thus Eq.  \eqref{nonparspins} is discarded and the contradiction is avoided.
 \newline
 (c) There is, however, a third way, namely to let both \eqref{parspins}  and \eqref{nonparspins}  stand and to deny the distributive  law \eqref{clogdist}.
 This is the choice made by \textcite{bvn}, which they termed \lq quantum logic'.  As  explained below, we shall be influenced by this choice in approaching the formulation of quantum mechanics, though we shall not in fact embrace it, and we end up instead with the  essential elements of Consistent Histories.

2. Quantum logic?

It is essential to understand that the difficulty in choice (c) is not simply a formal one, as though the distributive law in logic were an arcane rule that could be brushed aside.  Anyone can see, without referring to \eqref{distrib}, that there is no way of assigning truth values to $[S_{i+}]$, $[S_{i-}]$,
$[S_{j+}]$,  and $[S_{j-}]$ without violating at least one instance of \eqref{parspins} or \eqref{nonparspins}.  Therefore the disease that needs curing has to do with the assignment of truth
values to propositions.

One's first tendency is to suggest that truth values should be made \lq fuzzy' or dispersive, thus abandoning either one or both of Aristotle's \lq Law of Contradiction' (an entity cannot both have and not have the same property) and his \lq Law of Excluded Middle' (an entity either has a property or fails to have it).  Again, however, these are not just scholarly dicta but express the starting point of rational thought as it has been understood through the ages. A mind that persists in violating one of these two laws even in the presence of the most careful clarification is considered irrational.

A different cure has been attempted by \textcite{put69}, in which auxiliary propositions are introduced which satisfy the laws of logic when the original ones do not.  This road has been eventually modified or abandoned by Putnam himself, see \textcite{put81}, or the review by Maudlin in \cite{hilary}, but it has been pursued by other authors, e.g. \cite{bacc}.

 We are inclined to think that the disease stems from the very concept of a \lq proposition', and so we are moved to examine the extent to which logic can be carried out without propositions, and thereby without truth values.
  We will conclude that without truth values there is no \emph{logic}, since truth and falsehood are necessary in order to define logical propositions that make contact with physics. We thus consider the designation \lq quantum logic' \cite{bvn} to be misleading,  and we shall continue to use the term \lq logic' exclusively to denote classical logic. However, we will arrive at logic in two distinct steps: the first following Birkhoff and von Neumann, to see how far one can go without the distributive law, and the second by adding this law in a way that avoids contradictions. We are aware of the broader definitions of logic in mathematics and philosophy, see, e.g. \textcite{modal}, but for the purpose of making contact with physics we equate logic with classical (i.e. Aristotelian) logic.

3. Doing without truth values

The propositions we are concerned with assert that the actual condition of a system
- that is, the one determined by the state at a given time - has or does not have a particular property.  Thus the state may be regarded as the \emph{subject} of any proposition, and
each property, represented mathematically by a projection operator, as the \emph{predicate} of that proposition.  Now if all predicates are considered in relation to some fixed subject, they can be
combined by the usual logical operations without mentioning the subject.  We understand perfectly
what is meant by \lq is a man and is mortal', by \lq is a man or is mortal', and by \lq is not a man', without reference to the subject of these phrases.  Moreover, predicates combined in this way obey the usual axioms of propositional calculus (see Eq. \eqref{clogic} in Appendix A), regardless of one's choice of the fixed subject. This
\lq property calculus' mirrors the abstract features of propositional calculus, but as long as the fixed subject (the state) has not been specified, the predicates
are not propositions and do not have truth values.

Going back to the three choices (a, b, c), we shall follow choice (c) in taking subspaces of Hilbert space to be properties which may violate the distributive law, but shall not associate truth values with them as is done in \eqref{parspins} and \eqref{nonparspins}.  When we finally combine the properties with states we shall do so in a way that avoids contradiction, by associating truth values only to the properties within appropriate \emph{collections} of properties, in which the distributive law is preserved.  These collections will turn out to be the Griffiths \lq frameworks', so that we wind up closest to choice (b).

\section{\label{4} Compatible Quantum Theory (CQT)}

We call our formulation of quantum mechanics \lq Compatible Quantum Theory (CQT)', a designation intended to suggest that it is a version of \lq Consistent Quantum Theory' \cite{grif1}, also abbreviated CQT, and often referred to as the \lq Consistent (or Decoherent) Histories' approach (\textcite{grif2,gh3,omnes1992}; see also \textcite{pch}) . There are, however, distinctions between our formulation and previous ones. We first provide a fully realistic formulation which is applicable to an arbitrary closed quantum system. This is what we call the \lq microscopic theory' (MIQM). We find, however, that this theory is \emph{physically indeterminate} and therefore \emph{incomplete}, in that it has no unique set of physical assertions. In order to complete the theory we require a \emph{macroscopic framework selection mechanism}, which can be either \lq external' or \lq internal' (see below). This mechanism is described by the macroscopic theory (MAQM). We stress that Compatible Quantum Theory does not contradict the tenets of the earlier histories formulations, but it employs a somewhat different language intended to clarify and justify the arguments and to relate the history formulations of QM more closely to other formulations. Note also that in our usage the terms \lq microscopic' and \lq macroscopic' are not meant to distinguish between small and large systems. Instead, the microscopic theory applies to any system regardless of size, whereas the macroscopic theory \emph{requires} a large (ideally infinite) system to define certain concepts used.

\subsection {\label{A} Hilbert Space Ontology}
The starting assumption of CQT is what we call the \emph{Hilbert Space Ontology}:
 the theory is based on properties and states, each of which has a formal representation in the Hilbert space.  The properties are subspaces of Hilbert space and their associated projection operators.  A state will be represented provisionally by a ray of vectors (later called a pure state). Ultimately, the ray will be replaced by a unit-trace self-adjoint positive operator called a density matrix.

Given these definitions we can create a \lq property calculus' by introducing the q-operations applicable to arbitrary properties (see Appendix B):
\begin{subequations}
\label{qops}
\begin{eqnarray}
A \wedge_{\text{q}} B = \text{Intersection} (A,B) = A \;\text{(q-and)}\; B,  \\
A \vee_{\text{q}} B = \text{Span} (A,B) = A \;\text{(q-or)}\; B,\\
\tilde{A}_{\text{q}} = \neg_{\text{q}}{A} = \text{Orthogonal Complement} (A) = \text{(q-not)}\; A,
\end{eqnarray}
\end{subequations}
\noindent
where the span of two subspaces is the set of all vectors obtained by linear combinations of vectors in the two subspaces, and the orthogonal complement of a subspace is the set of all vectors orthogonal to every vector in that subspace.

As discussed in Appendix B, the subspaces of Hilbert space form a \emph{lattice}, which is closed under the operations \eqref{qops} of the property calculus. For a given Hilbert space $\mathcal{H}$ we denote the lattice of all of its properties (subspaces) by  $\mathcal{L}^{\mathcal{H}}$. The all-important difference from the classical case is that this lattice is \emph{non-Boolean},
in that it does not satisfy the distributive law.  That is, in general
\begin{equation}\label{nodist}
A \wedge_{\text{q}} (B \vee_{\text{q}} C)  \ne (A \wedge_{\text{q}} B)\vee_{\text{q}}(A\wedge_{\text{q}}C),
\end{equation}
a feature that is at the root of the \lq quantum conundrum' of Sec. III. This failure will have dire consequences for our formulation of the assertions of CQT.

\subsection {\label{B} The search for quantum assertions: static frameworks}

So far we are not considering
any dynamics, i.e. the state and the properties are
all defined at the same time. This is what we call the \lq static theory'. In attempting to find a set of assertions for static QM, we first note that according to what was stated in Sec. III above, we shall think of properties (subspaces of Hilbert space) as predicates, and the state as the subject, of logical or probabilistic statements. Also, because of the non-Boolean nature of the lattice of quantum properties, it is not possible to define logic universally over $\mathcal{L}^{\mathcal{H}}$, i.e. to satisfy the
truth table relations
 \eqref{tfbool} simultaneously for all pairs $\{A,B\}$ of properties drawn from  $\mathcal{L}^{\mathcal{H}}$. This is why we do not favor the term quantum logic for the operations in Eqs.\eqref{qops}. We prefer the term \lq property calculus'. For a \emph{Boolean} lattice, on the other hand (one satisfying the distributive law), the q-operations  \eqref{qops} are equivalent to the operations \eqref{setlog} that lead to the standard relations \eqref{clogic} of classical logic.

We now ask whether one can find a truth function $\mathcal{T}(A)$, defined over all of
$\mathcal{L}^{\mathcal{H}}$, but required to satisfy the truth table relations
only for pairs $\{A,B\}$ that belong to the same Boolean sublattice. Such a function, uniquely defined over all of $\mathcal{L}^{\mathcal{H}}$, is said to be \emph{noncontextual}. It turns out that this is impossible for a Hilbert space with dimension $D\geq 3$, as first shown by \textcite{bell1966}. A striking counterexample was given independently by \textcite{koc}, and a simpler one for $D\geq 4$ was found by \textcite{mermin}.  These results are often referred to as \lq no-go theorems'.

We therefore turn to probability, and follow \textcite{grif1} in taking the term \lq probability function' in the classical sense as referring to a \emph{sample space} -- that is, a classical set whose subsets, called \lq events', are the arguments of the probability function (see Appendix A). An essential concept in constructing probability functions for QM is that of \emph{compatible} properties. Denoting by $[A]$ the projection operator associated with the property (subspace) $A$, we say that two properties $A$ and $B$ are compatible iff $[A]$ and $[B]$ commute. Since there is a one to one correspondence between subspaces and their associated projection operators, we shall freely refer to properties in terms of either concept. We also use the terms \lq subspace' and \lq property' interchangeably.

 Let us call a sublattice $\mathcal{L}$ \emph{internally compatible} if all of its properties are compatible with one another. On the other hand, we call two lattices of subspaces  \emph{mutually compatible} if each subspace of one is compatible with each subspace of the other.  The two notions are independent; in particular, one can easily construct two mutually compatible lattices, neither of which is internally compatible.  We shall use the simpler term \lq compatible' for either notion if we think the meaning is clear.

As explained more fully in Appendix B, the definition of a probability function of quantum properties begins by defining a \emph{sample space} $\mathcal{S}$ as any set of subspaces of Hilbert space that are mutually orthogonal (each vector of one is orthogonal to each vector of the other) and complete (they span the full Hilbert space $\mathcal{H}$). Next, we define the all-important concept of a (static) \emph{framework}, $\mathcal{E_S}$, as the so-called closure of the sample space, namely the set of subspaces obtained from $\mathcal{S}$ by linear combinations of its projectors with coefficients $0$ and $1$. In Appendix B we show that a framework is both an algebra of projection operators and a Boolean lattice of subspaces.  Following Griffiths we sometimes refer to the framework $\mathcal{E_S}$ as an
\lq event algebra', to its properties as \lq events', and to the members of the sample space $\mathcal{S}$ as \lq elementary' events or properties (also known as \lq atoms') of the framework. When we wish to call attention to the framework as a lattice we shall use the notation $\mathcal{L_S}$, but this is the same object as $\mathcal{E_S}$. Note that a given property can be an event in different incompatible frameworks, since it can be obtained from a variety of different sample spaces by algebraic closure (see Appendix B).

To construct a probability function we thus consider a sample space $\mathcal{S}$ and the associated framework $\mathcal{E_S}=\mathcal{L_S}$ obtained by lattice closure. We then need only assign to each member of $\mathcal{S}$ (i.e., each atom of $\mathcal{L_S}$) a real nonnegative probability so that the sum over all the atoms is 1.  The probability of any \emph{set} of atoms, i.e. of any event, is then the sum of the individual atomic probabilities, as given in \eqref{kolminf}. This probability function is closely tied to the chosen framework; we say that it is \emph{contextual} to that framework, and denote it as $\mathcal{P}_{\mathcal{S}}$ or $\mathcal{P}_{\mathcal{E_S}}$ or $\mathcal{P}_{\mathcal{E}}$.

It is here that we shall diverge somewhat from the point of view of Griffiths.  Like him, we require that the theory be \emph{formulated} without reference to empirical findings.
He, however, introduces the Born rule as a fundamental precept of the theory, whereas we wish to derive it from another principle.   Our starting principle will be that the theory
must conform to classical logic as far as possible, given the Hilbert space ontology.  Since truth values cannot be assigned to all properties simultaneously because of no-go theorems (see \textcite{bell1966,koc}), we do the next-best thing by requiring that \emph{probability values} be assigned simultaneously to all properties - in other words, that probability values, unlike truth values, be \emph{noncontextual}.  The remainder of this section will be devoted to showing that this can be done only if the probability values satisfy the Born rule.

We do not wish to depart from the principle that a \emph{probability function} can be defined only within a particular static framework.  For this reason we used the term
\lq probability value' at the end of the previous paragraph.  By noncontextuality of probability values we mean that the \emph{value} of a probability function in one framework is equal to that of a \emph{different} probability function defined in a different framework when both functions are evaluated for the same property, if that property belongs to both frameworks.  This would seem obviously true if we were working with a classical system, and we
postulate that it is still true in quantum physics, namely that given two mutually incompatible frameworks $\mathcal{E}$ and $\mathcal{E'}$ with at least one property $A$ in common, the probability functions $\mathcal{P_E}$ and $\mathcal{P'_{E'}}$ satisfy the relation
\begin{equation}\label{noncon}
\mathcal{P_E}(A)=\mathcal{P'_{E'}}(A), \;\;\;\forall\;A\;\in\;\mathcal{E}\cap\mathcal{E'}.
\end{equation}

A lengthy argument presented in Appendix B demonstrates that if $\rho$ is any unit-trace nonnegative operator (such an operator is called a \emph{density matrix}), then noncontextual probability values can be found for the full lattice $\mathcal{L^H}$ of quantum properties, based on the \emph{normalized lattice measure}
\begin{equation} \label{latmeas}
\mathcal{W}_{\rho}(A)=\text{Tr}(\rho[A]).
\end{equation}

\noindent The probability function $\mathcal{P}_{\rho,\mathcal{E}}$ associated with the framework $\mathcal{E}$ can be defined as the \emph{restriction} of $\mathcal{W}_{\rho}$ to that framework, considered as a sublattice of $\mathcal{L^H}$:
\begin{equation}\label{pdef}
\mathcal{P}_{\rho,\mathcal{E}}(A)=\mathcal{W}_{\rho}(A)=\text{Tr}(\rho[A]),\;\;\;\text{for}\;\;A\in\mathcal{E}.
\end{equation}

\noindent It is easily verified that the function $\mathcal{P}_{\rho,\mathcal{E}}$ defined by Eq.\eqref{pdef} satisfies the Kolmogorov relations \eqref{kolminf},\eqref{probmeas}, which imply the crucial \lq overlap equation' \eqref{kolm2}, for all properties belonging to the framework $\mathcal{E}$. It is thus appropriate to refer to $\mathcal{P}_{\rho,\mathcal{E}}$ as a probability function. The noncontextuality of probability values then follows immediately since such values are all derived from a single measure $\mathcal{W}_{\rho}$ defined over the whole lattice $\mathcal{L^H}$.

Gleason's theorem is the highly nontrivial statement that in a Hilbert space with dimension $>2$, \emph{any} normalized lattice measure on the lattice $\mathcal{L}^{\mathcal{H}}$ has the form \eqref{latmeas}, for some density matrix $\rho$. We take \eqref{latmeas} and \eqref{pdef} as consequences of Gleason's Theorem, rather than \emph{postulating} them from the outset.

From the above argument it follows that we must generalize the assumption of Sec.A above that states are represented by rays of vectors, and assume that they are represented by density matrices $\rho$, otherwise known as \lq mixed states'. A ray $|\psi\rangle$ then corresponds to the special case of a \lq pure state', where $\rho$ has a single nonzero eigenvalue and is given by $\rho = |\psi\rangle\langle\psi|$. The analogue of the classical statement in Sec. II.B above, that the state $x$ is the \lq source of truth' for properties, is the quantum statement that a pure state $|\psi\rangle$ defines a property through its projector $[\psi] = |\psi\rangle\langle\psi|$, whose probability is 1, i.e. a true property.  Substituting
$|\psi\rangle\langle\psi|$ for $\rho$ in \eqref{pdef}, we obtain the pure-state Born formula


\begin{equation}\label{purebornprob}
\mathcal{P_E}(A)= \mathcal{W}_{\psi}(A)=\langle\psi |[A]|\psi\rangle, \;\;\;\text{for} A \in \mathcal{E}.
\end{equation}
\noindent We thus see that in QM the state (mixed or pure) is in general the \lq source of probability', and for pure states it is the source of truth and falsehood for some properties and the source of probability for others.

In Griffiths's presentations, the Born rule is considered to be fundamental, and
the fact that it is noncontextual is a byproduct. In establishing the microscopic
theory of CQT as a deductive system, we take the noncontextuality of probabilities as
a first principle and the microscopic Born rule \eqref{pdef} is then
derived  by the use of Gleason's Theorem, for some density matrix $\rho$. Similar arguments have been put forward by other authors in the past, e.g. \textcite{cassinello}.

As stated above, a quantum property $A$ is associated with a subspace of the Hilbert space, or equivalently with the projection operator $[A]$ onto that subspace. The state
$\rho$ confers an intrinsic noncontextual probability value on $A$ via the Born rule \eqref{pdef}. In many formulations of QM (e.g. \textcite{david2012}, Sec. 4.2.1), what we have called a property is referred to as \lq an ideal projective measurement', since it has a probability value associated with it. We consider this language misleading and we shall reserve the term \lq measurement' for macroscopic (irreversible) operations employing a classical measuring device (see below).

Let us now consider probability from an intuitive point of view.  As discussed in Appendix A, when we think of probability we necessarily think of truth and truth values.  We think that the probability of a proposition is the likelihood (in some sense) of its being true.  How far can this intuition be maintained in CQT? The answer is given by the \lq Single Framework Rule' of Griffiths, which we restate as follows:

\noindent a) With respect to truth values, they are in general (that is apart from special properties having $\psi$ as eigenvector) unknowable, but if they are postulated to exist then one must
not assume that a property belonging to two incompatible frameworks has the same truth value in both.

\noindent b) With respect to probabilities, the probability value assigned to a property is independent of framework (assuming a fixed state),
but a full probability function can be defined only within a single framework.

In our formulation, the single framework rule applies to propositions concerning the truth or falsehood of a property, or concerning probability functions and their domains. These can only be combined by logical connectives drawn from the same framework: they are contextual. Propositions about the value of a probability function applied to a property, on the other hand, are able to transcend the barrier separating one framework from another: they are noncontextual.

As emphasized by Griffiths, the existence of mutually incompatible frameworks is the single most salient feature of the microscopic theory that distinguishes quantum mechanics from classical mechanics. Before exploring its consequences  we shall expand  the treatment to take into account the time dependence of quantum states and/or properties.

\subsection{ \label{C}  	Generalization to dynamics:  families and frameworks}

1. Extension from static theory: histories and families

In the static description given in the previous subsection, a framework is a complete compatible (Boolean) sublattice of $\mathcal{L}^{\mathcal{H}}$. It can be regarded as an interrogation, and a basis element (one of the members of the sample space that generate the sublattice) can be regarded as one of a complete set of mutually exclusive answers to the interrogation.  In CQT the interrogation is not a measurement but a proposed point of view, represented by a projector, in which the system must have one of the elementary properties corresponding to the basis elements, also known as atoms of the sublattice (these need not be one-dimensional subspaces). We may say that this is the elementary property selected to be \lq true' provided that we remember that \lq truth' is contextual to the particular framework.

In the dynamic description, a property is generalized to a \emph{history}, and a static framework is generalized to what we shall call a \emph{family} or candidate framework.  A family is generated by its elementary histories, defined as follows:  At each of $N$ times $t_1,...t_N$, a one-time sample space is identified, having $m_n$ members.  The sample space at time $t_n$ is the set $\{A_n\} = \{A_n^{1},...,A_n^{m_n}\}$.   These $N$ one-time sample spaces (we may call them \emph{static event spaces}) determine the family $\mathcal{F}_N$.  The elementary histories of the family are sequences of $N$ one-time properties, in which at time $t_n$ the projector $A_n^{j_n}$ has been selected from the set
$\{A_n\}$.  A particular elementary history is defined by the indices $j_1,...,j_N$, and can be notated as
\begin{equation}\label{elem}
C_N^j = (A_1^{j_1}, A_2^{j_2},..., A_N^{j_N}),
\end{equation}

\noindent where the superscript $j$ on the left side stands for the sequence $j_1,...,j_N$.  If all the $m_n$ ($j_n$ runs from 1 to $m_n$) are finite, the number of elementary histories is $\Pi_n^N{m_n}$.  (We are assuming no branch dependence, see end of Appendix C.1.)

We also introduce the \emph{dynamic event space} of the family $\mathcal{F}_N$.  It is generated from its elementary members (i.e., the elementary histories) by linear combination in the way described in Appendix C.1.  This produces $2^{\Pi_n^N{m_n}}$ \emph{dynamic events} in all, of which $\Pi_n^N{m_n}$ are elementary events and one is the \lq zero' event.  The dynamic events are subspaces in the tensor space $\mathcal{H}^N$, made up of $N$ copies of $\mathcal{H}$, not subspaces in $\mathcal{H}$.

An important concept is that of a \emph{homogeneous} event \cite{isham1994}.  Among dynamic events of $\mathcal{F}_N$, the homogeneous ones are those that can be written
as histories
\begin{equation}\label{Chom}
C_N = (B_1, B_2,..., B_N)
\end{equation}

\noindent where each $B_n$ is a static event at time $t_n$ - that is, $[B_n]$ is some linear combination of the members of $\{[A_n]\}$, with coefficients 1 or 0.  All elementary events are clearly homogeneous, but the converse is not so because $B_n$ is not necessarily elementary in the static event space at time $t_n$.  There are  $\Pi_n 2^{m_n} = 2^{\Sigma_n m_n}$ homogeneous events, but only $\Pi_n{m_n}$ elementary ones.  But there are in general many inhomogeneous events since the number of dynamic events is $2^{\Pi_n m_n}$.  We shall call any homogeneous event a history, but inhomogeneous events will be called \emph{history complexes}.

To review the terminology introduced above: an elementary history is the same as an elementary event, but the meaning of \lq elementary' depends on
the prior identification of a family to which the history/event belongs.  On the other hand, homogeneous events are histories and inhomogeneous events are history complexes; the meaning of the terms \lq history' and \lq history complex' is independent of any family containing the object.

As explained below and more fully in Appendix C.4, only families whose members obey an additional \emph{consistency} or \emph{decoherence condition}, dependent on the state, lead to a proper definition of probability and thus lead to \emph{frameworks}.

2. Physical time development between interrogation times

As mentioned above, members of $\mathcal{F_N}$ can also be considered to be subspaces in the tensor product Hilbert space
$\mathcal{H}^N$, where  the elementary histories  such as \eqref{elem} form a \emph{history sample space}. That is, they are all mutually orthogonal and together they span
$\mathcal{H}^N$.  Being mutually orthogonal, they are mutually
compatible and therefore $\mathcal{F}_N$ is internally compatible
as well as complete, as a sublattice of $\mathcal{L}^{\mathcal{H}^N}$.

We will say that two elementary histories having the same projection times $t_1,...,t_N$ are (mutually) compatible if for each $n$ the $n$th property of one
history is compatible with the $n$th property of the other. Then by construction all the elementary histories of the family are mutually compatible.   But this means that
the elementary histories are all mutually compatible as subspaces in $\mathcal{H}^N$.  From this it follows that all the histories and history complexes - that is, all dynamic events - in the family are mutually compatible as subspaces in $\mathcal{H}^N$.

Repeating the reasoning developed in Appendix B.4 for static events, we can define probability functions $\mathcal{P}_N(C_N)$ applicable to all histories and history complexes, that is to all dynamic events,  in the family $\mathcal{F}_N$,  considered as subspaces of $\mathcal{H}^N$. Such a function satisfies the relation analogous to \eqref{mackinf}
\begin{equation}\label{csum}
\mathcal{P}_N(C_N \vee_q C_N' \vee_q C_N'' +...) = \mathcal{P}_N(C_N) + \mathcal{P}_N(C_N')
+ \mathcal{P}_N(C_N'') + ....,
\end{equation}

\noindent where the q-operations on the lhs of \eqref{csum} refer to the property calculus relevant to $\mathcal{H}^N$.

If, however, we wish to relate the probabilities of histories to the state, as was done in the static case, then we must take into account the time development of the state, and must change our perspective from regarding the successive bases $\{A_1\}$, $\{A_2\}$, ..., $\{A_N\}$ as belonging to \emph{separate} copies of
$\mathcal{H}$, to seeing them as existing together in the \emph{same} Hilbert space, so that the relationship
of $A_n$ to $A_{n+1}$ in the same history can be treated algebraically.  For this purpose it will be helpful to think of a history $C_N$ as built up step by step out of its initial \emph{subhistories} $C_1$, $C_2$,...,$C_{N-1}$, rather than coming into being all at once. This is accomplished by introducing the Heisenberg projectors
\begin{equation}\label{adeftext}
\bar{A}_n \equiv U(t_n,t_0)^{-1} [A_n] U(t_n,t_0),
\end{equation}
where
\begin{equation} \label{udeftext}
U(t,t') = \text{exp}[-iH(t-t')],
\end{equation}
\noindent ($H$ is the Hamiltonian), as well as the \emph{chain operator}
\begin{equation}\label{chainH}
\hat{C}_N = \bar{A}_N \bar{A}_{N-1} ... \bar{A}_2 \bar{A}_1,
\end{equation}
\noindent an operator in $\mathcal{H}$, in contrast to $C_N$, which generates an operator in $\mathcal{H}^N$.

3. Probabilities and the Born rule

In order to generalize the Born rule to histories we first note that since $[A]$ is a self-adjoint projection operator, the linear relation \eqref{pdef} in the static case can be rewritten as

\begin{equation} \label{quadtext}
\mathcal{P}_{\rho,\mathcal{E}}(A)=\text{Tr}(\rho [A]^{\dagger}[A]) \;\; \text{for} \;\;A\in \mathcal{E}.
\end{equation}
In the dynamic case it is the quadratic relation, Eq.\eqref{quadtext}, that provides the generalization of the probability formula: specifically,
given the pure state $|\psi_0\rangle$ at $t=t_0$, Griffiths's rule for the probability of an elementary history is
\begin{equation}\label{chprob}
\mathcal{P}_{\psi_0,\mathcal{E}}(C_N) = \langle\psi_0|\hat{C}_N^{\dagger}\hat{C}_N|\psi_0\rangle \;\; \text{for}\;\; C_N \in \mathcal{E}.
\end{equation}

\noindent Here again, this quadratic expression was simply postulated by Griffiths as a natural generalization
of the Born rule \eqref{purebornprob}.  We would like instead to derive it from Gleason's Theorem and a noncontextuality argument, as we did in the static case.  The trouble is that $\hat{C}_N$, though an operator on $\mathcal{H}$, is not
a projector, and Gleason's Theorem involves a linear function on projectors, Eq.\eqref{latmeas}.  Following \textcite{nist}, this obstacle is overcome by considering the \emph{conditional probabilities} that take us from $C_n$ to $C_{n+1}$.  By requiring that these conditional probabilities themselves be noncontextual, one can express them in terms of probability measures on single projectors and so apply Gleason's Theorem.  An argument of some complexity developed in Appendix C.3 then leads to the conclusion that \eqref{chprob} is indeed the only possible probability formula.
In particular, competing formulas such as that of  \textcite{GP},
\begin{equation}\label{GPprob}
\mathcal{P}_{\psi_0}(C_N) = \text{Re} \langle\psi_0|\hat{C}_N|\psi_0\rangle,
\end{equation}

\noindent are ruled out.

We note that our argument does not involve any generalization of Gleason's theorem itself, only a varied \emph{application} of the original theorem, sometimes to probability measures on $\mathcal{H}$ and sometimes on $\mathcal{H}^N$.  In contrast,  \textcite{isham1994} prove a new theorem applicable directly to histories. Although the proof of that theorem requires an impressive amount of care and ingenuity and the result may have significant value for some purposes, the theorem yields, for our purposes, a result \emph{weaker} than the one derived in our Appendix C, in that to obtain our result \textcite{isham1994} would need to \emph{assume} that probabilities are given by a quadratic expression of the form  $\mathcal{D}(C_N,C_N^{\dagger})$,  where $\mathcal{D}(C_N,{C'}_N^{\dagger})$ is a bilinear functional on \emph{two} histories called a \lq decoherence functional', and only then, using their theorem, could they deduce that this functional must be such as to yield Eq. \eqref{chprob} for the probabilities.

Something similar may be said of  \textcite{sorkin1994}, who shows that classical and quantum mechanics arise from the first and second, respectively, of a hierarchy of elegantly connected equations that may be postulated to relate the probabilities of histories. Sorkin shows that the second of these equations, applied to Hilbert space, implies \eqref{chprob}, and also that this equation is equivalent to the assumption that probabilities arise from a decoherence functional.  But he provides no justification for the decoherence functional assumption itself. Thus, within the scope of the present paper,  the argument for \eqref{chprob} given in \textcite{sorkin1994} is no stronger than that of \textcite{isham1994}.

On the other hand, it should be said that the above remarks do not do justice to the far-reaching vision of either  Sorkin or Isham \emph{et al.}, both of which seek generalizations of nonrelativistic QM that go beyond the scope of Hilbert space or even of a description in terms of Euclidean space-time. The same can be said for the program of Gell-Mann and Hartle.

4. Families, frameworks and the consistency conditions

In the static case any orthogonal decomposition of the identity yields a Boolean sublattice and consequently a framework. For dynamics, on the other hand, although we were able to derive Eq.\eqref{chprob} and rule out Eq.\eqref{GPprob}, it turns out that an additional condition must be imposed in order to avoid contradictions. Specifically, consider a history belonging to two \emph{different} families. Then, as demonstrated in Appendix C.4, by requiring the noncontextuality of probability values for such histories, we are led to a necessary \emph{consistency condition}

\begin{equation}\label{nointerf}\
\text{Re}(\mathcal{D}(C_N,C'_N))=\text{Re}(\langle\psi_0|\hat{C}_N'^{\dagger}\hat{C}_N|\psi_0\rangle) = 0,
\end{equation}

\noindent where $\hat{C}_N$, $\hat{C}_N'$ refer to any two distinct elementary histories in the family, and $\mathcal{D}(C_N,C'_N)$ is the \lq decoherence functional' whose diagonal value is given in \eqref{chprob}.  Equation \eqref{nointerf} is called the \emph{weak decoherence} condition.

One may easily suspect that this condition, involving only the real part of the decoherence functional, lacks robustness in some respect, so that it should be
replaced by \emph{medium decoherence}

\begin{equation}\label{griffmedcc}
\mathcal{D}(C_N,C'_N)=\langle\psi_0|\hat{C}_N'^{\dagger}\hat{C}_N|\psi_0\rangle = 0.
\end{equation}

\noindent Such a suspicion may account  for the preference that \textcite{grif1} expresses for medium decoherence.

In fact, \textcite{diosi2004} has justified this suspicion in a convincing way by considering a system consisting of two subsystems with no entanglement between them, so that the decoherence functional of the whole system is a product of  \lq partial decoherence functionals' of the two subsystems.  Then medium decoherence of the full decoherence functional implies medium decoherence of the partial decoherence functionals,
but the same does not hold for weak decoherence since Re($\mathcal{D}_1$) Re($\mathcal{D}_2$) is not the same as Re($\mathcal{D}_1\mathcal{D}_2$). We shall therefore consider medium decoherence \eqref{griffmedcc} to be the applicable consistency condition.

It is important to remember that the consistency conditions involve the state $\rho$ or $\psi$, in addition to the family. We will call a pair $\mathcal{E}=(\psi,\mathcal{F}_N)$, consisting of  a state and a family a \emph{candidate framework}, and one whose elementary histories satisfy Eq.\eqref{griffmedcc}, simply a \emph{framework} (we continue to use the notation $\mathcal{E}$ for frameworks in the dynamic case, since they still constitute an event algebra). Then the Single Framework Rule discussed at the end of the previous subsection holds when histories are substituted for properties. Note that in the static case families are replaced by Boolean sublattices and every \lq candidate framework' satisfies the consistency conditions, so the distinction between candidate framework and framework was unnecessary. The noncontextuality of static probability values, on the other hand, only holds for a \emph{fixed} state. Thus if we think of a static framework as also  including both the state and the Boolean sublattice, then the noncontextuality of static probability values refers to \emph{conditional} probabilities, conditioned upon the state.

Another important point is that \emph{single-time} histories ($N=1$) can be shown (Appendix C.3) to be directly mapped onto the static theory, so they automatically satisfy the consistency condition \eqref{griffmedcc} for any state $\psi$ or $\rho$, even though they involve two times, $t=t_0$ at which the state is defined and $t=t_1$ at which the probabilities are evaluated. This means that the simpler static theory, without the need for consistency conditions, is actually of great relevance in practice, since single-time histories cover the most common applications of QM.

5. Two-slit diffraction

One familiar application of the consistency conditions is seen in two-slit diffraction.  Suppose that at time $t_0$ a particle is in a superposition of two states, one pertaining to the right-hand slit and the other to the left-hand slit.  At $t_1$ a decomposition of the identity is imposed, in which one projector selects the right-hand slit and the other
the left-hand slit. At $t_2$ the particle reaches the interference zone, and a new decomposition is chosen in which different projectors select different points in the zone.  One then finds two elementary histories $C_N$, $C_N'$, both selecting the same point $x$ at time $t_2$, but making different selections (right or left) at $t_1$.  One can also construct a history, $C_N''$ say, by adding together the two projectors (not their probabilities) at $t_1$.  This history belongs to the family under consideration but it is not an elementary member of it.  Nevertheless (see Appendix C.4) there is good reason to apply to it the
formula \eqref{chprob}.  This application yields the correct probability for the particle to pass through the point $x$ at time $t_2$, that is, the probability according to the interference pattern.  This probability contradicts the one calculated by addition from the separate
(right and left) probabilities, and so the consistency condition is not satisfied; the family containing $C_N, C_N'$  and $C_N''$ is not
a framework.  The contradiction is easily traced to the failure of \eqref{nointerf}. In fact the histories $C_N$ and $C_N'$ can be shown to be what are termed \lq intrinsically inconsistent histories' in Subsection 11.8 of \textcite{grif1}.

The above discussion should not be taken to mean that the two-slit experiment does not admit a description in CQT.  For the simplest two-slit experiment the appropriate framework contains the history $C_N''$, but not the other two histories. In order to discuss the probability of passage through one or the other of the two slits, the physical setup considered must be more elaborate than the one considered up to now: either the state at $t_1$ is not a coherent superposition, or additional degrees of freedom (e.g. the flip of a spin) must be added to the system at the slits, to mark the passage of the particle at $t_2$. In that case one can define different (incompatible) frameworks, one to describe the passage through the slits and the other to describe the interference pattern.

6. The work of Chisholm, Sudarshan and Jordan

As elucidated in a little known paper by \textcite{chisolm1996}, the consistency conditions put severe restrictions on many-time frameworks, so that by far the majority of imaginable families are ruled out.  This may give the impression that the CQT description of Nature is not rich enough to be interesting.  The results of Chishom \emph{et al.}, however, may be put in better perspective by reexamining the static case discussed in Subsection B above.

A static description has two components: the state and the sample space of properties (subspaces).  The two need have no relation to each other; any state can be combined with any sample space.
However, the sample space itself is severely restricted in that its members form a complete orthogonal decomposition of the identity.  This means that if a set of subspaces of $\mathcal{H}$ is chosen at random, it is almost surely not an admissible sample space. Yet this \lq impoverishment' is only what we expect when we seek a domain of interrogation within which one can operate with classical logic and define classical probability functions.  Necessarily, the richness of quantum physics will not appear in a single domain of this kind, but will be lodged in the availability of multiple domains each of which is a classical sample space by itself, but which are not compatible with one another. When both the state and the sample space have been specified, the nontrivial Born probabilities emerge from the degree of mismatch between the two.

Now consider a family of histories of length $N=1$.  We have a sample space imposed at $t_1$, preceded at $t_0$ by an 'initial state'.  As mentioned above, this situation is mathematically equivalent to the static situation (see Appendix C.3, second paragraph) and leads to the same Born probabilities.

If $N = 2$, one can construct inconsistent families such as exemplified in the double slit experiment, where the initial state is assigned to time $t_0$.  If $N= 3$, the inconsistency can be lodged in the projections chosen at $t_1$, $t_2$,
$t_3$ so that it is independent of the initial state.  What Chisholm \emph{et al.} show is that all inconsistencies at multiple times are due to the occurrence of this \lq double slit configuration' at some triplet of times not necessarily consecutive.  To put it baldly,
one cannot evade quantum incompatibility by distributing the inconsistent information through multiple times.  But this fact is already understood, and it arises from Nature itself, not from a limitation of CQT.

The result of Chisholm \emph{et al.} amounts to the statement that a framework with medium decoherence must consist of two parts and no more.  (We pass over the more complicated result for weak decoherence.) The earlier part consists of projection operators compatible with the state.  The later part consists of projection operators compatible with the final projections at $t_N$. A framework must not contain a projector at an intermediate time that is compatible neither with what precedes it nor with what follows it. Such a projector would activate the double-slit inconsistency. Thus the consideration of many times does not make the world of frameworks essentially richer than it is in the static case, except in the following respects:

\noindent (a) The transition from \lq earlier' to \lq later' may take place at different times in different parts of the Hilbert space.

\noindent (b) Within one part, if one allows projectors to be more than 1-dimensional, there can be unlimited topological complexity due to successive refinements and coarsenings, provided that there exists a maximal refinement (not necessarily present at any one time) that is a refinement of all the projectors at various times in that part.

\noindent (c) This picture is not essentially altered by branch dependence.

We see, therefore, that CQT does not attempt to evade or do away with quantum incompatibility, any more than it attempts to do away with the \lq fundamental conundrum' discussed in Section III for a single time.  These are features of the quantum world. CQT spells out just how far we can go in applying classical thought to a world possessing these features, using the Hilbert space ontology.

\subsection{\label{D} c-assertions and \lq physical' assertions}

We view the above theory as the full microscopic formulation of QM with its multiplicity of
frameworks, within each of which the theory makes assertions about probabilities and
permits the supposition of mostly unknown c-truths, not constant across incompatible frameworks, where the prefix c is meant to signify \lq contextual'.  This theory has been
summarized by \textcite{grifconsistent} with the slogan \emph{Liberty, Equality, Incompatibility!}  to highlight the presence
of incompatible frameworks of equal status and the freedom to choose among them. We refer to this freedom as \lq microscopic framework symmetry'. It is important to note at this stage that the microscopic theory does not possess a unique concept of truth which we might refer to as \lq physical' truth. It is simply not part of the
theory.  We thus consider the microscopic theory to be \emph{physically indeterminate} and hence \emph{incomplete}, even though it is logically coherent.

In order to arrive at a notion of physical truth we must therefore find a mechanism for \emph{selecting} a physical framework from the set of equivalent frameworks of the microscopic theory, a process that necessarily involves \emph{macroscopic} concepts. Every so-called \lq interpretation' of quantum mechanics corresponds in some sense to a different physical selection mechanism. We shall discuss these in the next section, but we can already anticipate that depending on the physical situation or on the question asked, the answer might not be unique.

As mentioned earlier, in distinguishing between the \lq microscopic' and \lq macroscopic' theories we do not mean the difference between small systems (with few degrees of freedom) and large (ideally infinite) systems. Instead, the microscopic theory applies to all systems, large and small, whereas the macroscopic theory \emph{requires} the system to be large in order to define certain concepts, e.g. measurement.

\subsection{\label{E} Framework Selection Mechanisms}

In discussing framework selection mechanisms we shall distinguish between what we call \lq external mechanisms', involving coupling the system \textbf{S} to another physical system external to it,  and \lq internal mechanisms'
that seek to identify special physical frameworks describing selected properties of the system \textbf{S} itself. The distinction reminds one of the difference between operationalist and realistic formulations of the theory, but it is different: here \emph{both} mechanisms are being discussed within the realistic formulation we call CQT, whereas in the operationalist (Copenhagen) view, measurements are what lends \lq reality' to properties.

1. External Mechanisms

Here we are dealing with a phenomenological approach since it posits the existence of a classical measurement apparatus without inquiring into its physical origin. The standard textbook approach (see, e.g. \textcite{ll6}) was initiated by \textcite{vN1932,vn}. Our description follows that of \textcite{grif1} and \textcite{o4}. We consider the system 
 \textbf{S} and couple it to a (classical) apparatus
 \textbf{M}, which is often accompanied by an environment 
 \textbf{E}, representing \lq the rest of the world'.

Let us consider the measurement of a physical quantity (an observable) represented by the Hermitian operator $\hat{A}$ in the Hilbert space of \textbf{S}. This operator defines a basis of orthogonal eigenstates $|A_i\rangle$, which for simplicity we take to be complete. This basis defines a family and hence a framework $\mathcal{E}_A$ in the Hilbert space of \textbf{S}. The discussion now proceeds in the same way as in the textbook accounts of quantum measurements. The \lq classical' apparatus \textbf{M} is physically coupled to the quantum system \textbf{S} and through the quantum dynamics the \lq pointer states' of \textbf{M} become entangled with the states $|A_i\rangle$, in such a way that each quantum state $|A_k\rangle$ is associated with a distinct pointer state $\textbf{M}_k$. The probability of obtaining the state $\textbf{M}_k$ in the experiment, from among all the different pointer states of the apparatus, is then the probability associated with the state $|A_k\rangle$ it is coupled to, which according to CQT is given by the Born weight $\mathcal{W}_{\psi}(A_k)$, for a system in the state $|\psi\rangle$. Measuring the full probability would involve repeating the experiment to obtain the appropriate statistics. The detailed argument leading to the above result can be called a \lq derivation of the macroscopic Born rule' \cite{hartle68,weinbergqm}. It \emph{assumes} the microscopic Born rule \eqref{purebornprob} and deduces from it a rule for measurement outcomes $\textbf{M}_k$.

We thus see that from the point of view of CQT the role of the external measurement is essentially to identify the framework $\mathcal{E}_A$ associated with the observable $\hat{A}$ and to link the members of that framework to the pointer states of the apparatus. In the ontology of CQT the operator $\hat{A}$
 is equivalent to the set of beables comprising its basis in the Hilbert space, together with the nonzero eigenvalues which do not matter for the purpose of determining the framework.  Frameworks of \textbf{S}  that are incompatible with $\mathcal{E}_A$, for example $\mathcal{E}_B$, say, will be  perturbed by the physical interaction of the system \textbf{S} with \textbf{M}, and the  properties $B_i$ of $\mathcal{E}_B$ will bear no simple relationship to the pointer  states of the apparatus \textbf{M}, which is specifically designed to couple to the observable $\hat{A}$ and  thus to choose the framework $\mathcal{E}_A$.

 One more remark is pertinent in describing the effect of the apparatus in the process of framework selection. We stated that the apparatus selects one among the multiplicity of frameworks of the system \textbf{S}, but this is only true in the simplest case. There are many possible physical interactions which one might call \lq measurements', which single out properties belonging to more than one framework of \textbf{S}, in which case one is still left with multiple sets of c-assertions even after the \lq measurement'. A classic case is the two-slit diffraction experiment discussed in the previous section, in which the detection of the particle at a particular point on the screen can belong either to a history in which the particle passed through one slit or the other, or to another history, belonging to an incompatible framework, in which the \lq particle' diffracted from both slits on its way to the detector. There are also system-apparatus interactions that fall short of being full measurements, often referred to as \lq weak measurements'. As emphasized by Griffiths, all of these cases can be analyzed within the Consistent Histories formulation, but the upshot is that one does not end up with what we would term a unique physical truth, without resort to some additional macroscopic criterion.

 2. Internal Mechanisms

 In the spirit of a fully realistic formulation of quantum mechanics we can ask how to select the physical framework, starting from the microscopic theory with no phenomenological assumption, a question that is explicitly considered in the \lq Decoherent Histories' formulation of CQT \cite{gh3, jh1} and also by \textcite{omnes1992}. A careful and detailed discussion of a simplified microscopic model of measurement has recently been provided by \textcite{bal}. In all of these approaches the microscopic system \textbf{S} is considered to be a subsystem of a larger quantum system $\textbf{S}'$, which the authors often consider to be the whole universe.

 In the Decoherent Histories formulation, see e.g. \textcite{jh1}, the histories of the larger system $\textbf{S}'$ are assembled, by a coarse-graining operation, into equivalence classes that can themselves be considered to be histories of $\textbf{S}'$. In a first step let us not treat the subsystem \textbf{S} as special and consider an appropriate set of coarse-grained histories of $\textbf{S}'$, consisting entirely of what Gell-Mann and Hartle refer to as \lq quasiclassical properties'. These histories constitute the \lq quasiclassical realm' and we shall refer to them as \lq classical histories', suppressing the prefix \lq quasi' which is implied. The above construction, which relies heavily on the physical effects of decoherence for macroscopic properties, justifies the statement that under appropriate circumstances classical mechanics \emph{emerges} from QM in the macroscopic limit.

 Having established the existence of the quasiclassical realm (or framework) and of classical histories, we are ready to discuss the internal (i.e. fundamental) mechanism for selecting a physical framework. Consider the quantum system \textbf{S} to be a subsystem of the macroscopic system $\textbf{S}'$ and now apply a coarse graining to the histories of $\textbf{S}'$ that leaves the properties of the subsystem \textbf{S} unchanged, averaging only over the properties of the complement $\tilde{\textbf{S}} = \textbf{S}'-\textbf{S}$. In this way one obtains what we may call \emph{physical histories}, histories whose properties are quantum in nature for the degrees of freedom of \textbf{S} and classical in nature for the degrees of freedom of the complement $\tilde{\textbf{S}}$.

 In particular, consider a simple case in which \textbf{S} is a spin-$\mbox{\textonehalf}$ prepared in the state $|S_z\rangle = + \mbox{\textonehalf}$ and the classical degrees of freedom describe the measurement of the $x$-component with a Stern-Gerlach apparatus with pointer states $\textbf{M}_x$. The analysis then proceeds analogously to the phenomenological case of an external selection mechanism considered above. The choice of measurement selects a particular framework of \textbf{S} whose properties are correlated to the pointer states $\textbf{M}_x$, say, to form a \emph{physical framework} of the full system $\textbf{S}'$, with a well-defined concept of \emph{physical truth}. A different choice of measurement (say the $y$-component) selects a different physical framework with pointer states $\textbf{M}_y$. Although the $x$ and $y$ frameworks are incompatible (they refer to different sample spaces), the physical truths associated with $\textbf{M}_x$ and $\textbf{M}_y$ respectively, can be unified by considering them to arise from \emph{conditional} probabilities, conditioned on pointing the apparatus in the $x$ or $y$ directions.

 It should be clear from the above highly condensed discussion of external and internal mechanisms of framework selection that there is no unique \lq physical' mechanism, since depending on the question asked and the physical circumstances, one could legitimately consider physical truth to be definable by different choices of physical framework. Moreover, the internal and external mechanisms should not be seen as mutually exclusive, but rather as different perspectives on the same question. What remains invariant in CQT are the fundamental tenets of the theory applicable to any closed system, i.e. the microscopic theory with its Hilbert space ontology and its multiple incompatible sets of histories, which are the starting point for defining a macroscopic physical framework selection mechanism. Note also that our notion of selection differs from the \lq set selection problem' defined by \textcite{kent} as the search for \emph{the} preferred framework.

 According to MIQM any closed quantum system obeys what one may call \lq framework symmetry', whereby no single framework yields truth values that are to be preferred over those of any other framework. It is only in MAQM that this symmetry is \emph{broken} and a particular framework acquires the characteristic we call \lq physical truth'. Such macroscopic symmetry breaking is standard in classical and quantum statistical mechanics, but here we encounter it in the very formulation of quantum theory for any system \textbf{S}, even for a single spin.

Let us comment briefly on the issue of state preparation alluded to in the preceding discussion. Just as in classical mechanics (see Sec. II above), the designation of the state of the system as an initial value in MIQM has a counterpart in MAQM in the physical operation of state preparation using a macroscopic apparatus. A precise quantum description of such a procedure has all of the same difficulties as a theory of measurement, but it can be carried out in analogous fashion. We note, in addition, that the preparation of a pure state for a given quantum system is at most an idealization, since interaction and entanglement with adjacent systems, or with the environment, will inevitably turn the initial pure state into a mixed state. Fortunately, the formulation of CQT is based on a general Born rule \eqref{pdef} based on the density matrix, rather than on pure states only.

 To conclude this description of CQT we note that the feature of quantum mechanics that distinguishes it from classical mechanics and makes it \lq weird' or \lq mysterious' has to do with its assertions, not with its ontology, although the former are a consequence of the latter. Contrary to the view expressed by \textcite{merint}, the histories formulation does not reinvent \lq reality', it reinvents \lq truth'.

 Figure 2 illustrates the CQT formulation, both the microscopic and macroscopic theories, in such a way as to emphasize the similarities and differences with classical mechanics depicted in Fig.1.

\begin{figure}
\begin{center}
\includegraphics[width=6in]{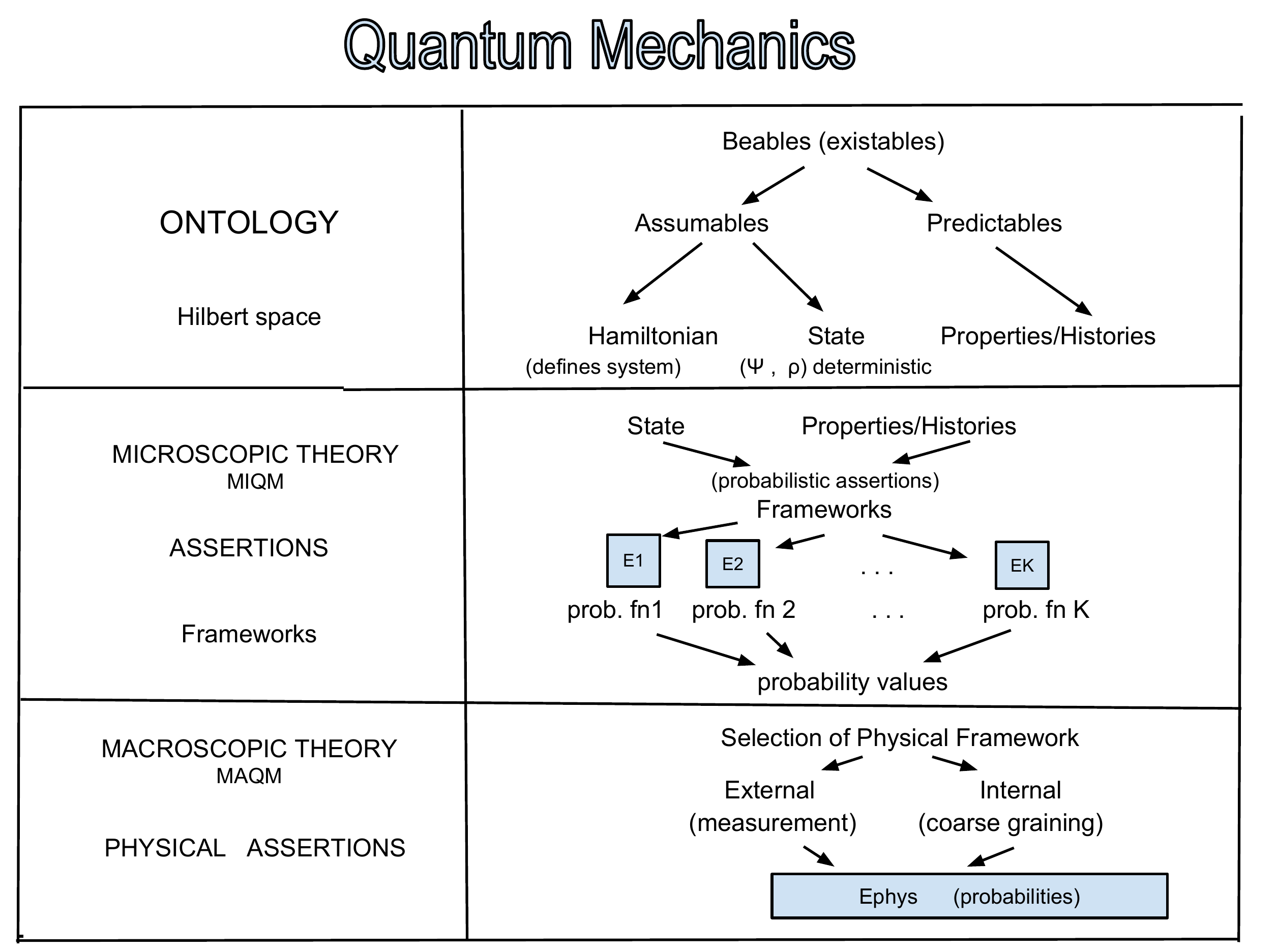}
\end{center}

\caption{Schematic representation of quantum mechanics. The top row describes the ontology, which consists of a Hamiltonian, states, properties and histories, all of which are objects in Hilbert space. The state is either a ray of vectors $\psi$ (pure), or a unit-trace operator $\rho$ (mixed), the properties are subspaces of Hilbert space and histories are sets of properties defined at a sequence of times. The assertions of the microscopic theory (MIQM, middle row) consist of probabilities of properties and histories, conferred by the state which is the source of probability. For a given state there are a multiplicity of contextual probability functions, each one associated with a subset of properties or histories, called a framework, denoted by $E_1, E_2,...E_K$ in the figure. In order to arrive at physical assertions, a macroscopic framework selection mechanism is required (MAQM, bottom row), which can proceed either phenomenologically via external measurements, or more fundamentally via coarse graining and decoherence.}

\end{figure}

\section { \label {5} Other Formulations and Interpretations}

With the expanded view of CQT presented above, many of the traditional formulations or interpretations of quantum mechanics can to some extent already be seen to be contained within the theory, rather than being considered as logical alternatives. Let us discuss the major interpretations in turn.

\subsection { Copenhagen Operationalism}

Consider a quantum system \textbf{S} prepared in a state $|\psi \rangle$ represented by a ray in Hilbert space. Since there is no way for the state to confer determinateness on an arbitrary property of the system, i.e. to label the property as true or false in an absolute sense, Bohr's operationalist point of view denies intrinsic determinateness to any quantum property. Instead, the Born weight is interpreted as the probability that the corresponding eigenvalue will be observed when a (classical) \emph{measurement} of a physical observable is carried out. Thus the property $A_i$, say, acquires determinateness (some would say \lq existence') only by virtue of the coupling to the measurement instrument, as described in the previous section. In Bohr's radical operationalism (which we refer to as \lq Copenhagen'), there is no quantum ontology. Quantum properties become physically real only when they are entangled with particular (classical) measurement instruments. A particularly cogent description of the operationalist point of view may be found in the textbook by \textcite{peres}.

The Copenhagen view has been criticized, most notably by John Bell, as being \lq unprofessionally vague and ambiguous' (see \textcite{b8}, p. 173), since it introduces the undefined notion of measurement into the basic formulation of the theory, and since it assumes the prior existence of classical mechanics (to describe the apparatus), whose origin and ontology are not explained.

While we agree that the formulation appears strangely incomplete, we consider it to be logically coherent. In fact, as indicated in the previous section, Copenhagen can be seen as the \emph{specialization} of CQT to the case of externally measured systems, leading to \lq recorded histories', a point which has been emphasized by Hartle (2011). Thus Copenhagen is included in CQT, but the reverse is not true. As mentioned above, moreover, the essential role of external measurements in the Copenhagen approach can be viewed as providing the selection of an appropriate physical framework, since the predictions themselves, in both CQT and Copenhagen, involve a Born weight which refers only to the system \textbf{S} and not to the apparatus. This, in our view, explains why the Copenhagen interpretation has been considered adequate to the majority of physicists over the eighty-year history of QM, as they investigate the vast richness of quantum phenomena without extensively probing its philosophical implications. There is a sense in which the move from Copenhagen to CQT is a short step, replacing the role of external measurement in choosing frameworks by a simple interrogation among a set of available possibilities. It is in this sense that we can quote \textcite{grif19} and refer to CQT as \lq\lq Copenhagen done right!". From our point of view, apart from the restriction to externally measured systems noted above, we agree with Bell that the main philosophical flaw in Bohr's operationalism is that measurement instruments and the classical world are posited at the outset without further elucidation and without explicit recognition that the instruments are after all made of atoms.

\underline{Quantum Information}

A more sophisticated, or at least more modern, version of operationalism is the information-theoretic approach, which considers quantum states as states of belief about the quantum system, see for example \textcite{bruk-zeil}, \textcite{bayes}, \textcite{qbism}. It is also an operationalist point of view, according to our definition, because the belief is held by observers or agents external to the quantum system. It too recognizes no quantum ontology.  One can consider the information-theoretic formulation, like Copenhagen, to be a specialization of CQT, so there is no need for proponents of CQT to \lq refute' it. The only part of its creed that CQT denies is the assertion that such an approach is \emph{necessary} because quantum states are \lq nothing but' states of belief.

\subsection{\label{B.}The Orthodox or Dirac-von Neumann-Born formulation}
Although the distinction is not often made, we shall follow \textcite{bub} in distinguishing between Bohr's strict operationalism on the one hand, and the \lq orthodox' or \lq textbook' version of quantum mechanics on the other hand, which we associate with Dirac, von Neumann and Born, see for example \textcite{ll6}. In contrast to what we have called Copenhagen, the orthodox approach does have a quantum ontology, in that the wave function $\psi$ is considered to represent the system \textbf{S}, and it selects the set of determinate properties as those properties $A$ for which the Born weight $W_{\psi}(A)$ is either unity (true) or zero (false), i.e. properties that are compatible with $[\psi]$.

 Any other property is indeterminate and therefore not part of the ontology. Indeterminate properties are given the same interpretation as in the Copenhagen approach, namely that the Born weight of an indeterminate property $A$ in the state $\psi$ is the probability that the corresponding eigenvalue will be observed when an external  measurement is made. The orthodox approach is thus seen to be a combination of realism (the dynamic state $\psi$ and properties compatible with it are \lq real') and operationalism (other properties are given their meaning by reference to measurements). As is well known, however, attempts to describe the measurement process dynamically within this theory encounter the notorious measurement problem (the one \textcite{grif3} refers to as the \lq first' measurement problem), which both Dirac and von Neumann resolve by introducing a projection postulate, also known as collapse of the wave function.

\subsection{\label{C.}The de Broglie-Bohm hidden-variables formulation}
This approach was introduced very early by \textcite{debroglie} and abandoned because of heavy criticism from Pauli and other defenders of the new quantum orthodoxy, and then rediscovered by \textcite{bohm15} as a concrete refutation of the prevailing view that no hidden-variable formulation was consistent with QM. This view was based on the combination of Gleason's Theorem and the Bell-Kochen-Specker Theorem referred to above, but as was later emphasized by \textcite{bell1966}, the application of the theorem depends on accepting the full algebraic structure of Hilbert space in the ontology and assertions of the theory, including the noncontextuality of the probability values associated with quantum properties.

We shall follow the practice of referring to this formulation as \lq Bohmian mechanics' (see \textcite{durr2013quantum} for a modern version). Its ontology consists of a set of $N$ \lq particles', with coordinates $\{Q_i\}$ in three-dimensional space, evolving in time. In addition, the ontology can also be considered to comprise the wave function
$\psi(Q_1,...,Q_{3N})=\psi(\{Q_i\})$
 whose role is two-fold: it tells the particles how to move and it determines the initial statistical distribution of the ensemble of particles with density

\begin{equation}
\rho(\{Q_i\},t_0) = |\psi(\{Q_i\}, t_0)|^2,
\end{equation}

\noindent a relation that is preserved in time if it is satisfied at the initial time $t_0$ and if $\psi$ satisfies the Schroedinger equation.

It is important to emphasize that no other part of the Hilbert space structure belongs to the ontology of Bohmian mechanics. For example, the energy, spin, or momentum are derived quantities inherited from classical mechanics and they are given an operationalist interpretation in terms of external measurements. Even though one can define an intrinsic momentum variable in terms of the beables of the theory, namely the $\{Q_i\}$ and $\psi$, a measurement of momentum will be unrelated to this quantity; instead, it will be related to the eigenvalues of the corresponding operator. Indeed, as noted by Bohm himself \cite{bohm2}, a free particle confined to a box of width $L$ can be in a state of energy $E=(1/2m)(nh/L)^2$ for some integer $n$, even though it is at rest $(p=0)$. A measurement of momentum, however, will disturb the state and yield the values $p=\pm nh/L$, given by the Born rule, as in the Copenhagen approach.

This combination of realism and operationalism seems to us to be a major weakness of Bohmian mechanics, no less than the well-known feature of nonlocal dynamics. Actually, we see no physical basis for either feature. In principle, an equivalent formulation could be constructed with momenta $\{P_i\}$ as the fundamental beables, or indeed with eigenstates of any hermitian operator (see below). Thus, although Bohmiam mechanics does contain a microscopic theory based on the ontology of particle coordinates, together with a guiding wavefunction, this theory is so impoverished, lacking as it does quantities like momentum, energy, spin, as to be physically useless without the operationalism that restores the latter quantities.

\textcite{b8} (p. 77) has drawn an interesting analogy between Bohmian mechanics and a constructive theory of \textcite{lorentz} that retains the aether and explains the experimental consequences of Lorentz transformations through a modification of the dynamics (see also \textcite{bub}). This contrasts with Einstein's reformulation of the basic geometry of space-time that incorporates Lorentz invariance at the outset. In a similar way, Bohmian mechanics retains some of the classical ontology of local beables, and relies on nonlocal dynamics to obtain agreement with experiment, whereas QM, at least in its CQT formulation, changes the geometry of phase space and then draws the necessary epistemological consequences.

The nonlocal character of Bohmian mechanics led John Bell to ask whether \emph{any} theory agreeing with the predictions of QM must be nonlocal. His celebrated theorem \cite{bell14} states that a theory satisfying a condition he termed \lq local causality' (see \textcite{b8}, p.232) must violate QM, a result which has led to the frequent statement that \lq QM is nonlocal' (e.g. \textcite{goldstein2011bell}). It is important to note, however, that Bell's condition of local causality can be \emph{formulated} only within a classical ontology. In the Hilbert-space ontology the probability functions appearing in Bell's local causality condition are undefined. We therefore suggest that the proper conclusion to draw from Bell's theorem is that QM violates \lq classical locality', but preserves a notion of \lq quantum locality', referred to by \textcite{griffiths2011quantum}, Sec. 6,  as \lq Einstein locality'.

\subsection{\label{D.}	Modal Formulations}
A natural generalization of Bohmian mechanics consists in selecting an arbitrary hermitian operator (or \lq observable') $\hat{R}$ and designating it as determinate, given the state $\psi$. This variable defines at any time a determinate sublattice $\mathcal{D}(\psi,\hat{R})$ which is Boolean, in contrast to the full non-Boolean lattice of q-properties $\mathcal{L}^{\mathcal{H}}$. (The sublattice $\mathcal{D}$ consists essentially of the eigenspaces of $\hat{R}$, but with special treatment given to those eigenspaces that have no overlap with $\psi$.)  Thus all of the properties belonging to $\mathcal{D}$ can be considered determinate, i.e. they are either true or false in \lq reality', but with a probability determined by $\psi$.
(This is a simplified account. For details see \textcite{bub}). Any other property of $\mathcal{L}^{\mathcal{H}}$ not belonging to $\mathcal{D}(\psi,\hat{R})$ is indeterminate in the system \textbf{S} and must thus be given an operationalist interpretation. The modal formulations have more flexibility than Bohmian mechanics, since any observable can be designated as determinate and $\hat{R}$ can even vary in time, but they share with the Bohmian view the feature that in any instantiation a single set of properties is determinate, i.e. can be considered \lq real'. CQT in its microscopic version, in contrast, considers all quantum properties on an equal footing (Equality) and simply assigns a contextual meaning to the probability functions in its assertions.

\subsection{\label{E.}	Many Worlds Formulations}
The original Many Worlds formulation of \textcite{ev1} (he called it the \lq\lq Relative State" theory) was motivated by a critique of Copenhagen operationalism similar to those of Einstein and Bell. The critique is well illustrated by Everett's statement  quoted by \textcite{byrne}:

\begin{quote} The Copenhagen interpretation is hopelessly incomplete because of its a priori reliance on classical physics as well as a philosophic monstrosity with a \lq reality' concept for the macroscopic world and denial of the same for the microcosm.
\end{quote}

In the Many Worlds theory the quantum ontology consists exclusively of the wavefunction $\psi_U$ of the whole universe, which evolves deterministically according to the Schroedinger equation (or a suitable relativistic generalization). This assumption is unsurprising and it is consistent with CQT. The implications of this fact, however, are stated in language that differs for different proponents of the theory, and that leads to surprising and at times extravagant claims, see, e.g. \textcite{dew}.

In certain versions of the theory the wave function $\psi_U$ splits into diverse branches (referred to as \lq worlds'), each one of which is supposed to be \lq real', in the course of the unitary evolution. No direct recourse to operationalism, as in the Orthodox formulation of subsection B above, is possible here, since there are no external observers for the universe as a whole, and all properties of individual systems must emerge from an analysis of the quantum dynamics. Thus basic notions such as physical properties of individual systems, determinateness, or probabilities of measurement results, that are at the heart of the assertions of QM, are the subjects of detailed analysis and vary significantly among different versions of the theory.

If one goes back to the basic tenets of the theory, namely the deterministic and unitary evolution of the state, then the Everett formulation has many points in common with CQT (see, for example, the minimal formulation of \textcite{tegmark}, in which he distinguishes between the \lq inside view' and the \lq outside view'). This explains why \textcite{hartle2005} refers to Decoherent Histories as ``an extension and completion of the Everett formulation".

The problem with most \lq Many Worlds' formulations subsequent to Everett's, in our opinion, is that their ability to make any physical predictions depends from the outset on concepts such as \lq our experiences', \lq branches', \lq worlds', \lq observers', \lq belief' or \lq real', that are often imprecisely defined and whose meaning often differs between different authors. The root of the problem seems to us to lie in a confusion between the task of formulating QM for arbitrary physical systems on the one hand, and that of applying the formulation to a particular system, the universe as a whole, on the other hand. The distinction between the two tasks exists for classical as well as quantum mechanics, but the cost of confounding them is much greater in the quantum case, at least in CQT, where the distinction between MIQM and MAQM plays such an important role. Quantum cosmology can of course be formulated in the histories approach, the Decoherent Histories version being the best suited for that purpose, but the particular focus on the universe as a whole can obscure important features of the more general theory of histories.

\subsection{\label{F.}	Spontaneous Collapse Theories}
In contrast to all other \lq interpretations', which assume the correctness of the physical predictions of QM, spontaneous collapse theories (see, e.g. \textcite{ab5}) belong to the category that give the answer \lq no' to the first question in the Introduction: they correct the Schroedinger equation by adding a stochastic force which ensures the physical collapse of the wavefunction. The theories have a classical ontology, consisting exclusively of the state $\psi$, whose time evolution is adjusted to agree with known quantum properties. Our point of view is that such theories (they are really different theories, not just different formulations) are logically coherent, but that until there is experimental evidence for departures from standard QM we see no convincing motivation for modifying its physical content. A major shortcoming of spontaneous collapse theories, which they share with the Bohmian approach, is their extreme nonlocality, which makes them very difficult, perhaps impossible, to render Lorentz invariant.

\section{\label{6}	Conclusion}

CQT asserts that the basic microscopic formulation of QM for a closed system is relatively simple and involves a minimum of assumptions, the essential one being the Hilbert-space ontology. The result is what we have called MIQM, which can be summarized by the existence of a multiplicity of mutually incompatible sets of c-assertions (arising from incompatible frameworks), each one of which embodies its own notion of c-truth. We thus arrive at the novel conclusion that the quantum mechanics of closed systems is \emph{physically incomplete}, but logically coherent. The necessary completion entails some macroscopic mechanism for selecting a physical framework from the multiplicity of incompatible frameworks of the microscopic theory. This is what we have called MAQM and it can be accomplished in a variety of different ways, which we classify as external (e.g. measurement) or internal (e.g. coarse-graining to construct a physical framework consisting of histories composed of both quantum and classical properties). Most of the well-known paradoxes and mysteries of quantum mechanics appear only when one asks about measurement results and the transmittal of quantum information, i.e. when one asks about macroscopic phenomena that require apparatus, observers and agents external to the system. In CQT these questions are treated in the macroscopic selection phase of the theory, not as part of the microscopic formulation.

There is an analogy between the selection of a physical framework in quantum mechanics and the treatment of the arrow of time in statistical mechanics. A phenomenological formulation simply posits the second law and derives the irreversible hydrodynamic equations by appealing to conservation laws and general symmetry principles. This corresponds to our external framework selection mechanism. A more fundamental approach (the internal mechanism) starts from the microscopic description of a large system and derives the equations satisfied by macrovariables via a coarse-graining procedure. This latter program was initiated by Maxwell and Boltzmann and its full realization remains a subject of study (and controversy!) to this day, but the essential correctness of the Maxwell-Boltzmann point of view is generally accepted (see e.g. \textcite{leb}).

In his last lecture (delivered in 1989), entitled \lq Against Measurement', \textcite{b8} expressed the following view regarding QM:
\begin{quote}
Surely, after 62 years, we should have an exact formulation of some serious part of
quantum mechanics? By \lq\lq exact" I do not of course mean \lq\lq exactly true". I mean only that the
theory should be fully formulated in mathematical terms, with nothing left to the discretion
of the theoretical physicist,... until workable approximations are needed in applications... Is it not good to know
what follows from what, even if it is not really necessary for all practical purposes (\lq FAPP')?
\end{quote}
\noindent The answer provided by CQT is that MIQM is the desired full mathematical formulation, and that it is only in attempting to select a macroscopic physical framework, i.e. to find \lq physical truth', that \lq FAPP' arguments become necessary.

We conclude by listing the two principal advances which in our opinion Compatible Quantum Theory (CQT) makes over earlier histories formulations. The first is the distinction we have drawn between the microscopic theory (MIQM) on the one hand and the macroscopic theory (MAQM) on the other. From the point of view of CQT, moreover, the Copenhagen viewpoint, which involves a phenomenological framework selection mechanism via external measurement, can be viewed as a version of MAQM.

A second, more concrete advance made in the present paper is the derivation of the principal result of the microscopic theory, the quadratic Born formula \eqref{chprob} for the probability of a history. This result was postulated by earlier workers, as was the consistency or decoherence condition \eqref{nointerf}, but here they are derived starting from the Hilbert space ontology and the assumption of noncontextuality of probability values for histories and subhistories. In these derivations, essential roles are played by Gleason's Theorem, by an important \lq quadratic' theorem based on it due to Cassinelli and Zanghi, and by Nistico's extension of this theorem to histories, drawing on ideas of Omn\`es.

\section*{Acknowledgements}
The authors wish to express their sincere appreciation to Robert Griffiths and James Hartle for many fruitful interchanges and particularly for their comments on a preliminary version of this work, as well as to David Mermin for early discussions and encouragement.

\vspace{10mm}
\numberwithin{equation}{section}
\appendix
\section {Lattices, Set Theory, Classical Logic and Probability Theory}

In this appendix we provide a brief summary of set theory, classical (Aristotelian) logic and classical probability theory, and we show how the three are formally related.

\noindent
\underline{Lattices}

In pure mathematics, a \emph{lattice} is a set with a partial ordering $\leq$, and two binary operations, \lq meet' and \lq join', that satisfy the relations
\begin{subequations}
\label{lattice}
\begin{eqnarray}
\text{meet}(A,B) \leq A, \;\;\;\;\; \text{meet}(A,B) \leq B,\\
\text{If}\;\; C\leq A \;\;\;\; \text{and}\;\; C\leq B,\;\;\;\text{then}\;\; C\leq \text{meet} (A,B),\\
\text{join}(A,B) \geq A, \;\;\;\;\; \text{join}(A,B) \geq B,\\
\text{If} \;\;C\geq A \;\;\;\; \text{and} \;\;C\geq B,\;\;\;\text{then}\;\; C\geq \text{join} (A,B),
\end{eqnarray}
\end{subequations}
	
\noindent yielding the
greatest lower bound  and the least upper bound
of the two operands. A \emph{bounded} lattice has a universal lower bound ${\bot}_b$ and a universal upper bound ${\top}^t$ (we have placed subscripts and superscripts on these symbols to avoid confusion with the $\perp$ symbol signifying orthogonality, or the T symbol signifying truth, to be used later). A
\emph{self-dual} lattice has also a unary operation  $A \rightarrow\;\sim A$. 
The 
join and meet 
of $A$ and $\sim A$ are $\top^t$ and $\bot_b$, respectively, and the operation $\sim$ induces an automorphism of the lattice in which the ordering is reversed.

\noindent
\underline{Set Theory}

For simplicity we consider a discrete set $\Omega$ of $N$ elements $x\in \Omega$. The subsets $A,B,...$ of $\Omega$ form a set $\mathcal{L}(\Omega)$ 
of sets (in set theory, a \emph {field} of sets) 
for which the operations of union $\cup$, intersection $\cap$ and complement $\sim$ obey the axioms  of set theory:

\begin{subequations}
\label{sets}
\begin{eqnarray}
A\cup \oslash = A & A\cap \Omega = A, \\
A\cup \sim A = \Omega & A\cap \sim A = \oslash,\\
A\cup B = B\cup A  &  A\cap B = B\cap A, \\
A\cup (B\cup C) = (A\cup B)\cup C, & A\cap (B\cap C) = (A\cap B)\cap C,\\
A\cup (B\cap C) = (A\cup B)\cap (A\cup C), & \;\; A\cap (B\cup C) = (A\cap B)\cup (A\cap C),\label{setdist}
\end{eqnarray}
\end{subequations}	
where $\oslash$ is the empty subset of $\Omega$.

Clearly, $\mathcal{L}(\Omega)$ is a self-dual lattice if one interprets $\cup$ as the 
join 
and $\cap$ as the 
meet, 
and $\Omega, \oslash$ as ${\top}^t$ and ${\bot}_b$. The first four lines of \eqref{sets} are satisfied by all self-dual lattices.  But $\mathcal{L}(\Omega)$ has an additional property (the distributive law
\eqref{setdist}) not shared by all self-dual lattices, which makes it a \emph{Boolean} lattice.

\noindent
\underline{Classical Logic}

The subsets $A,B,...$ can also be considered as logical propositions,  in which case the operations of set theory become logical operations

\begin{subequations}
\label{setlog}
\begin{eqnarray}
\cup\;\; \longrightarrow \;\; \vee\;\; \text{disjunction\;\;(or)},\label{q-1}  \\
\cap\;\; \longrightarrow \;\; \wedge\;\; \text{conjunction\;\;(and)},\label{q-2}  \\
\sim\;\; \longrightarrow \;\; \neg\;\; \text{negation\;\;(not)},\label{q-3}\\
\Omega\;\; \longrightarrow \;\; \text{T}\;\; (\text{true}),\label{q-4}\\
\oslash\;\; \longrightarrow \;\; \text{F}\;\; (\text{false}).\label{q-5}
\end{eqnarray}
\end{subequations}

\noindent
Under the replacements \eqref{setlog}, Eqs.\eqref{sets} become the usual axioms of propositional calculus
\begin{subequations}
\label{clogic}
\begin{eqnarray}
A \vee \text{F} = A & A \wedge \text{T} = A, \\
A \vee\sim A = \text{T} & A \wedge\sim A = \text{F},\\
A \vee B = B \vee A  &  A \wedge B = B \wedge A, \\
A \vee (B \vee C) = (A \vee B) \vee C, &\;\; A \wedge (B \wedge C) = (A \wedge B) \wedge C,
\end{eqnarray}
\begin{eqnarray}\label{clogdist}
A \vee (B \wedge C) = (A \vee B) \wedge (A \vee C), & \;\; A \wedge (B \vee C) = (A \wedge B)\vee (A \wedge C).
\end{eqnarray}
\end{subequations}	
\noindent In particular \eqref{setdist} becomes the distributive law \eqref{clogdist}. The set ${\mathcal{L}(\Omega)}$ of $2^N$ propositions forms a Boolean lattice under the logical operations.
On this lattice we can define truth functions $\mathcal{T}(A)$ with values 1 (True) and 0 (False).
Such truth functions must agree with the standard truth tables for the logical functions,
which imply the algebraic relations

\begin{subequations}
\label{tfbool}
\begin{eqnarray}
\mathcal{T}(\neg A) = 1 - \mathcal{T}(A),  \\
\mathcal{T}(A\wedge B) = \mathcal{T}(A) \mathcal{T} (B),  \\
\mathcal{T}(A\vee B) = \mathcal{T}(A) + \mathcal{T}(B) - \mathcal{T}(A) \mathcal{T} (B), \label{dist2}
\end{eqnarray}
\end{subequations}

\noindent and must also satisfy  $\mathcal{T}(\Omega)=1$,  $\mathcal{T}(\oslash)=0$.
We shall refer to these equations as \lq truth table relations'.

Let us consider in particular those subsets of $\Omega$ containing only one member, that is sets of the form $\{x\}$ where $x\in\Omega$.  We may call them \emph{atomic} sets, and the corresponding logical propositions atomic propositions.  Then by applying \eqref{tfbool} we find that any truth function must assign the value 1 to some atomic proposition and 0 to all the others.  We shall denote the truth function that assigns 1 to a particular $\{x\}$ by ${\cal T}_x$.  Then for $x, y \in \Omega$

\begin{equation}
{\cal T}_x(\{y\}) = 1  \;\;\; \text{if} \;\;\;  y=x, \;\;\; \text{otherwise}\;\;\; 0.
\end{equation}

\noindent Again  applying \eqref{tfbool}, we see that for any $A\in \mathcal{L}(\Omega)$,

\begin{equation}\label{Txdef}
{\cal T}_x(A) = 1  \;\;\; \text{if} \;\;\;  x\in A, \;\;\; \text{otherwise}\;\;\; 0.
\end{equation}

\noindent We may say that $x$ is the \lq\lq source of truth" for the atomic truth function ${\cal T}_x$.

\noindent
\underline{Probability Theory}

Truth functions can be generalized by introducing a \emph{probability function}
$\mathcal{P}(A)$ from $\mathcal{L}(\Omega)$ to the unit interval $[0,1]$.  One first defines a \emph{measure} as a function from $\mathcal{L}(\Omega)$ to the interval $[0,\infty]$, which satisfies the linearity condition for countable sets of \emph{disjoint}
subsets,

\begin{equation} \label{kolminf}
\mathcal{P}(A^{(1)}\vee A^{(2)}\vee ... ) =  \mathcal{P}(A^{(1)})+ \mathcal{P}(A^{(2)})+..., \; \;\; \text{where}\;\; A^{(i)}\wedge A^{(j)} = \oslash \;\;\text{for}\;\;i\neq j.
\end{equation}

 \noindent A \emph{probability measure} or \emph{probability function} (classically, these two ideas can be identified) is a measure which satisfies the additional condition

\begin{equation} \label{probmeas}
\mathcal{P}(\Omega)=1.
\end{equation}

\noindent (The relation $\mathcal{P}(\oslash)=0$ is already implied by \eqref{kolminf}).
In the context of probability theory, the elements $A^{(i)}$ are called \lq events' and the set
$\Omega$ is the \lq sample space' of the probability measure. It can be shown that for any two events $A,B$, \eqref{kolminf} implies the relation

\begin{equation} \label{kolm2}
\mathcal{P}(A \vee B) = \mathcal{P}(A) + \mathcal{P}(B) - \mathcal{P}(A \wedge B).
\end{equation}
\noindent The converse is true for a finite $\Omega$.  We shall sometimes refer to Eqs. \eqref{kolminf} and \eqref{probmeas} as the \lq Kolmogorov conditions', and to \eqref{kolm2} as the \lq Kolmogorov overlap equation'.

One notices a similarity between \eqref{kolm2} and \eqref{dist2}.  This suggests that even on a formal level there is a relationship between probabilities and truth values.  The relationship can be displayed explicitly by starting not with the probability function $\mathcal{P}(A)$, where $A$ ranges over $\mathcal{L}$, but with a function $p(x)$, where $x$ ranges over $\Omega$, and $p$ satisfies

\begin{subequations}
\begin{eqnarray}
p(x)\geq 0, \;\;\;\; x\in \Omega, \\
\sum_x p(x) = 1.
\end{eqnarray}
\end{subequations}

\noindent This $p$ can be understood naturally as a \emph{probability distribution} over $\Omega$.

We can now construct from $p$ a probability function $\mathcal{P}$ on $\mathcal{L}(\Omega)$ by setting

\begin{equation}\label{Pp}
\mathcal{P}(A) = \sum_{x\in\Omega}\; p(x) {\cal T}_x(A) = \sum_{x\in A}\; p(x).
\end{equation}

\noindent It is easily seen that $\mathcal{P}$ defined in this way satisfies the conditions \eqref{kolminf}
and \eqref{probmeas} as well as the overlap equation corollary \eqref{kolm2}.  The subtracted term
$\mathcal{P}(A\wedge B)$ in \eqref{kolm2} has the same origin as that in \eqref{dist2}:  those
points $x$ that belong to both $A$ and $B$ contribute twice to $\mathcal{P}(A)+\mathcal{P}(B)$, as well as to ${\cal T}(A)+{\cal T}(B)$.  This explains the formal resemblance between \eqref{dist2} and \eqref{kolm2}. If $\Omega$ is finite, one may alternatively take $\mathcal{P}$ satisfying \eqref{kolm2} as fundamental, and derive the atomic probability function $p$ satisfying \eqref{Pp} by setting

\begin{equation} \label{defp}
p(x) = \mathcal{P}(\{x\}),
\end{equation}

\noindent for each $x\in\Omega$. We have presented the detailed argument leading to \eqref{Pp} in order to justify the statement that the probability function $\mathcal{P}$ can be thought of as a \lq distributed truth function' with $p$ controlling the distribution of weights to various choices of the source of truth that determines ${\cal T}_x$.

The above ideas have been extended to a continuous universe $\Omega$ of points $x$ bearing infinitesimal probability, by replacing the sum in \eqref{Pp} with an integral.  The technical details are well known and will not be described here.

It should be noted, moreover, that our definitions of probability and truth are formal ones, and they are thus consistent with either a frequentist or a Bayesian approach to probabilities. At this stage we are not inquiring into the relationship of probabilities to the \lq real world', which is where such distinctions arise. The connection between truth and probability explored above exists already on the formal level and is therefore independent of any real-world interpretation of probability.

\vspace{10mm}

\numberwithin{equation}{section}
\section {Static CQT}

This appendix presents the detailed mathematical derivation of the principles of CQT in the so-called \lq static' case, i.e. without taking time dependence into account. These principles are the interpretation of probability \emph{functions} as being contextual to (static) frameworks, and the derivation of the Born rule assuming only the Hilbert space ontology and the noncontextuality of probability \emph{values}. The mathematical formalism we use, the so-called \lq lattice' and \lq algebraic' approaches, goes back to von Neumann and collaborators, but our argumentation is novel in some respects.

In the abstract study of what \textcite{bvn} called quantum logic,
two independent traditions have grown up, the algebraic and the logical.  In the algebraic approach \cite{vN1932,vn} one starts with an abstract operator algebra and imposes various restrictions on it until the resulting structure can be modeled by the operators in a Hilbert space, including the projection operators in particular.
In the logical approach \cite{bvn} one starts with an abstract lattice and imposes restrictions until the lattice can be identified as $\mathcal{L}^{\mathcal{H}}$ (see below) derived from a Hilbert space $\mathcal{H}$.  These two approaches are admirably summarized in a paper by \textcite{david2012}, in which the algebraic (chapter 3) and logical (chapter 4) approaches are presented separately and independently, each step by step culminating in the Hilbert space model.  The successive restrictions in each approach are of course postulated and not derived from any logical foundation; they are justified in terms of the desired consequence.

We, on the other hand, are starting from the assumption of a Hilbert space ontology and finding  within it the lattice $\mathcal{L}^{\mathcal{H}}$ as well as the projectors
$[A]$.  Hence we shall freely mix the lattice and the operator concepts, drawing on each to help prove theorems relating to the other.

\subsection*{B.1. The property calculus and the search for quantum logic in $\mathcal{L}^{\mathcal{H}}$}

We start with the logical, or lattice, point of view.  As mentioned in Sec. IV.A, quantum properties are represented by closed linear subspaces of Hilbert space. Although these subspaces are sets of vectors, they do not form a field of sets in that the set-theoretical union of two subspaces is in general not a subspace, i.e. it is not a vector space. Therefore the logical relations of properties cannot be carried over from the relations of subspaces as sets of vectors, and the procedure used in Appendix A, relevant to the classical case, is inapplicable
to the totality of subspaces.

We shall follow \textcite{bvn} in defining \lq q-logical' operations on the \emph{subspaces} of the Hilbert space $\mathcal{H}$, and only afterwards examine the extent to which this
\lq quantum logic' reproduces classical logic.  (To ensure that all statements apply if $\mathcal{H}$ has infinite but countable dimension, we note that by definition a Hilbert
space is topologically complete, and we specify that the word \emph{subspace} is to mean a topologically closed linear subspace of $\mathcal{H}$, so that each subspace is itself a Hilbert space.)

Thus we define
\begin{subequations}\label{qlog}\begin{eqnarray}
\text{q-not }A = \neg_q A = \text{orthogonal complement of }A, \\
A \text{ q-and }B = A\wedge_q B = (A\cap B), \\
A \text{ q-or }B = A\vee_q B = \text{span}(A, B),
\end{eqnarray}\end{subequations}

\noindent and note that the q-not ($\neg_q$) and q-or ($\vee_q$) operations are different from the corresponding ones in \eqref{setlog}, since the orthogonal complement of a subspace $A$ contains only those vectors orthogonal to the vectors in $A$, and the span of two subspaces $A$ and $B$ contains all linear combinations of vectors in
$A$ and $B$, including those not belonging to either $A$ or $B$.

With these definitions the subspaces form a \emph{lattice}
 $\mathcal{L}^{\mathcal{H}}$, defined by taking the ordering operation $A\leq B$ to mean that $A$ is a subspace of $B$, and $\sim A$  to be $\neg_q A$.   As in the lattice of classical logic \eqref{clogic}, the glb and lub turn out here to be $\wedge_q$ and $\vee_q$ as defined in \eqref{qlog}, and ${\top}^t$ and
${\bot}_b$ turn out here to be $\mathcal{H}$ and $\text{O}$, the latter defined as the \lq zero subspace' containing only the vector $0$.  It can be shown that this lattice is self-dual under the operations $A\leftrightarrow\neg_q A$, $\vee_q\leftrightarrow\wedge_q$, and in addition we have $A\perp\neg_q A$ (see below).  Moreover, $\neg_q A$ is the unique subspace $A'$ orthogonal to $A$ and satisfying $A\vee_q A' = \mathcal{H}$. Careful proofs of many properties of $\mathcal{H}$ and its subspaces are given in \textcite{Driver}. In particular the important lemma
\begin{equation}\label{orthspan}
A\vee_q\neg_q A = \mathcal{H}
\end{equation}

\noindent is proved in detail with illuminating comments. In the mathematical literature a lattice $\mathcal{L}^{\mathcal{H}}$ possessing
 the above properties is referred to as an \emph{orthocomplemented} lattice, or ortholattice for short.

The operations defined in \eqref{qlog} can be used to express certain familiar ideas connected with subspaces:

\noindent (i) orthogonality:   $A$ and $B$ are orthogonal ($A{\perp}B$) iff every vector of $A$ is orthogonal to every vector of $B$. Obviously $A\perp B$ is equivalent to
$B\perp A$.  In terms of lattice operations,
\begin{equation}
A{\perp}B \text{ iff } A\leq\neg_q B.
\end{equation}

\noindent (ii) projectors:  The projector of a subspace $A$ shall be written as $[A]$;  it is the unique self-adjoint operator such that $[A]v=v$ if $v$ is a vector $\in A$, and $[A]v=0$ if $v\in\neg_q A$. Given $[A]$, $A$ is determined. There is thus a one to one correspondence between subspaces of Hilbert space and projectors. Note that projectors, which in general are not additive, \emph{are} additive among orthogonal subspaces:  if $A{\perp}B$ then $[A\vee_qB] = [A] + [B]$.  An operator $\hat{A}$ is the projector of some subspace if and only if ${\hat{A}}^2 = \hat{A}$. Moreover, Eq.\eqref{orthspan} implies the relation $[\neg_q A] = \text{I} - [A]$, where I is the identity operator on $\mathcal{H}$. These properties will be referred to without comment hereafter.  For relevant theorems and proofs, see \textcite{Driver}.

Since, as mentioned above, the span is not the set theoretical union, there is no guarantee that the q-operations will satisfy laws analogous to \eqref{sets}.  In fact, the operations $\neg_q$, $\wedge_q$, $\vee_q$ satisfy all but one of the laws of logic, namely
\begin{subequations}\label{qlogic}\begin{eqnarray}
A\vee_q\text{O} = A & A \wedge_q \mathcal{H} = A, \\
A\vee_q\neg_q A = \mathcal{H} & A\wedge_q\neg_q A = \text{O}, \\
A\vee_q B = B\vee_q A  &  A\wedge_q B = B\wedge_q A,\\
A\vee_q (B\vee_q C) = (A\vee_q B)\vee_q C &\;\;\;\; A\wedge_q (B\wedge_q C) = (A\wedge_q B)\wedge_q C.
\end{eqnarray}\end{subequations}

\noindent  However, the distributive law \eqref{clogdist} does not hold :
\begin{eqnarray}\label{qlognodist}
A\vee_q (B\wedge_q C) \neq (A\vee_q B)\wedge_q (A\vee_q C)\nonumber, \\
A\wedge_q (B\vee_q C) \neq (A\wedge_q B)\vee_q (A\wedge_q C), 
\end{eqnarray}

\noindent in general, as shown, for example, by the counterexample of three distinct coplanar one-dimensional subspaces. It follows that the property calculus \eqref{qlogic} does not constitute a proper logic, and the term \lq quantum logic' frequently used for these relations is apt to lead to confusion, so we shall avoid it.

\subsection*{B.2. Sublattices of $\mathcal{L}^{\mathcal{H}}$, sample spaces and event algebras}

We have appended the subscript \lq $q$' to the operations in \eqref{qlog} as a reminder that these are not true logical operations on the whole lattice $\mathcal{L}^{\mathcal{H}}$ of Hilbert subspaces since they do not satisfy the distributive law. This is expressed by saying that $\mathcal{L}^{\mathcal{H}}$ is not a Boolean lattice. We shall see, however, that there are so-called \lq Boolean sublattices', within which the distributive law is satisfied, thus allowing the definition of a sublattice-dependent (contextual) logic, which we refer to as \lq c-logic'.

Rather than trying to identify all Boolean sublattices of  $\mathcal{L}^{\mathcal{H}}$, we shall follow Griffiths in starting from a stronger requirement, that the desired lattice should be able to support a probability function.  We adhere to the principle voiced in \textcite{GrQL}, that a probability function must be based on a sample space, that is, a set of mutually exclusive alternatives which together exhaust all possibilities.  In the language of subspaces, this means that a sample space $\mathcal{S}$ is a set of mutually \emph{orthogonal} subspaces $\{D_1,D_2,...\}$ that together span $\mathcal{H}$. (Despite the name \lq sample space', $\mathcal{S}$ itself is not a subspace of Hilbert space, but a set of subspaces. Within Appendix B alone, we use the letter $D$ to denote the members of $\mathcal{S}$, since we have other uses for the letter $A$.)

It is useful, at this point, to slide from the logical to the algebraic mode by using the fact noted above that the subspaces $D_i$ are in one-to-one correspondence with their projectors $[D_i]$.  The projectors of
$\mathcal{S}$
form an orthogonal decomposition of the identity operator $\text{I}$;  that is,
\begin{equation}\label{SJ}
\mathcal{S} = \{D_1,D_2,...,\}= \{D_i|i\in \mathcal{J}\},
\end{equation}

\noindent where
\begin{equation}\label{orthdec}
\sum_{i\in\mathcal{J}}[D_i] = \text{I}, \text{ and } D_i\perp D_j \text{ if } i\neq j.
\end{equation}

\noindent Here $\mathcal{J}$ is the finite or countably infinite set of indices $i$ used in the definition of $\mathcal{S}$.

The condition $D_i\perp D_j$ translates to $[D_i][D_j] = [D_j][D_i] = 0$, and for the case $i=j$ we have $[D_i][D_i] = [D_i]$, so that \eqref{orthdec} can be written entirely in terms of projectors as

\begin{equation}\label{orthdecproj}
\sum_{i}[D_i] = \text{I}, \text{ and } [D_i][D_j] = [D_i] \delta_{ij}.
\end{equation}

\noindent We may sometimes use the notation $[\mathcal{S}]$ for the set of projectors $\{[D_i]\}$.

Now let an \lq event' $E_J$ be a subspace determined by a subset $J$ of $\mathcal{J}$.  This can be any subset, including the empty set or the whole set  $\mathcal{J}$,
so that if  $\mathcal{J}$ has only a finite number $n$ of members, there will be $2^n$ possible events.  The subspace $E_J$ is defined as
\begin{equation}\label{event}
E_J = (\vee_q)_{i\in J}D_i,
\end{equation}

\noindent and its algebraic representation is

\begin{equation}\label{evproj}
[E_J]  = \sum_{i\in J} [D_i].
\end{equation}

\noindent We note that \eqref{evproj} can also be written as

\begin{equation}\label{evTc}
[E_J] = \sum_{i\in\mathcal{J}}c_i[D_i],
\end{equation}

\noindent where $c_i = 1$ if $i\in J$, otherwise $c_i=0$.  We see that $c_i^2=c_i$ for all $i$, so that $(\Sigma_i c_i[D_i])^2 =  \Sigma_{i,j} c_ic_j[D_i][D_j] = \Sigma_i c_i[D_i]$ on account of \eqref{orthdecproj},
and therefore $[E_J]^2 =[E_J]$, as expected for a projector.

Now we introduce the \emph{event space} $\mathcal{E_S}$ consisting of the events $E_J$ corresponding to \emph{all} the subsets $J$ of $\mathcal{J}$.  (We briefly postpone the use of Griffiths' term \lq event algebra' for $\mathcal{E_S}$.)  The event space, like the sample space, is not a subspace of $\mathcal{H}$ but a set of such subspaces. The sample space $\mathcal{S}$ shall be called the \emph{basis} of the event space $\mathcal{E_S}$. Note, however, that the word basis here does not refer to a basis of a vector space; the members of $\mathcal{S}$ may be many- or even infinite-dimensional subspaces of $\mathcal{H}$.

We now show, in Theorem B1, that  $\mathcal{E_S}$ is closed under the operations \eqref{qlog} and is therefore a sublattice of $\mathcal{L^H}$, which we can therefore denote $\mathcal{L_S}$. In Theorem B2 we show moreover that $\mathcal{E_S} = \mathcal{L_S}$ is a Boolean lattice, meaning that its members obey not only the first four equations of \eqref{clogic} but also the q-analogue of the distributive law \eqref{clogdist}. This enables one to treat the operations \eqref{qlog} as true logical operations as long as one draws propositions only from
$\mathcal{L_S}$. We shall refer to this restricted logic as contextual logic (c-logic).

{\bf Theorem B1:} The event space $\mathcal{E_S}$ is a sublattice of $\mathcal{L^H}$, since it is closed under the operations \eqref{qlog}.

\noindent {\bf Proof:} In the following we note that the $J$ are classical sets to which the operators $\sim$, $\cup$, $\cap$ of Eq. \eqref{sets} can be applied.

(i) Let the event $A=E_J$ be some member of $\mathcal{E_S}$, which satisfies the relation $[A] = \Sigma_i c_i[D_i]$ in accordance with \eqref{evTc}. Then $[\neg_q A] = \text{I} -[A] = \Sigma_i (1-c_i)[D_i]$ since $\Sigma_i [D_i] = \text{I}$ by \eqref{orthdec}.
But the definition of $c_i$ makes $1-c_i = 1$ if $i\in\;\sim\!\! J$, otherwise 0. Therefore applying \eqref{evTc} again we have  $[\neg_q A] = \Sigma_{i\in\sim J}[D_i] = E_{\sim J}$, which belongs to $[\mathcal{E_S}]$.  Hence $\neg_q A$ belongs to $\mathcal{E_S}$.

(ii) Let $A=E_J$, $B=E_{J'}$.  Then using \eqref{evTc} we have $[A\wedge_q B] =  \Sigma_{i,i'} c_ic'_{i'}[D_i\wedge_q D_{i'}] = \Sigma_i c_ic'_i[D_i]$.  Applying \eqref{evTc} again
we get $[A\wedge_q B] = \Sigma_{i\in J\& i\in J'}[D_i] = [E_{J\cap J'}]$, which belongs to $[\mathcal{E_S}]$.  Hence $A\wedge_q B$ belongs to $\mathcal{E_S}$.

(iii)  Let $A=E_J$, $B=E_{J'}$.  By \eqref{event} we have $A\vee_q B = (\vee_q)_{i\in J} D_i \vee_q (\vee_q)_{i\in J'} D_i =  (\vee_q)_{i\in (J\cup J')} D_i = E_{J\cup J'}$, which belongs to $\mathcal{E_S}$.

(iv) In part (iii), instead of $A$ and $B$ we could have had any sequence $A^{(1)}, A^{(2)}, ..., A^{(k)}, ...$, finite or infinite, with $A^{(k)}= E_{J^{(k)}}$ for each $k$.  Then we would have $(\vee_q)_k A^{(k)} = (\vee_q)_k (\vee_q)_{i\in J^{(k)}} D_i  =  (\vee_q)_{i\in (\cup_k J^{(k)})} D_i = E_{\cup_k J^{(k)}}$.

It follows from the above reasoning that $\mathcal{E_S}$ is a sublattice of $\mathcal{L^H}$, which we can call the \emph{lattice closure} of $[\mathcal{S}]$. We express this fact by the relation

\begin{equation}\label{evlat}
\mathcal{E_S} = \mathcal{L_S},
\end{equation}

\noindent referred to above. We now show that $\mathcal{E_S}$ is a Boolean lattice - that is, the q-analog of the distributive law \eqref{clogdist} holds.

{\bf Theorem B2:} If $A$, $B$, $C$ belong to $\mathcal{E_S}$, then they satisfy the distributive laws:
\begin{equation}\label{distvee}
(A\wedge_q B) \vee_q (A\wedge_q C) = A\wedge_q (B\vee_q C),
\end{equation}

\noindent and
\begin{equation} \label{distwedge}
(A\vee_q B) \wedge_q (A\vee_q C) = A\vee_q (B\wedge_q C).
\end{equation}

\noindent {\bf Proof}:  We refer to the formulas developed in the proofs of Theorem B1 (ii) and (iii):  if $A= E_J$ and $B=E{_J'}$ then
\begin{equation}\label{evwedge}
A\wedge_q B = E_{J\cap J'},
\end{equation}

\noindent and
\begin{equation}\label{evvee}
A\vee_q B = E_{J\cup J'}.
\end{equation}

\noindent Now, if $A$, $B$, $C$ all belong to $\mathcal{E_S}$, then applying \eqref{evwedge} and \eqref{evvee} to arbitrary pairs of events we arrive at
\begin{equation}\label{AB-AC}
(A\wedge B) \vee (A\wedge C) = E_{(J\cap J') \cup (J\cap J'')},
\end{equation}

\noindent and
\begin{equation}\label{A-BC}
A\wedge_q (B\vee_q C) = E_{J\cap (J' \cup J'')}.
\end{equation}

\noindent Since the $J$'s are classical sets, we have in accordance with \eqref{clogdist}

\begin{equation}\label{Tdistvee}
(J\cap J') \cup (J\cap J'') = J\cap (J' \cup J''),
\end{equation}

\noindent from which \eqref{distvee} follows. Equation \eqref{distwedge} is proved similarly.

\textcite{GrQL} uses the term \lq event algebra' rather than \lq event space'.  In modern mathematics, an \lq algebra' is a set of entities with (i) an addition and a multiplication, both commutative and associative, (ii) multiplication distributive over addition, with zero and one having the usual properties, and (iii) an associative multiplication by the elements, called scalars, of a field.  Thus one can always divide a member of an algebra by any nonzero scalar, but not necessarily by a nonzero algebraic element.

In lattice theory, however, there exists the term \lq Boolean algebra', which denotes not a modern algebra with a special \lq Boolean' feature, but simply a lattice that is Boolean in the sense of Theorem B2. A Boolean algebra in this sense does not necessarily have anything to do with an algebra in the modern sense, nor conversely.  The reason for this linguistic inconsistency is that Boole's work actually preceded the rise of the modern concept of an algebra. We shall use the term Boolean algebra in its lattice theory sense.

Given a sample space  $\mathcal{S}$, it is evident from Theorems B1 and B2 that its event space  $\mathcal{E_S}$ is a Boolean algebra.  But also, the space  $[\mathcal{E_S}]$ of its projectors forms an algebra in the modern sense, provided that one takes the underlying scalar field to be the 2-member field  $Z/2Z$ (the integers modulo 2).  Griffiths does not draw a distinction between these two spaces, which after all are in 1-1 correspondence.  Consequently his term \lq event algebra' could be taken either in the lattice or the modern sense.
In fact, since the appellation in either sense is correct provided one starts with a sample space, there is no need to quibble about which sense is meant.  In the following subsections, though, we shall lean more to the lattice point of view, so that unless otherwise stated the event algebra of $\mathcal{S}$ shall mean $\mathcal{E_S}$, rather than the set  $[\mathcal{E_S}]$ of corresponding projectors.

\subsection* {B.3. Static frameworks}

In order to define probability functions we shall need Griffiths's notion of a \lq framework'.   What we call a \emph{static framework} is nothing other than the event algebra $\mathcal{E_S}$ of a sample space $\mathcal{S}$.  In accordance with \eqref{evlat} we may also say it is the lattice closure
$\mathcal{L_S}$ of $\mathcal{S}$. Thus any sample space $\mathcal{S}$ defines a framework $\mathcal{E_S}=\mathcal{L_S}$, which has $\mathcal{S}$ as its basis. For the rest of Appendix B we shall freely say \lq framework' meaning \lq static framework'. Let us delve somewhat further into the lattice properties of $\mathcal{L}_\mathcal{S}$.  In particular we ask whether the basis $\mathcal{S}$ is unique.  That is, would it be possible that two different projective decompositions of I have the same lattice closure? In that case we could have
$\mathcal{L_S}=\mathcal{L_{S'}}$ even though $\mathcal{S}\neq \mathcal{S'}$.

To answer this we draw on the concept of atoms in a lattice. In any sublattice $\mathcal{L}$ of  $\mathcal{L}^H$ an \emph{atom} or atomic subspace is a nonzero member of
$\mathcal{L}$ that has no proper nonzero subspace in $\mathcal{L}$.  In other words,  $A$ is an atom of $\mathcal{L}$ iff when $A\in \mathcal{L}$, any $B\in \mathcal{L}$ that is $\leq A$ is either $A$ or O. It is important that the same subspace can be atomic in one sublattice but not in another.  In the whole lattice $\mathcal{L}^{\mathcal{H}}$, the atomic subspaces are just the 1-dimensional ones.  But
a 2-dimensional subspace, for example, is atomic in a sublattice that contains it but none of its
1-dimensional subspaces. Even an infinite-dimensional subspace can be atomic in a sufficiently coarse sublattice.  Any subspace $A$ will be atomic in the lattice composed of
$A$, $\neg_q A$, $\mathcal{H}$, and $\text{O}$.

This situation is quite analogous to that of the lattice of sets of a space
$\Omega$ considered in Appendix A.  The 1-dimensional subspaces of $\mathcal{H}$ are analogous to sets $\{x\}$ containing a single point of $\Omega$.  Sets containing many points can be atomic in a sufficiently coarse field of sets. Any set is atomic in the field (lattice) composed of that set, its complement,  $\Omega$, and
$\oslash$.  In classical mechanics, to be sure, we are primarily concerned with the full
$\mathcal{L}^{\Omega}$ so that only the sets $\{x\}$ are atomic.

We now answer the question posed above about the uniqueness of $\mathcal{S}$, given its lattice closure.

{\bf Theorem B3:}  A static framework has only one basis, which consists exactly of all of its atoms.

\noindent {\bf Proof:}  By examining the definition of \lq atom', and using Theorem B1 to replace $\mathcal{L_S}$ by $\mathcal{E_S}$, the reader may verify that if $\mathcal{S}$ is a sample space, all of its members are atoms of $\mathcal{L}_\mathcal{S}$, and that $\mathcal{L_S}$ has no other atoms.  The theorem follows.

We now see that the events called elementary by \textcite{GrQL} are precisely the atoms of the event algebra considered as a lattice.  (In the above reasoning we have not appealed to the property of  \lq atomic covering', as proving it would take us deeper into lattice theory than we intend to go.) Inasmuch as $\mathcal{S}$ and $\mathcal{L}_\mathcal{S}$ determine one another uniquely, Griffiths applies the term framework indiscriminately to both.  Our usage is stricter:
the static framework is $\mathcal{L}_\mathcal{S}$, which is the same as $\mathcal{E_S}$, and we refer to $\mathcal{S}$ as its basis.

\subsection*{ B.4. Probabilities: Mackey's generalization of measure and Gleason's Theorem}

We now approach the question of probabilities.  As mentioned above, we follow Griffiths in taking the term \lq probability function' in the \emph{classical} sense, as being a function from some lattice $\mathcal{L}(\Omega)$ of subsets of a universe $\Omega$, to nonnegative real numbers such that \eqref{kolminf} and \eqref{probmeas} are satisfied.  Any  static framework $\mathcal{L_S}$, being the lattice closure of a sample space  $\mathcal{S}$,  can be regarded as such a lattice because, in accordance with Theorem B1, the members of $\mathcal{L_S}$ are in one-to-one correspondence with the subsets of the sample space $\mathcal{S}$.

To construct a classical probability function over this sample space, one need only assign to each member of $\mathcal{S}$ (i.e., each atom of $\mathcal{L_S}$) a real nonnegative probability so that the sum over all the atoms is 1.  The probability of any \emph{set} of atoms is then the sum of the individual atomic probabilities, as given in \eqref{kolminf}. Through the correspondence between subsets of $\mathcal{S}$ and events $E$ of the framework $\mathcal{E_S} = \mathcal{L_S}$, this classical probability function may be regarded as acting on subspaces $E$ of the Hilbert space. In this way the probability function is inexorably tied to the sample space $\mathcal{S}$ and to the framework $\mathcal{E_S}$, so we shall denote it as $\mathcal{P_S}$, or as $\mathcal{P_{E_S}}$, or $\mathcal{P_E}$.

According to the above construction the probability assigned to any atom is freely chosen, provided they are all nonnegative and sum to 1. As explained more fully in Subsection IV.B of the main text, however, it is physically appropriate to constrain the atomic probabilities so that any event has the \emph{same} probability, independent of the framework to which it belongs - a constraint we refer to as the \emph{noncontextuality} of probability values. Before we show how to satisfy this constraint, we need to discuss the ideas of Mackey.

As mentioned earlier, \textcite{bvn} drew attention to the lattice of subspaces of a Hilbert space as possibly analogous to the lattice of sets that is closely related to classical logic. \textcite{mackey1957quantum} pursued that analogy into measure theory by proposing that a measure could be defined as a function on the elements of \emph{any} lattice, not just on a lattice of \emph{sets}. The key idea is that in Kolmogorov's additivity condition \eqref{kolminf} for the definition of a probability measure, the provision $A^{(i)}\wedge A^{(j)} = \oslash$ could just as well have been written as
$A^{(i)} \leq \neg A^{(j)}$, where $\leq$ denotes set inclusion or logical implication.  In set theory, the two statements are equivalent.  But in a general (non-Boolean) lattice, with $\oslash$ replaced by ${\bot}_b$, they are not, as we have seen in the previous subsection when the lattice is taken to be $\mathcal{L^H}$. The first statement, then, which in the non-Boolean case we write as  $A^{(i)}\wedge_q A^{(j)} = \oslash$, is what we are calling \lq disjointness' of the subspaces, and the second is orthogonality
$A^{(i)}\perp A^{(j)}$ (we write $\perp$ without a q subscript since it is defined directly in terms of an inner product of vectors).  Some authors take \lq disjoint' to mean
\lq orthogonal',  even when speaking of subspaces, which we consider a waste of a good adjective.

 Mackey proposes to write Kolmogorov's additivity condition Eq.\eqref{kolminf} as
\begin{equation} \label{mackinf}
\mathcal{P}(A^{(1)}\vee A^{(2)}\vee ... ) = \mathcal{P}(A^{(1)}) +  \mathcal{P}(A^{(2)}) + ..., \text{when}\;\; A^{(i)} \perp A^{(j)}\;\; \text{for} \; i \neq j.
\end{equation}

\noindent In this form, with $A \perp B$ taken to mean $A\leq\neg B$, he generalizes it to all lattices, and in particular to $\mathcal{L}^{\mathcal{H}}$ .   The relation $A\perp B$ thus defined turns out in
$\mathcal{L}^{\mathcal{H}}$ to be orthogonality of subspaces defined in the usual way.  (We note that  this relation is symmetric.)

Specializing to $\mathcal{L^H}$, we follow Mackey in defining a \emph{lattice measure} $\mathcal{W}$ as satisfying the q-version of \eqref{mackinf}, namely
\begin{equation} \label{Wmackinf}
\mathcal{W}(A^{(1)}\vee_{q} A^{(2)}\vee_{q} ... ) = \mathcal{W}(A^{(1)}) +  \mathcal{W}(A^{(2)}) + ..., \text{when}\;\; A^{(i)} \perp A^{(j)}\;\; \text{for} \; i \neq j.
\end{equation}

\noindent A \emph{normalized lattice measure} $\mathcal{W}$ on a lattice $\mathcal{L}$ is a function from members of $\mathcal{L}$ to nonnegative real numbers, satisfying \eqref{Wmackinf} and in addition, mapping the ${\top}^t$ element of $\mathcal{L}$ into 1 (and therefore the
${\bot}_b$ element into 0).  Since we shall  only be interested in lattice measures on
$\mathcal{L}^{\mathcal{H}}$, it is $\mathcal{H}$ that is mapped into 1 and $\text{O}$ into 0.  We shall never be concerned with a classical measure on $\mathcal{H}$ itself; that would be a function on all \emph{subsets} of $\mathcal{H}$.  (In Appendix C, a similar treatment will be applied to $\mathcal{H}^N$, the $N$-fold tensor product of $\mathcal{H}$ with itself, instead of to $\mathcal{H}$.)

Mackey does not use the term \lq lattice measure', but rather \lq measure on the questions';  his term \lq question' is defined to be what we call \lq projector'.  Since subspaces are in 1-1 correspondence with projectors, one may as well speak of a normalized measure on projectors, defined as a function
$W$ from projectors to real nonnegative numbers, mapping the projector I (the identity operator on
$\mathcal{H}$) to the number 1 and satisfying
\begin{equation}\label{projmeas}
W([A^{(1)}] + [A^{(2)}] + ... ) = W([A^{(1)}]) + W([A^{(2)}]) + ..., \;\text{when}
\;\;\; [A^{(i)}][A^{(j)}]=0\;\text{  for } i\neq j.
\end{equation}

\noindent The span on $A$ in \eqref{Wmackinf} has been replaced by a sum on $[A]$ in \eqref{projmeas}, which is correct when the
$A$'s are mutually orthogonal, and the provision $A^{(i)} \perp A^{(j)}$ has been replaced by
 $[A^{(i)}][A^{(j)}]=0$.  Note that these replacements would not work if in \eqref{Wmackinf} we had written $A^{(i)}\wedge_{q} A^{(j)} = {\bot}_b$ instead of $A^{(i)} \perp A^{(j)}$. The connection between \eqref{Wmackinf} and \eqref{projmeas} is rendered explicit by setting
\begin{equation}\label{WW1}
W([A]) = \mathcal{W}(A),
\end{equation}

\noindent where we use a script letter for a function on properties and a Roman letter for a function on the corresponding projector.

Since \eqref{Wmackinf} is defined entirely in terms of lattice operations, its meaning does not depend on being able to interpret the lattice elements as sets.  A certain confusion may arise (and should be avoided) from the fact that in the case of $\mathcal{L}^{\mathcal{H}}$ the lattice elements actually are sets of vectors in $\mathcal{H}$.  Even though we have used set inclusion in this sense to define the lattice ordering relation $A\leq B$, from which all other lattice relations are deduced, a \lq lattice measure' satisfying the condition \eqref{Wmackinf} is not a function on arbitrary sets of vectors, but only on subspaces treated as lattice elements.   Therefore it is not a measure on $\mathcal{H}$ in the usual sense.  For this reason, mathematicians who have long been accustomed to understanding a measure as a function on sets may resist the application of the term \lq measure' to a function on subspaces, or on projectors as in \eqref{projmeas}.  We point out, though, that this resistance might have been considerably diminished if Kolmogorov's  additivity condition for classical measures had traditionally been expressed in terms of $A^{(i)} \subseteq\neg A^{(j)}$ instead of $A^{(i)}\wedge A^{(j)} = \oslash$, as it perfectly well could have been.

Do there exist functions $W$ as described above?  One sees that \eqref{projmeas} is just a linearity condition, but restricted to orthogonal projectors.  Now, of course there exist functions $\tilde{W}$ from \emph{all} linear operators $\hat{A}$ on $\mathcal{H}$ to complex numbers that satisfy an ordinary linearity condition, namely
\begin{equation}\label{oplin}
\tilde{W}(\lambda_1 \hat{A}_1 + \lambda_2 \hat{A}_2) = \lambda_1 \tilde{W}(\hat{A}_1) + \lambda_2 \tilde{W}(\hat{A}_2),
\end{equation}

\noindent where the $\lambda_i $ are complex scalars, and in addition give real values to self-adjoint operators and map the identity operator to 1.  Such  functions can always be written as $\hat{A} \rightarrow \text{Tr}(\rho \hat{A})$, where the information coded by $\tilde{W}$ is now contained in a positive (i.e., all expectation values are $\geq 0$) self-adjoint operator $\rho$  of unit trace.  In physics such an operator is called a density matrix, but if one does not wish to call attention to a specific basis one may call it a density operator.  We shall use the former term.

For any fixed density matrix $\rho$, let us define the function
\begin{equation}\label{wtilde}
\tilde{W}_{\rho}(\hat{A}) = \text{Tr}(\rho \hat{A}),
\end{equation}

\noindent where $\hat{A}$ is \emph{any} linear operator in the Hilbert space. The restriction of $\tilde{W}_{\rho}$ to projectors $\hat{A} = [A]$ yields a function $W_{\rho}$ from projectors to nonnegative real numbers, satisfying \eqref{projmeas}, and  consequently a normalized lattice measure $\mathcal{W}_{\rho}$ on $\mathcal{L}^{\mathcal{H}}$, defined by \eqref{WW1}, that satisfies Eq.\eqref{Wmackinf}.

Mackey then asked whether \eqref{projmeas} has solutions not of the form \eqref{wtilde}.  This question was answered in the negative by \textcite{glea}, who proved a difficult and celebrated theorem named after him.  Gleason's Theorem tells us that with the exception of
$\mathcal{H}$ having dimension 2, \emph{all} normalized lattice measures on $\mathcal{L}^{\mathcal{H}}$ are of the form $A\rightarrow \text{Tr}(\rho [A])$ where $\rho$ is some density matrix.  We shall ignore the exception, although counterexamples do exist \cite{bell1966}, and simply exclude such counterexamples from consideration as lattice measures.

\subsection*{ B.5. Probability and noncontextuality}

We now apply the Mackey-Gleason ideas to our problem as stated at the beginning of B.4: how can the choice of probability values for the atoms of each static framework be constrained to satisfy noncontextuality, i.e. so that any subspace of
$\mathcal{H}$ (any property) will have the same probability value in every static framework of which it is a member?
The answer is given in terms of normalized lattice measures as defined in B.4.

Let us first recall that in Appendix B.2 we represented the members of a sample space $\mathcal{S}$ by indices $i$ belonging to a classical set $\mathcal{J}$. The events $E$, defined in \eqref{event} in terms of subsets $J$ of $\mathcal{J}$, are thus correlated with subsets of $\mathcal{S}$. But $\mathcal{S}$ is itself a perfectly good classical set, although its members are quantum mechanical beables. Therefore the proofs of Theorems B1 and B2 might have been given directly in terms of subsets of  $\mathcal{S}$ instead of subsets of  $\mathcal{J}$.  We found, however that the presentation was clearer in the \lq $J$-language' than in the \lq $S$-language'.  Presently, though, we shall deal with arguments for which the \lq $S$-language' works better. (We are aware of the fact that the symbols for the set (script) $\mathcal{S}$ and its subsets (Roman) $S$ are quite close in appearance and could be confused, especially when they occur in subscripts, but we have run out of convenient notations and we trust that the context will allow the reader to distinguish between the two symbols. In particular, subscripts for the probability function $\mathcal{P}$ are always (script) $\mathcal{S}$ and those for events $E$ are (Roman) $S$).

Corresponding to $S$ there is a subspace $E_S$
defined as $(\vee_q)_{D \in S} D$, which is a transcription of \eqref{event} to the $S$-language.  This definition is noncontextual: given only the set $S$, then $E_S$ is determined uniquely, independent of $\mathcal{S}$. However, the inverse
relation is contextual: the notation $S_E$ has only the meaning that $S_E$ is the unique subset $S$ of $\mathcal{S}$
such that $E_S=E$. Thus one must know both $E$ and $\mathcal{S}$, to determine $S$. Strictly, we should write
$S_{E,\mathcal{S}}$.

{\bf Theorem B4:}  Any normalized lattice measure acting on all of $\mathcal{L^H}$ will yield a noncontextual network of probability functions on the various static frameworks.

\noindent {\bf Proof:}  Let  $\mathcal{W}_{\rho}$ be such a normalized lattice measure, and let $\mathcal{S}$ be a sample space.  As $\rho$ is to remain fixed in what follows, we shall omit the subscript $\rho$. Let  $\mathcal{P_S}$ be the restriction of $\mathcal{W}$ to the static framework $\mathcal{L_S}$.  Then
$\mathcal{P_S}$ is essentially a function on subsets $S$ of $\mathcal{S}$.  Formally,
\begin{equation}\label{PW}
\mathcal{P_S}(S) = \mathcal{P_E}(S)= \mathcal{W}(E_S),
\end{equation}

\noindent where we have labelled the probability function either with the subscript $\mathcal{S}$ designating the sample space or with the subscript $\mathcal{E}$ designating the corresponding framework, since the two are in one to one correspondence. If any two subsets $S$ and $S'$ are disjoint in the classical sense, $S\cap S' = \oslash$, then (since $\mathcal{S}$ is a sample space) $E_S\perp E_{S'}'$.
Therefore, if $S^{(1)}$, $S^{(2)}$,...  are any finite or countably infinite list of mutually disjoint subsets, we have from \eqref{Wmackinf}
\begin{equation}\label{Wvee}
\mathcal{W}(E_{S^{(1)}} \vee_q E_{S^{(2)}} \vee_q ...) = \mathcal{W}(E_{S^{(1)}}) + \mathcal{W}(E_{S^{(2)}}) +...\;.
\end{equation}

\noindent But part (iv) of the proof of Theorem B1 tells us (replacing $J$'s by $S$'s) that
\begin{equation}\label{B1iv}
(\vee_q)_k (E_{S^{(k)}}) = E_{\cup_k(S^{(k)})},
\end{equation}

 \noindent where $k = 1, 2, ...$, so that \eqref{Wvee} becomes
 \begin{equation}
 \mathcal{W}(E_{S^{(1)}\cup S^{(2)}\cup ...}) = \mathcal{W}(E_{S^{(1)}}) + \mathcal{W}(E_{S^{(2)}}) +...\;.
 \end{equation}

 \noindent Applying \eqref{PW} we have
\begin{equation}\label{kolminfS}
\mathcal{P_S}(S^{(1)}\cup S^{(2)} \cup ... ) = \mathcal{P_S}(S^{(1)}) + \mathcal{P_S}(S^{(2)}) + ..., \;\text{when}
\;\;\; S^{(i)}\cap S^{(j)}=\oslash\;\text{  for } i\neq j.
\end{equation}

\noindent But this is just \eqref{kolminf} with $S$ replacing $A$, and since $\mathcal{W}$ is normalized, \eqref{probmeas} holds with $\mathcal{S}$ replacing $\Omega$.  Therefore $\mathcal{P_S}$ is a classical probability function on  $\mathcal{S}$. Hence by restricting $\mathcal{W}$, in the manner described, to each static framework $\mathcal{E_S}$, one obtains a network of classical probability functions.  The values of these probability functions are noncontextual since the right-hand side of \eqref{PW} does not depend on $\mathcal{S}$.  This proves Theorem B4. We now prove its converse.

{\bf Theorem B5:}  Suppose that we are given a noncontextual network of probability functions belonging to the static frameworks of $\mathcal{L^H}$. This means that for each sample space
$\mathcal{S}$ a classical probability function $\mathcal{P_S}$ is defined as acting on the subspaces in $\mathcal{L_S}$, or equivalently on the subsets $S$ of
$\mathcal{S}$, such that the values of the $\mathcal{P_S}$'s are noncontextual;  that is, if $S$, $S'$ are respectively subsets of $\mathcal{S}$, $\mathcal{S'}$ such that $E_S=E'_{S'}$,  then $\mathcal{P_S}(S) = \mathcal{P_{S'}}(S')$. Under these conditions, says the theorem, there exists a normalized lattice measure $\mathcal{W}$ such that each $\mathcal{P_S}$ is given by \eqref{PW}.

\noindent {\bf Proof}:  If $A$ is any subspace of $\mathcal{H}$, we can define a sample space
$\mathcal{S}_A = \{A, \neg_q A\}$ and let  $\mathcal{W}(A) = \mathcal{P}_{\mathcal{S}_A}(\{A\})$.  With $\mathcal{W}$ defined in this way for every $A$, it follows that \eqref{PW} will hold for any
$\mathcal{S}$ and any subset $S$ of $\mathcal{S}$.  Indeed, let us consider  two sample spaces, the first being $\mathcal{S}$ and the second $\mathcal{S'}=\mathcal{S}_{A}$ where $A = E_S$. Then $A$ is an event in both associated frameworks $\mathcal{L_S}$ and $\mathcal{L_{S'}}$,   and if we let $S' = \{A\}$ (the set of subspaces whose only member is $A$), then $E_{S'} = A$ also, so by noncontextuality $\mathcal{P_S}(S) = \mathcal{P_{S'}}(S')=\mathcal{P}_{\mathcal{S}_A}(\{A\}) = \mathcal{W}(A) = \mathcal{W}(E_S)$, which is just \eqref{PW}.

To show that $\mathcal{W}$ so defined is a lattice measure, we suppose that the events $A^{(1)}$,
$A^{(2)}$, ... are all orthogonal. Then we must show that  \eqref{Wmackinf} holds.   Let $A^{tot} = A^{(1)} \vee_q A^{(2)} \vee_q ...$ and $A_0 = \neg_q A^{tot}$.
Then $\mathcal{S} = \{A^{(0)}, A^{(1)}, A^{(2)}, ...\}$ is a sample space.  Therefore a probability function $\mathcal{P_S}$ exists and satisfies \eqref{kolminfS} for any list of (classically) disjoint subsets of $\mathcal{S}$.  Define $S^{(k)}=\{A^{(k)}\}$ for $k=1,2,....$, and $S^{tot} = \{A^{(1)},A^{(2)},...\}$; then $E_{S^{tot}} = A^{tot}$ and $E_{S^{(k)}} = A^{(k)}$
for each $k$.  Hence by \eqref{PW}, we have $\mathcal{W}(A^{tot}) = \mathcal{P_S}(S^{tot})$, and $\mathcal{W}(A^{(k)}) = \mathcal{P_S}(S^{(k)})$ for each $k$.
Consequently \eqref{kolminfS} becomes
\begin{equation}
\mathcal{W}(A^{tot}) = \mathcal{W}(A^{(1)}) + \mathcal{W}(A^{(2)}) + ...\;,
\end{equation}

\noindent which is just \eqref{Wmackinf}. Finally we observe that \eqref{PW}, applied to $\mathcal{S} = \{\mathcal{H},\oslash\}$ and $S=\{\mathcal{H}\}$, yields
$\mathcal{W(H)} = \mathcal{P_S}(S) =\mathcal{P_S}(\mathcal{H})= 1$, so that
$\mathcal{W}$ is normalized.  This completes the proof.

Putting Theorem B5 together with Gleason's Theorem, we infer that any network of probability functions
with noncontextual values must be (disregarding the exception in 2 dimensions) the restriction to the various static frameworks of a function of the form
\begin{equation}\label{GW}
\mathcal{W}_{\rho}(A) = \text{Tr}(\rho [A]),
\end{equation}

\noindent for some density matrix $\rho$, which we refer to as the \lq state' of the system (see below).  Hence, if we want the probability functions on all the frameworks to form a noncontextual network, they must be given by
\begin{equation}\label{GP}
\mathcal{P}_{\rho,\mathcal{S}}(A) = \mathcal{P}_{\rho,\mathcal{E}}(A)=\text{Tr}(\rho [A]), \;\;\; \text{for} A \in \mathcal{E},
\end{equation}

 \noindent where $\mathcal{E}$ is any framework containing $A$ (either as a member of the sample space $\mathcal{S}$ or as a compound event) and $\rho$ is a single density matrix applying to \emph{all} frameworks.  Equation \eqref{mackinf} will still be satisfied because of the linearity of \eqref{GP}.  (This equation should be compared with
 \eqref{purebornprob} in the text.)

Although $\mathcal{W}_{\rho}$ and $\mathcal{P}_{\rho,\mathcal{S}}$ have the same \emph{value} wherever their domains overlap, each has properties the other lacks: $\mathcal{W}_{\rho}$ is defined over
$\mathcal{L}^{\mathcal{H}}$ whereas any $\mathcal{P}_{\rho,\mathcal{S}}$ is defined only for a particular sublattice, and on the other hand $\mathcal{P}_{\rho,\mathcal{S}}$
satisfies certain rules such as the Kolmogorov overlap equation \eqref{kolm2} everywhere in its domain, while
$\mathcal{W}_{\rho}$ violates that equation when applied to two incompatible subspaces.

\subsection*{B.6. States, truth values and probability values}

In Section II we suggested that classical properties (subsets of phase space) can be regarded as predicates of propositions of which the phase point $x$  is the subject. The subsets are then given the truth value 1 if $x\in A$ and 0 if not, which makes the point $x$ the source of truth for properties. To carry the procedure over to QM, we might wish to make the state analogous to the phase point $x$, and the subspaces $A$ analogous to the Borel subsets of classical phase space. To discuss truth values we would then represent a state by a wave function (normalized ket) $|\psi\rangle$, which we could consider as the \lq source of truth'.  We have seen above, however, that to discuss probabilities we must represent a state by a density matrix $\rho$, in accordance with \eqref{GW} and \eqref{GP}. To blend these two approaches, we proceed as follows:

We generalize the Hilbert space ontology of Sec. IV.A and represent the quantum state by a \emph{density operator} $\rho$, which is a basic input to the theory, just as the state $x$ (or its probabilistic counterpart $\rho$) is an input to classical mechanics. In the special case of a \emph{pure state}, where the density operator has a single nonzero eigenvalue (which must be equal to unity), we designate the corresponding eigenvector by $\psi$ and write it as $|\psi\rangle$. The density operator can then be expressed in terms of the \emph{projector}

\begin{equation} \label{pureproj}
\rho= \rho_{\psi}= |\psi\rangle\langle\psi|,
\end{equation}

\noindent which could also be written as $[\psi]$. Then \eqref{GW} becomes

\begin{equation}\label{purebornrho}
\mathcal{W_{\rho_{\psi}}}(A) = \langle\psi|[A]||\psi\rangle,
\end{equation}

\noindent and similarly \eqref{GP}.  In particular any property $A$ satisfying  $[A]|\psi\rangle = |\psi\rangle$ is given probability 1, and those projectors that annihilate $|\psi\rangle$ have probability 0.  But probabilities of 1 and 0 imply truth values T and F respectively.  Thus, for pure states the density operator yields truth values for particular properties (those compatible with $[\psi]$), that are the same as the truth values inferred by taking $|\psi\rangle$ as primary and making it the source of truth.  These limited assignments of truth values do not violate no-go theorems because other properties have only probability values, not truth values.

\section{Families, Histories, Frameworks, and Probabilities}

In this appendix we shall furnish the details of extending CQT from the static case to a sequence of times.

\subsection*{C.1. Histories and families}

We define a \emph{history} of length $N$ as a set
\begin{equation} \label{homogdef}
C_N = (B_1, B _2,...., B_N)
\end{equation}
of properties (subspaces of $\mathcal{H})$, to be associated with an ordered sequence of times $t_1, t_2,...,t_N$. A \emph{family}  $\mathcal{F}_N$ of length $N$ is a set of entities, referred to as \emph{dynamic events}, which will turn out to be either histories (the homogeneous events) or history complexes (the inhomogeneous events). They are generated by the \emph{elementary histories} of the family, which are a set of histories obtained in the following way:
(This construction is the same as given in Subsection IV.B.1 of the text, after \eqref{elem}, except that here we go into detail.  We revert now to our usual practice of using the letter $A$, rather than $D$ as in Appendix B, for members of a static sample space.)

At each time $t_n$ we choose a sample space in the sense of Appendix B, that is a set
$\{A_n\} = \{A_n^1, A_n^2, ..., A_n^{m_n}\}$ of mutually orthogonal subspaces, which together span
$\mathcal{H}$.  We have denoted by $m_n$ the number of subspaces in $\{A_n\}$; each $m_n$ may be either finite or infinite.   The set $\{A_n\}$ also corresponds to a set            $\{[A_n]\}$ of projectors, which forms an \emph{orthogonal decomposition} of the identity operator on $\mathcal{H}$, that is relevant to the time $t=t_n$. The elementary
events of $\mathcal{F}_N$, henceforth also referred to as \emph{elementary histories}, are obtained by selecting from each sample space $\{A_n\}$ a particular member $A_n^{j_n}$, where $1\leq j_n\leq m_n$.  One thereby forms a particular history
\begin{equation}\label{elemhist}
C_N^{\{j\}}=(A_1^{j_1},...,A_N^{j_N}),
\end{equation}

\noindent which we call an elementary history of $\mathcal{F}_N$. (We denote by $j$ the sequence $j_1,...,j_N$.)   Clearly the family has $M$ elementary histories, where $M(\mathcal{F}_N) = \Pi_1^N m_n$ if all the $m_n$ are finite.

To construct the whole family, we imagine that each time $t_n$ is associated with a separate copy of $\mathcal{H}$.  Then an elementary member $C_N^{\{j\}}$ of $\mathcal{F}_N$ may be regarded as a subspace of $\mathcal{H}^N$, the $N$'th tensor product of $\mathcal{H}$ with itself. These $M$ subspaces are mutually orthogonal and together they span $\mathcal{H}^N$. Thus they form a sample space $\mathcal{S}_N$ in $\mathcal{H}^N$, and their projectors $[A_1^{j_1}] \times [A_2^{j_2}]\times...\times [A_N^{j_N}]$ are an orthogonal decomposition of the identity in $\mathcal{H}^N$.   The family $\mathcal{F}_N$ is then the event space, as defined in Appendix B.3, of $\mathcal{S}_N$,
and by Theorem B1, applied now to $\mathcal{H}^N$,  it is also the lattice closure.  We see that $\mathcal{F}_N$ is determined by the choice of the $N$ sample spaces $\{A_n\}$.  As in Subsection IV.C, we call the members of $\mathcal{S}_N$  \emph{dynamic events}, as opposed to the individual projectors $[A_n^{j_n}]$, which are \emph{static events}.

As in Appendix B, the events of the family $\mathcal{F}_N$ correspond 1-1 with the subsets $J$ of its $M$ elementary histories.  Therefore they are subject to a \lq property calculus' obtained by the replacements \eqref{qlog}, and each event can also be represented by a projection operator

\begin{equation}\label{tensor}
[C_N^J] = \sum_{j\in J}[C_N^j]= \sum_{j\in J} [A_1^{j_1}] \times [A_2^{j_2}] \times...\times [A_N^{j_N}].
\end{equation}

\noindent In this way the property calculus can be replaced by algebraic operations involving projectors. It is easily seen that the family $\mathcal{F}_N$ has
$2^M$ events if $M$ is finite.  Of these, one event is the null history corresponding to the empty set, and $M$
events, corresponding to the sets containing just one member, are the elementary histories themselves.  The remainder are \emph{compound events} or \emph{compound histories}, obtained by taking the disjunction, represented by the summation in \eqref{tensor}, in the property calculus belonging to the tensor space $\mathcal{H}^N$.

We now observe that some dynamic events, given by \eqref{tensor}, can be factored into the form \eqref{homogdef}, where each $B_n$ belongs to the static event space at time $t_n$.  These are \emph{homogeneous events} (terminology of \textcite{isham1994}) and we accept them as histories.  They are not all elementary histories, because $B_n$ is not necessarily an elementary member of the static event space, as are the $A_n^{j_n}$'s in \eqref{elemhist}.  Those events of $\mathcal{F}_N$ that cannot be factored in this way are \emph{inhomogeneous events} and we call them \emph{history complexes} rather than histories.

As an illustration, let $N=2$, $m_1 = 3,$ $m_2=2$, and let
$C_2 = (A_1^2, A_2^1)$, $C_2' = (A_1^3, A_2^1)$, $C_2'' = (A_1^3, A_2^2)$.  Then
$C_2\vee_q C_2' = (A_1^2\vee_q A_1^3, A_2^1)$ is a nonelementary homogeneous event with $B_1 = A_1^2\vee_q A_1^3$,
$B_2 = A_2^1$, but $C_2\vee_q C_2''$ is inhomogeneous. In this family there are $M = 3\times 2 = 6$ elementary histories (events) and altogether $2^6 = 64$ dynamic events, of which $2^3\times 2^2 = 8\times 4 = 32$ are homogeneous.  In general, for finite but not small $N$,
$m_1,....,m_N$, the $\Pi_1^N 2^{m_n} = 2^{\tilde{M}}$ homogeneous events of $\mathcal{F}_N$ (including the elementary events and the null event), where $\tilde{M} = \Sigma_1^N m_n$, constitute only a small fraction of all the $2^M$ events, where $M = \Pi_1^N m_n$.

In identifying an elementary history of $\mathcal{F}_N$ there are two steps at each time $t_n$: the choice of a static framework $\{A_n\}$ and the selection of an index $j_n$, which determines a subspace $A_n^{j_n}$ of $\mathcal{H}$. The whole family is identified by the first step at each time.  If we were talking about measurement (but we are not) we could say that the first step is choosing what measurement to make and the second is selecting a possible outcome.  To continue that analogy, we might let the choice of a measurement depend on the outcome of a prior measurement.

So here, it is in principle possible (see \textcite{grif1}, Eq. (8.37)) to allow each choice $\{A_n\}$ to depend on the prior selections $j_{i<n}$. \textcite{gell2007quasiclassical} have stressed the importance of this option, called \emph{branch dependence}, in cosmology.   For simplicity we shall exclude branch dependence in constructing families.  (With branch dependence, the formula for $M$ would have to be discarded since the numbers $m_i$ could also depend on prehistory.)  We leave it to the interested reader to verify that all the reasoning to be presented in subsequent parts of this Appendix would hold as well in the presence of branch dependence, although the notation would be fearfully encumbered.

\subsection*{C.2. Projectors and probabilities}

As mentioned above, the elementary histories of $\mathcal{F}_N$ form a complete, mutually orthogonal set in
$\mathcal{H}^N$,  i.e. a \emph{history sample space}.  The whole family corresponds 1-1 with subsets of this set.  Hence, treating $\mathcal{H}^N$
as we have treated $\mathcal{H}$, we can write q-logical operations within $\mathcal{F}_N$, which will turn out to be set operations (see Eqs.\eqref{evwedge} and \eqref{evvee}). The atomic members of $\mathcal{F}_N$ are just its elementary
histories (see Theorem B4).

As was done for $\mathcal{H}$ at the end of Appendix B.3, a classical probability function $\mathcal{P}_N$ can be defined on the elementary histories of  $\mathcal{F}_N$ by choosing any set of real nonnegative numbers that sum to 1.  The probabilities of compound events (homogeneous or not) are then determined, in analogy to \eqref{mackinf}, by the relation
\begin{equation}\label{Csum}
\mathcal{P}_N(C_N \vee_q C_N' \vee_q C_N'' +...) = \mathcal{P}_N(C_N) + \mathcal{P}_N(C_N')
+ \mathcal{P}_N(C_N'') + ....,
\end{equation}

\noindent where the q-operations on the left-hand side of \eqref{Csum} refer to the q-logic relevant to $\mathcal{H}^N$. At this point in Appendix B, we called the analogue of the family a \lq static framework'.  Here, however, $\mathcal{F}_N$
does not yet qualify as a framework: there is an additional condition that will be explained in Appendix C.4. In view of \eqref{tensor}, Eq.\eqref{Csum} may be rewritten as

\begin{equation}\label{[C]sum}
P_N([C_N] + [C_N'] + [C_N''] +...) = P_N([C_N]) + P_N([C_N']) + P_N([C_N'']) + ....,
\end{equation}

\noindent where, in analogy to the notation adopted in \eqref{WW1}, we use (Roman) $P$ for a function on projectors and other operators and
(script) $\mathcal{P}$ for a function on subspaces (properties). The symbols $P_N([C_N])$ and $\mathcal{P}_N(C_N)$ represent the same quantity.

So far nothing has been said that would prevent us from assigning arbitrary probabilities to each elementary history, as long as they add up to 1, and determining the probabilities of compound events by summation.  We wish, however, to introduce noncontextual conditional probabilities relating each time to those before.   To do this, it will be advantageous to change our perspective from regarding the successive bases $\{A_1\}$, $\{A_2\}$, ..., $\{A_N\}$ as belonging to separate copies of $\mathcal{H}$, to seeing them as existing together in the same Hilbert space, so that the relationship of $A_n$ to $A_{n+1}$ in the same history can be treated algebraically.  For this purpose it will be helpful to think of a history $C_N$ as built up step by step out of its initial subhistories $C_1$, $C_2$,...,$C_{N-1}$ rather than coming into being all at once.  Let us review this process.

To form an elementary history $C_N\in \mathcal{F}_N$, we choose $N$ successive \emph{projection times} $t_1<t_2< ....< t_N$, and select $N$ projectors $([A_1], ...., [A_N])$ belonging to the respective static sample spaces at these times.  We also fix the state at a time $t_0<t_1$.  (Griffiths sometimes regards the state as part of the history, but we do not.)  We shall also need to represent the time evolution of the state under the action of the Hamiltonian $H$.  As explained in Subsection B.6, the state of the system is in general described by a density matrix $\rho$, whose time development between projection times is

\begin{equation}\label{Schrho}
\frac{d}{dt}\rho(t) = i[\rho(t),H].
\end{equation}

\noindent For the rest of this appendix, however, we shall restrict ourselves to the special case of a pure state, described by a wave function, which we write as $|\psi\rangle$ (or simply as $\psi$), in which case Eq.\eqref{Schrho} becomes the Schroedinger equation

\begin{equation}\label{Schreq}
\frac{d}{dt} |\psi(t)\rangle = -iH |\psi(t)\rangle.
\end{equation}

\noindent This is closer to Griffiths's own presentation and it is intended to make the equations more transparent.  Strictly speaking, this means that our proofs will thus apply only to the pure state case, but it is a simple matter to generalize them by replacing \eqref{Schreq} with \eqref{Schrho}.  All the reasoning, including variants and deeper comments, is unaffected by the change from $|\psi\rangle$ to
$\rho$.

Independently of the state evolution, in constructing an elementary history we suppose that at each time $t_{n>0}$ the system \lq has' the property associated with $[A_n]$.  This supposition will of course receive a probability determined by the state and (if $n>1$) the previous $[A_{n'<n}]$.  The evolution will then be followed through \emph{conditional probabilities} at each $t_n$ based on the supposition made for $t_{n-1}$.  It is as though there were a \lq collapse' at each projection time, but in the microscopic theory we have no physical collapse.  There is only a system of conditional probabilities mathematically resembling a sequence of collapses.  For any $n=1,..,N-1$, we have defined
the  \emph{subhistory} $C_n$ to be $(A_1, ...., A_n)$.  By arresting the procedure at $t_n$ we obtain
a family $\mathcal{F}_n$ of length $n$, whose elementary histories are the possible $C_n$'s.

The time evolution associated with an elementary history thus consists of a continuous Schroedinger development given by \eqref{Schreq}, controlled by the Hamiltonian $H$, punctuated by projections at times $t_n$.  Fortunately the effect of \eqref{Schreq} can be removed by a simple transformation.  It turns out that the properties
$[A_n]$ enter into probabilities through products of the form
$[A_n]U(t_n,t_{n-1})[A_{n-1}]...[A_2]U(t_2,t_1)[A_1]U(t_1,t_0)$, which can also be written as
$U(t_n,t_0)\bar{A}_n \bar{A}_{n-1}...\bar{A}_2 \bar{A}_1$, where
\begin{equation}\label{adef}
\bar{A}_n \equiv U(t_n,t_0)^{-1} [A_n] U(t_n,t_0),
\end{equation}
and
\begin{equation} \label{udef}
U(t,t') = \text{exp}[-iH(t-t')].
\end{equation}
\noindent We shall also use the notation

\begin{equation} \label{histdefbar}
\bar{C}_N = (\bar{A}_1, \bar{A} _2,...., \bar{A}_N),
\end{equation}

\noindent for the same history $C_N$, expressed in terms of the $\bar{A_n}$. (We shall freely switch between the $C_N$ and the $\bar{C}_N$ notations in this appendix.) All calculations can be done in terms of
the $\bar{A}_n$, which are also projectors since $U(t,t')$ is unitary. We prefer, however, not to introduce the subspaces associated with these projectors because they are physically artificial and confusing to the intuition.  The $\bar{A}_n$ will be recognized as the projection operators that enter into the \lq Heisenberg picture' of QM.  Our choice is to think physically in the \lq Schroedinger picture', where the subspaces (hence their projectors) are fixed in time and the wave function evolves, but to calculate in the Heisenberg picture, using the projectors $\bar{A}_n$ and the
fixed wave function $|\psi_0\rangle = |\psi(t=t_0)\rangle$.

The relations among the terms family, homogeneous and inhomogeneous event, history and history complex, elementary and nonelementary history, remain unchanged when we replace the projectors $[A]$ by their Heisenberg counterparts $\bar{A}$, as do the ideas of tensor products and addition of histories as in \eqref{[C]sum}.  The sequence of projectors $(\bar{A}_1,...)$, however, has a special feature that will be essential to our reasoning.  If two successive projectors $\bar{A}_n, \bar{A}_{n+1}$ happen to be equal, then the second projection might as well not have happened and the time $t_{n+1}$ can be deleted from the history.  (In the Schroedinger picture, we would need to say that the system, supposed to have the property $[A_n]$ at $t_n$, evolves so as to have the property $[A_{n+1}]$ at $t_{n+1}$.)  We shall refer to this as \emph{redundancy} of two equal $\bar{A}$'s in succession.  The redundancy concept will enter into the argument behind Theorem C1, which is a central part of the reasoning of the next part of this Appendix.

\subsection*{C.3. The extended Born rule}
In the static case the Born rule \eqref{GP} for probabilities is linear in the projector, but since it involves self-adjoint projectors it can equally well be written as
\begin{equation} \label{quadratic}
\mathcal{P}_{\rho}(A)=\text{Tr}(\rho [A]^{\dagger}[A]),
\end{equation}
\noindent where we have suppressed the subscript $\mathcal{E}$ indicating the framework.
In the dynamic case, according to Griffiths, it is \eqref{quadratic} that provides the generalization of the probability formula. Specifically, since the dynamic generalization of a property is a history,
given the wave function $|\psi_0\rangle$ at $t=t_0$ (corresponding to the density operator $\rho_0= |\psi_0\rangle\langle\psi_0|$), Griffiths assumes that the probability of an elementary history is given by the Born rule for histories

\begin{equation}\label{chainprob}
\mathcal{P}_{\rho_{0},N}(C_N) = \text{Tr}(\rho_0\hat{C}_N^{\dagger} \hat{C}_N) = \langle\psi_0|\hat{C}_N^{\dagger} \hat{C}_N|\psi_0\rangle,
\end{equation}

\noindent where the \emph{chain operator}
\begin{equation}\label{chSch}
\hat{C}_N = U(t_N,t_0)[A_N] U(t_N,t_{N-1})[A_{N-1}]...[A_2]U(t_2,t_1)[A_1]U(t_1,t_0)
\end{equation}

\noindent is a single operator acting on $\mathcal{H}$, not a sequence of subspaces like $C_N$, or an operator on $\mathcal{H}^N$ like $[C_N^j]$ defined in Eq. \eqref{tensor}.  In view of \eqref{adef} and \eqref{histdefbar}, the chain operator can also be written in terms of the Heisenberg projectors as
\begin{equation}\label{chHeis}
\hat{C}_N = \bar{A}_N \bar{A}_{N-1} ... \bar{A}_2 \bar{A}_1.
\end{equation}
\noindent We shall usually suppress the subscript $\rho_0$ of \eqref{chainprob} in writing probability functions.

\noindent \underline{Derivation of the extended Born rule \eqref{chainprob}}

It is our intention in this part of the Appendix to \emph{derive} Eq.\eqref{chainprob} for elementary histories of $\mathcal{F}_N$, rather than postulating it.  The function $\mathcal{P}_N$ will then be determined on all of $\mathcal{F}_N$ by
\eqref{Csum}.  In the next subsection we shall present an additional derivation proving \eqref{chainprob} for nonelementary histories.  This will raise the danger of a contradiction with the values of $\mathcal{P}_N$ already determined by \eqref{Csum}.  To avoid such a contradiction, one must
impose Griffiths's \lq consistency condition', which also involves the wave function. Only if this condition is satisfied may the pair ($\psi,\mathcal{F}_N$) be called a \emph{framework}.

Let us discuss first a family of length $N=1$.  The determination of a probability function on the sample space $\{\bar{A}_1^1,...,\bar{A}_1^{m_1}\}$ is to be accomplished through an interplay of the (Heisenberg) properties associated with $t_1$ and the wave function associated with $t_0$.  Proceeding in accord with Subsection IV.B, we wish the values of this probability function to be noncontextual. This implies, because of Gleason's Theorem, that it should be obtained by restriction from a normalized lattice measure over all of $\mathcal{L}^{\mathcal{H}}$.  This measure  must be independent of the choice of a system of projectors at $t_1$ and it must therefore be determined by the wave function
$|\psi_0\rangle$ alone.  For this reason we call the measure $W_0$ and consider it as belonging to the time $t_0$, although the process of restricting it to a probability function belongs to the time $t_1$, since that is when the sample space of projectors is chosen.   Now we proceed as in Appendix B to obtain
\begin{equation}\label{W0}
W_0(\hat{B}) = \langle\psi_0|\hat{B}|\psi_0\rangle,
\end{equation}

\noindent for any projector $\hat{B}\in\mathcal{L}^{\mathcal{H}}$.  At $t_1$ we choose the particular sample space of projectors $\{\bar{A}_1\}=\{\bar{A}_1^1,...,\bar{A}_1^{m_1}\}$ and deduce that the probabilities of the different
$\bar{A}_1^{j_1}$ are given
by the restriction of $W_0$, that is
\begin{equation}\label{P1}
P_1(\bar{A}_1^{j_1}) =  W_0(\bar{A}_1^{j_1}) = \langle\psi_0 |\bar{A}_1^{j_1}|\psi_0\rangle.
\end{equation}

\noindent The function $P_1$ can now be understood as the classical probability function associated with the family $\{\bar{A}_1\}$ of length 1. It is defined for general events of the family by the summation rule \eqref{Csum}.  Its sum over elementary events is 1 because $W_0$ is normalized.

It should be understood that the reasoning leading to \eqref{W0} and \eqref{P1} is the same single-time reasoning that was used in Appendix B.  The artificial separation between two times $t_0$ and $t_1$ is introduced only in order to display analogy between the step at $n=1$ and the later steps involving conditional probabilities.  The confrontation between state and projector takes place at the single time $t_1$; the state then is the same (in the Heisenberg picture) as it was at  $t_0$.

Now we consider $N>1$.   At time $t_{n-1}$ (with $n\leq N$) there exists an elementary history
$\bar{C}_{n-1} = (\bar{A}_1^{j_1}, ...., \bar{A}_{n-1}^{j_{n-1}})$, reflecting all the selections from sample spaces chosen at times $t_1,...,t_{n-1}$.  With respect to $t_n$ or to $\bar{C}_n$, we may call $\bar{C}_{n-1}$ a prehistory, i.e. it is the subhistory of a subhistory.  Given this prehistory, there must be a \emph{conditional probability distribution} governing the selection of $\bar{A}_n^{j_n}$ out of the sample space chosen at $t_n$.   Since the selection of $\bar{A}_n^{j_n}$ is determined by the selection of $j_n$, we shall abbreviate notation by writing the conditional probability as $Q_{\bar{C}_{n-1},\{\bar{A}_n\}} (j_n)$, where $\{\bar{A}_n\}$ denotes the whole sample space $\{\bar{A}_n^1, ...,\bar{A}_n^{m_n}\}$ at $t_n$.    The probability of the whole history
$C_n = (A_1^{j_1}, ...., A_n^{j_n})$ must then be given by
\begin{equation}\label{PQ}
\mathcal{P}_n(C_n) =  Q_{\bar{C}_{n-1},\{\bar{A}_n\}} (j_n)\mathcal{P}_{n-1}(C_{n-1}).
\end{equation}

\noindent Of course $Q$ is required to be normalized:  $\Sigma_{j=1,m_n}Q_{\bar{C}_{n-1},\{\bar{A}_n\}}(j_n) = 1$, so that the normalization of $\mathcal{P}_n$ will follow from that of  $\mathcal{P}_{n-1}$.

\textcite{nist} has introduced the important idea that \emph{the conditional probability values themselves should be noncontextual, with respect to the sample space at $t_n$.}  That is, if
$\{\bar{A'}_n\}$ is an alternative sample space $\{\bar{A'}_n^1, ...\bar{A'}_n^{\,m'_n}\}$, and
$\bar{A}_n^j=\bar{A'}_n^{\,j'}$ for a particular pair $(j,j')$, then we must have
$Q_{\bar{C}_{n-1},\{\bar{A}_n\}}(j) = Q_{\bar{C}_{n-1},\{\bar{A'}_n\}}(j')$.  Note that the same prehistory
$\bar{C}_{n-1}$ appears on both sides of this equation.

As in the static theory, the consequence of the values of $Q$ being noncontextual (remember that the argument $j$ stands for the projector $\bar{A}^j$) is that $Q$ must be the restriction of a normalized measure on all projectors in $\mathcal{L}^{\mathcal{H}}$.  Since this measure does not have to do with the choice of a sample space at $t_n$, we associate it with the time $t_{n-1}$  and call it $Z_{\bar{C}_{n-1}}$, dropping the subscript $\{\bar{A}_n\}$.  Equation\eqref{PQ} now becomes
\begin{equation}\label{PWdyn}
\mathcal{P}_n(C_n) =  Z_{\bar{C}_{n-1}} (\bar{A}_n^{j_n})\mathcal{P}_{n-1}(C_{n-1}).
\end{equation}

\noindent For convenience let us think of $\bar{C}_0$ as the unique history of zero length, and define
\begin{equation}\label{C0}
Z_{\bar{C}_0} = W_0, \;\;\;\;\;\; \mathcal{P}_0(C_0) = 1.
\end{equation}

\noindent Then the first equality of \eqref{P1} becomes a special case of \eqref{PWdyn}, so that the latter is now established for $n=1$ as well as for higher $n$.  We shall usually drop the superscript $j_n$ on the right-hand side of   \eqref{PWdyn} since the selection of $\bar{A}_n^{j_n}$ from the $n$'th sample space is implied by the history $C_n$ on the left-hand side. From the point of view of the candidate framework which includes the state as the \lq zero'th' member, we may regard $C_0$ as a \lq prehistory' inherited from time $t_0$, and $Z_{\bar{C}_0}$ as the conditional probability of $C_1$, conditioned on the state.

At this stage it is important to note that $\hat{C}_N$ is not in general a projection operator, so that it is problematic to derive the quadratic relation \eqref{chainprob} from Gleason's Theorem, which is linear in the projector.  The program suggested by \textcite{nist} is to construct such a derivation using \eqref{PWdyn}. In our derivation only the single-time version of Gleason's Theorem will be used in proving Theorem C3, once (above) at $t_1$ and again at each subsequent time. We first infer from \eqref{PWdyn} a recursion relation involving only the $Z$'s, as expressed in the following theorem:

{\bf Theorem C1}:  Consider an elementary history $C_N\in\mathcal{F}_N$.  Write $C_N$ as
$(A_1,..., A_N)$, dropping the superscripts $j_n$.  For $1\leq n<N$, if $\hat{B}$ is any projector $\leq\bar{A}_n$, then
\begin{equation}\label{WW}
Z_{\bar{C}_n}(\hat{B}) = \frac{Z_{\bar{C}_{n-1}}(\hat{B}) }{Z_{\bar{C}_{n-1}}(\bar{A}_n)}.
\end{equation}

{\noindent \bf{Proof}:}
Let $1\leq n<N$ and consider a projector $\hat{B}\leq\bar{A}_n$.  We can form a history
$\bar{C'}_{n+1} = (\bar{C}_n, \hat{B}) = (\bar{A}_1,...,\bar{A}_n,\hat{B})$.  Since $\hat{B}$ need not belong to the sample space $\{\bar{A}_{n+1}\}$, $\bar{C'}_{n+1}$ does not in general belong to
$\mathcal{F}_{n+1}$, the $(n+1)$st  subfamily  of $\mathcal{F}_N$, but its prehistory
$\bar{C}_n$ belongs to $\mathcal{F}_n$.  For definiteness let us say that $\bar{C'}_{n+1}$
belongs to a family $\mathcal{F}'_{n+1}$ which is identical to $\mathcal{F}_{n+1}$, except that in $\mathcal{F}'_{n+1}$ the sample space $\{\bar{A}_{n+1}\}$ is replaced by
$\{\bar{A}_{n+1}'\}$, a refinement of $\{\bar{A}_n\}$ in which the two projectors $\hat{B}$ and $\bar{A}_n-\hat{B}$ are substituted for
$\bar{A}_n$. (Recall that $\hat{B}\leq \bar{A}_n$.)  Then $\mathcal{P'}_{n+1}$ is a valid probability function, and by a double application of \eqref{PWdyn}, we obtain
\begin{equation}\label{PWdoub}
\mathcal{P'}_{n+1}(C'_{n+1}) = Z_{\bar{C}_n}(\hat{B}) Z_{\bar{C}_{n-1}}(\bar{A}_n) \mathcal{P}_{n-1}(C_{n-1}),
\end{equation}

\noindent where the history $\bar{C'}_{n+1}$ can be written as $(\bar{C_n}, \hat{B})$ or equivalently
as $(\bar{C}_{n-1}, \bar{A}_n, \hat{B})$.

On the other hand, $\mathcal{F}'_{n+1}$ contains a compound (i.e., nonelementary) history $(\bar{C}_n,\bar{A}_n)$, the disjunction of
$(\bar{C}_n,\hat{B})$ with $(\bar{C}_n,\bar{A}_n - \hat{B})$. Therefore, since $ \hat{B}\leq \bar{A}_n$, the selection of $\hat{B}$ out of the sample space $\{\bar{A}_{n+1}'\}$ in forming $\bar{C}_{n+1}'$ can be viewed in two steps: first select $\bar{A}_n$, then from within $\{\bar{A}_n\}$ select $\hat{B}$.  But the first step is \emph{redundant} with the selection of $\bar{A}_n$ at $t_n$.  Therefore the selection at $t_n$ can be dropped, and $\bar{C}'_{n+1}$ is equivalent to
$(\bar{C}_{n-1}, \hat{B})$ with $\hat{B}$ occurring at $t_{n+1}$. It follows that in \eqref{PWdyn} we may replace the left-hand side
by $\mathcal{P}'_{n+1}({C}'_{n+1})$, while retaining $n-1$ on the right-hand side but replacing $\bar{A}_n^{j_n}$ with $\hat{B}$.  We thus obtain
\begin{equation}\label{PWsing}
\mathcal{P}'_{n+1}(C'_{n+1}) = Z_{\bar{C}_{n-1}}(\hat{B})  \mathcal{P}_{n-1}(C_{n-1}).
\end{equation}

\noindent As the left sides of \eqref{PWdoub} and \eqref{PWsing} are identical, we may equate the right sides and drop the common factor $\mathcal{P}_{n-1}(C_{n-1})$.  After rearrangement one obtains \eqref{WW}.  Theorem C1 is proved.

We now introduce an important theorem due to \textcite{CZ}, Theorem C2 below. Let $V$ and $W$ be any two measures on the projectors in
$\mathcal{L}^{\mathcal{H}}$ (see \eqref{projmeas}).  Let $\hat{A}$ be a fixed nonzero projector.  Let us say that \lq $W \text{ quot}(\hat{A})\;V$' iff for every projector $\hat{B}\leq \hat{A}$, we have
$W(\hat{B}) = V(\hat{B})/V(\hat{A}).$  Let $\tilde{V}$ be the unique function defined from all linear operators on $\mathcal{H}$ to complex numbers, whose restriction to projectors is $V$, and which possesses the properties attributed to $\tilde{W}$ in \eqref{oplin}. (The existence and uniqueness of $\tilde{V}$ are guaranteed by Gleason's Theorem.)

{\bf Theorem C2:} The condition $W \text { quot}(\hat{A})\;V$ implies that
\begin{equation}\label{CZ}
W(\hat{B}) = \frac{\tilde{V}(\hat{A}\hat{B}\hat{A})}{V(\hat{A})},
\end{equation}

\noindent for \emph{all} projectors $\hat{B}$.  [Note that \eqref{CZ} determines
$W(\hat{B})$ for all projectors $\hat{B}$, although the condition $W \text { quot}(\hat{A})\;V$ without the theorem determines it only for $\hat{B}\leq \hat{A}$, in which case $\hat{A}\hat{B}\hat{A} = \hat{B}$.]

It is through this theorem that the quadratic dependence on $\hat{C}_N$ in \eqref{chainprob} will come about.  (A simplified proof of Theorem C2 for a restricted case will be given in Appendix D.) Using this theorem we are able to solve the recursion relation \eqref{WW} by proving:

{\bf Theorem C3:}  For $0\leq n<N$,  the projector measure $Z_{\bar{C}_n}$ is given by
\begin{equation}\label{chainW}
Z_{\bar{C}_n}(\hat{B}) =
	\frac{\langle\psi_0|\hat{C}_n^{\dagger}\hat{B} \hat{C}_n|\psi_0\rangle}
	{\langle\psi_0|\hat{C}_n^{\dagger}\hat{C}_n|\psi_0\rangle},
\end{equation}

\noindent for all projectors $\hat{B}\in\mathcal{L}^{\mathcal{H}}$. ($\hat{C}_n$ is the result of replacing $N$
by $n$ in \eqref{chHeis}; $\hat{C}_0 = 1$.)

\noindent \textbf{Proof} by induction on $n$ (the following argument follows the reasoning of \textcite{nist} with some changes of detail):

\noindent Initial step: let $n=0$.  Then replacing $\hat{C}_0$ by 1, \eqref{chainW} becomes
\begin{equation}
Z_{\bar{C}_0}(\hat{B}) = \langle\psi_0|\hat{B}|\psi_0\rangle,
\end{equation}

\noindent which follows from \eqref{C0} and \eqref{W0}.

\noindent Induction step:  For all $0 < n \leq N $, define
\begin{equation}\label{V}
V_n(\hat{B}) = \langle\psi_0|\hat{C}_{n-1}^{\dagger}\hat{B}\hat{C}_{n-1}|\psi_0\rangle,
\end{equation}

\noindent and
\begin{equation}\label{W}
W_n(\hat{B}) = Z_{\bar{C}_n}(\hat{B}),
\end{equation}

\noindent for all projectors $\hat{B}$ in $\mathcal{L}^{\mathcal{H}}$.  (Formally \eqref{W} is an extension of \eqref{C0}, but that equation furnished the definition of $Z_{\bar{C}_0}$ whereas \eqref{W} defines $W_n$.
The extension is unambiguous within this proof, because Theorem C3 pertains to only one full history
$\bar{C}_N$ so that for each $n$ there is only one subhistory $\bar{C}_n$.)

Assume that \eqref{chainW} holds when $n$ is replaced by $n-1$.  That is,
\begin{equation}\label{chainWpr}
Z_{\bar{C}_{n-1}}(\hat{B}) =
	\frac{\langle\psi_0|\hat{C}_{n-1}^{\dagger}\hat{B} \hat{C}_{n-1}|\psi_0\rangle}
	{\langle\psi_0|\hat{C}_{n-1}^{\dagger}\hat{C}_{n-1}|\psi_0\rangle}.
\end{equation}

\noindent Since \eqref{chainWpr} holds for all projectors $\hat{B}$, it holds with $\hat{A}_n$ in place of
$\hat{B}$.  But this replacement does not affect the denominator; therefore
\begin{equation}\label{TV}
\frac{Z_{\bar{C}_{n-1}}(\hat{B})}{Z_{\bar{C}_{n-1}}(\bar{A}_n)} =
	\frac{\langle\psi_0|\hat{C}_{n-1}^{\dagger}\hat{B} \hat{C}_{n-1}|\psi_0\rangle}
		{\langle\psi_0|\hat{C}_{n-1}^{\dagger}\bar{A}_n \hat{C}_{n-1}|\psi_0\rangle}
	=\frac{V_n(\hat{B})}{V_n(\bar{A}_n)},
\end{equation}

\noindent as defined in \eqref{V}. We now recall Theorem C1:  Eq.\eqref{WW} holds for all projectors $\hat{B}\leq \bar{A}_n$.
But in view of \eqref{W} and \eqref{TV}, \eqref{WW} becomes
\begin{equation}\label{WV}
W_n(\hat{B}) = \frac{V_n(\hat{B})}{V_n(\bar{A}_n)},
\end{equation}

\noindent so that Theorem C1 says that \eqref{WV} holds for all projectors $\hat{B}\leq \bar{A}_n$.

Now, $V_n$ is easily seen to be a measure on projectors, but not necessarily normalized.  By Gleason's Theorem, $V_n$ is the restriction to projectors of a function $\tilde{V}_n$, which can only have the form
\begin{equation}\label{VJ}
\tilde{V}_n(\hat{J}) = \langle\psi_0|\hat{C}_{n-1}^{\dagger}\hat{J}\hat{C}_{n-1}|\psi_0\rangle,
\end{equation}

\noindent for arbitrary operators $\hat{J}$ acting on $\mathcal{H}$.  Looking at \eqref{WV}, we see that the assertion of Theorem C1 is exactly the condition
\begin{equation}\label{bqa}
W_n\text{ quot}(\bar{A}_n)\;V_n,
\end{equation}

\noindent required by Theorem C2.  Therefore the conclusion of Theorem C2,
\begin{equation}\label{alphabet}
W_n(\hat{B}) = \frac{\tilde{V}_n(\bar{A}_n\hat{B}\bar{A}_n)}{V_n(\bar{A}_n)},
\end{equation}

\noindent is established. Finally, we substitute $\bar{A}_n$ for $\hat{B}$ in \eqref{V}, and use the fact that

\noindent $\bar{A}_n = \bar{A}_n^2 = \bar{A}_n^{\dagger}\bar{A}_n$ and $\bar{A}_n \hat{C}_{n-1} = \hat{C}_n$, obtaining
\begin{equation}\label{VA}
V_n(\bar{A}_n) = \langle\psi_0|\hat{C}_n^{\dagger}\hat{C}_n|\psi_0\rangle.
\end{equation}
 \noindent We also substitute $\bar{A}_n\hat{B}\bar{A}_n$ for $\hat{J}$ in \eqref{VJ} to obtain
\begin{equation}\label{VABA}
\tilde{V}_n(\bar{A}_n\hat{B}\bar{A}_n) = \langle\psi_0|\hat{C}_n^{\dagger}\hat{B}\hat{C}_n|\psi_0\rangle,
\end{equation}

\noindent since $\hat{C}_n = \bar{A}_n\hat{C}_{n-1}$.  Substituting \eqref{VA} and \eqref{VABA} into \eqref{alphabet}, we have \eqref{chainW}.
We have thus shown that \eqref{chainW} follows from \eqref{chainWpr}.  The proof by induction of Theorem C3 is complete.

A word about the part played by \eqref{C0} and \eqref{W0} in Theorems C1 and C3.  The definition \eqref{C0} is an artificial device that makes it possible, in Theorem C1, to allow $n=1$ in \eqref{WW}, \eqref{PWdoub}, \eqref{PWsing}.  Otherwise the initial step in the proof of Theorem C3 would have to be $n=1$ and the induction would have to start with $n=2$, making the proof more unwieldy.   The finding \eqref{W0}, on the other hand, is derived from our reasoning in the static case (Appendix B.6) and forms an essential part of our approach to the \lq single-time' case of the Born rule.

Griffiths postulates the Born rule from the start, but we wish to derive it rather than stating it didactically.  Since the probability of a history is deduced from that of its immediate prehistory via the conditional probability that relates them, the induction proof of Theorem C3 extends the Born rule step by step to histories of arbitrary length; but there must be a starting point, \eqref{W0} or the corresponding  equation with $\rho$ rather than $\psi$, that does not depend on conditional probability.  To justify this starting point on the basis of the empirical correctness of quantum mechanical predictions would sacrifice the important principle that (in the \lq microscopic theory' at least) the theory should be purely deductive
and not refer to actual measurements.  Therefore \eqref{W0} must be deduced, and this requires the appeal to Gleason's Theorem we make in Appendix B.6.

We are now able to prove Griffiths's extended Born rule, \eqref{chainprob}, by substituting \eqref{chainW}
into the recursion \eqref {PWdyn}.

{\bf Theorem C4:}  the probability of an elementary history of length $N$ is given correctly by \eqref{chainprob}.

\noindent \textbf{Proof:} Iterating \eqref{PWdyn}, we find
\begin{equation}\label{chainprod}
\mathcal{P}_N(C_N) = Z_{\bar{C}_{N-1}}(\bar{A}_{N-1})...Z_{\bar{C}_1}(\bar{A}_1)\mathcal{P}_1(C_1).
\end{equation}

\noindent Substituting $\bar{A}_{n+1}$ for $\hat{B}$ in \eqref{chainW}, we have
\begin{equation}\label{chainWE}
Z_{\bar{C}_n}(\bar{A}_{n+1}) =
	\frac{\langle\psi_0|\hat{C}_n^{\dagger}\bar{A}_{n+1}\hat{C}_n|\psi_0\rangle}
			{\langle\psi|\hat{C}_n^{\dagger}\hat{C}_n|\psi_0\rangle}
		= \frac{\langle\psi_0|\hat{C}_{n+1}^{\dagger}\hat{C}_{n+1}|\psi_0\rangle}
			{\langle\psi|\hat{C}_n^{\dagger}\hat{C}_n|\psi_0\rangle},
\end{equation}

\noindent so that in \eqref{chainprod} all the intermediate factors cancel and we are left with
\begin{equation}
\mathcal{P}_N(C_N) = \mathcal{P}_1(C_1) \;\frac{\langle\psi_0|\hat{C}_N^{\dagger}\hat{C}_N|\psi_0\rangle}
			{\langle\psi_0|\hat{C}_1^{\dagger}\hat{C}_1|\psi_0\rangle},
\end{equation}

\noindent which is identical to \eqref{chainprob},
in view of \eqref{P1} and the fact that $\hat{C}_1 = \bar{A}_1^{j_1}=\bar{A}_1$ is a projector.

We note that the content of Theorems C1-4 can be decomposed into two parts:

\noindent (a) For a given wave function $|\psi_0\rangle$, the probability of a history $C_N$ is given by an expression of the
form $\mathcal{D}(C_N,C_N)$, where for any two histories $C_{\alpha}$, $C_{\alpha'}$ the expression $\mathcal{D}(\alpha, \alpha')$, called a \emph{decoherence functional}, has certain algebraic properties.

\noindent (b) The decoherence functional referred to above has the specific form $\mathcal{D}(\alpha, \alpha') = \langle\psi_0|\hat{C}_{\alpha'}^{\dagger}\hat{C}_{\alpha}|\psi_0\rangle$.

 \textcite{isham1994}, as well as \textcite{sorkin1994} prove, effectively, that if  (a) is \emph{assumed} then (b) follows.  This is far short of proving both (a) and (b) from reasonable assumptions, which is done here. In Subsec. IV.C of the text, however, we comment on the wider scope of these papers, that goes beyond nonrelativistic QM.

\subsection*{C.4. Consistency: from families to frameworks}

So far we have determined only the probabilities of those histories that are elementary within a family.  Let us take the simple example described in part 1 of this Appendix, in which $N=2$, $m_1=3$, $m_2=2$.  There are  $3\times 2 = 6$ elementary histories. The probabilities of compound events must be determined by summing those of the elementary events composing them, as explained at the beginning of Appendix C.2.  In our example, we may let
$\bar{C}_2 = (\bar{A}_1^2,\bar{A}_2^1)$ and $\bar{C}_2'' = (\bar{A}_1^3,\bar{A}_2^2)$.  Since any two elementary histories are disjoint,  \eqref{[C]sum} yields
\begin{equation}
P_2(\bar{C}_2\vee_q \bar{C}''_2) = P_2(\bar{C_2}) + P_2(\bar{C}''_2).
\end{equation}

\noindent Likewise one may construct an event by disjunction (\lq q-or') from any subset of the set of 6 elementary histories.   Altogether there are $2^6 = 64$ such events in this family, including the 6 elementary histories constructed from a set with one event, and the null history constructed from the empty set.  The probability of any such event is the sum of the probabilities of the elementary histories that make it up.  In all of this reasoning there arises no problem of consistency; all the probabilities identified so far are consistent.

Of particular interest, however, are those compound events that we have called \emph{homogeneous} in Appendix C.1.  We naturally identify compound homogeneous events as those formed by disjunction as described above.  As illustration, take again the family described in the first paragraph above.  Let
a homogeneous event $\bar{C}_2^{B} = (\bar{B}_1,\bar{B}_2)$ be chosen by letting $\bar{B}_1 = \bar{A}_1^2 + \bar{A}_1^3$,
$\bar{B}_2 = \bar{A}_2^1$.  Then $\bar{B}_1$ \lq says' that either $\bar{A}_1^2$ or $\bar{A}_1^3$ is selected, and $\bar{B}_2$ \lq says' that $\bar{A}_2^1$ is selected.  In terms of the tensor space
$\mathcal{H^N}$, we may say that $\bar{C}_2^{B} = \bar{C}_2\vee_q \bar{C}'_2$, that is,
$\bar{C}_2^{B}$ says that either $\bar{C}_2$ or $\bar{C'}_2$ is selected.
Therefore the probability of $\bar{C}_2^{B}$ is given in accordance with \eqref{Csum} as
\begin{equation}\label{cpd}
P_2(\bar{C}_2^{B}) = P_2(\bar{C}_2) + P_2(\bar{C}_2')
	=  \langle\psi_0|\bar{A}_1^2 \bar{A}_2^1 \bar{A}_1^2|\psi_0\rangle + \langle\psi_0|\bar{A}_1^3 \bar{A}_2^1 \bar{A}_1^3|\psi_0\rangle,
\end{equation}

\noindent in accordance with \eqref{chainprob}.  (We have contracted $\bar{A}_2^1 \bar{A}_2^1$ to $\bar{A}_2^1$.)  So far there is still no inconsistency.

But suppose that we substitute the history $\bar{C}_2^{\bar{B}} = (\bar{B}_1, \bar{B}_2)$ directly into \eqref{chainprob}.  We then get
\begin{equation}\label{cpdint}
P_2(\bar{C}_2^{\bar{B}}) = \langle\psi_0|\bar{B}_1 \bar{B}_2 \bar{B}_1|\psi_0\rangle
	=   \langle\psi_0|(\bar{A}_1^2+\bar{A}_1^3) \bar{A}_2^1 (\bar{A}_1^2+\bar{A}_1^3)|\psi_0\rangle.
	\end{equation}

\noindent Because of the quadratic nature of \eqref{chainprob}, Eq.\eqref{cpdint} has four terms, two of which add up to \eqref{cpd} and the other two are both equal to 	
$\langle\psi_0|\bar{A}_1^2 \bar{A}_2^1 \bar{A}_1^3)|\psi_0\rangle$ (in more complicated cases the two interference terms are conjugate so that their sum is always real).  In order to ensure that \eqref{cpdint} will always agree with
\eqref{cpd}, \textcite{grif1} imposes a \emph {consistency condition}
\begin{equation}\label{griffcc}
\text{Re}(\langle\psi_0|\hat{C}_N^{\dagger}\hat{C}_N'|\psi_0\rangle) = 0,
\end{equation}

\noindent where $\hat{C}_N$ and $\hat{C}_N'$ are the chain operators for any two distinct \emph{elementary} histories.  Only under this additional condition, which involves the wave function as well as the histories,
does he admit that the pair $(\psi_0, \mathcal{F}_N)$ is a framework. Equation \eqref{griffcc} is known as the \lq weak decoherence' condition.

But now we ask, why is it necessary for \eqref{cpd} and \eqref{cpdint} to agree?  Griffiths takes it for granted that \eqref{chainprob} should apply directly to any history whether elementary or not.  But we are deriving \eqref{chainprob} rather than positing it, and we have derived it only for elementary histories.  Can we dispense with the consistency condition by simply disallowing \eqref{chainprob} unless the history is elementary?

We answer this question in the negative by another noncontextuality argument.  Consider a compound history $\bar{C}_N^{\bar{B}} = (\bar{B}_1,...,\bar{B}_N)$, belonging to the family we have described.  For each $n$, $\bar{B}_n$ is the sum of some subset $X_n$ of the projectors $\bar{A}_n^{j_n}$ belonging to the sample space at $t_n$.  Create a new family by deleting all members of $X_n$ from the $n'th$ sample space and replacing them by the single projector $\bar{B}_n$; do this at every $n$ from 1 to $N$.  This new family is built up out of valid new sample spaces at each $t_n$, and in it the history $\bar{C}_2^{\bar{B}}$ is elementary.  Therefore in the new family Eq.\eqref{chainprob} applies and the probability of $\bar{C}_2^{\bar{B}}$ is given  by \eqref{cpdint}.  But the concept of noncontextuality requires that
 $\bar{C}_2^{\bar{B}}$ have the same probability value in the new family as in the old.  Therefore \eqref{cpdint} must agree with \eqref{cpd}, and in general we must have \eqref{griffcc}.  So we have derived the weak decoherence condition for frameworks quite generally. (In Sec. IV.C of the text we discuss the more restrictive condition known as \lq medium decoherence', which is widely considered to be necessary as well.)

If one takes it for granted that all the steps of induction (corresponding to our Theorems C1 and C3), and therefore the final result \eqref{chainprob}, apply to all histories, then there is no need for the argument in the preceding paragraph.  We object, however, on the ground that (as stressed by Griffiths) all true probabilities must arise from a probability function, which can exist only in relation to a sample space.  In our proof of Theorem C1, there are frequent references to sample spaces.  When one deals with a history that is not elementary, there is no sample space and therefore the argument breaks down.  This is why we consider it necessary, as in the preceding paragraph, to introduce another, auxiliary, family, in which the history under consideration is elementary, and to apply noncontextuality to the histories in that family.

\section {Limited proof of the CZ theorem}

The theorem of \textcite{CZ} (Theorem C2), as we use it, states: Let $V$ be a lattice measure on the projectors of $\mathcal{H}$, and $\hat{A}$ a fixed nonzero
projector.  We seek a normalized lattice measure $W$ on the projectors such that for any projector $\hat{B}$ satisfying
\begin{equation}\label{cz1}
\hat{B}\leq \hat{A},
\end{equation}

\noindent $W$ satisfies
\begin{equation}\label{cz2}
W(\hat{B}) = \frac{V(\hat{B})}{V(\hat{A})}.
\end{equation}

\noindent Then the unique $W$ satisfying \eqref{cz2} under the condition \eqref{cz1} is
\begin{equation}\label{cz3}
W(\hat{B}) = \frac{\tilde{V}(\hat{A}\hat{B}\hat{A})}{V(\hat{A})},
\end{equation}

\noindent where $\tilde{V}$ is the linear extension of $V$ given by Gleason's Theorem, and
\eqref{cz3} is asserted for all projectors $\hat{B}$.

\noindent \textbf{Proof:} It is trivial that \eqref{cz3} satisfies \eqref{cz2} given \eqref{cz1}, since $\hat{A} = \hat{B}$ when $\hat{B}\leq \hat{A}$, and that \eqref{cz3} describes a normalized lattice measure since the rhs is 1 when $\hat{B}=1$ and the additivity condition is guaranteed by $\tilde{V}$ being linear.  Therefore the proof consists in showing that if \eqref{cz2} holds for all $\hat{B}$ satisfying \eqref{cz1}, then $W$ must satisfy \eqref{cz3} for all projectors
$\hat{B}$. Here we shall content ourselves with showing that $W$ satisfies \eqref{cz3} for projectors $\hat{B}$ such
that:
\begin{equation}\label{cz4}
\text{the eigenvalues of }\hat{A}\hat{B}\hat{A} \text{ form a discrete countable set.}
\end{equation}

\noindent First we prove a lemma applicable to all projectors $\hat{B}$, regardless of \eqref{cz1} or \eqref{cz4}.

\noindent \textbf{Lemma:}  for any projector $\hat{B}$,
\begin{equation}\label{cz5}
W(\hat{B}) = \tilde{W}(\hat{A}\hat{B}\hat{A}),
\end{equation}

\noindent where $\tilde{W}$ is the linear extension of $W$ given by Gleason's Theorem.

\noindent \textbf{Proof:}  Define
\begin{equation}\label{cz6}
\hat{A}^{\perp} = 1-\hat{A}.
\end{equation}

\noindent Since $(\hat{A}^{\perp})^2 = 1-2\hat{A}+\hat{A}^2 = 1-\hat{A} = \hat{A}^{\perp}$, $\hat{A}^{\perp}$ is a projector.  By additivity of $\tilde{W}$, we have $\tilde{W}(\hat{A}^{\perp}) =  \tilde{W}(1) - \tilde{W}(\hat{A})$.  But
$\tilde{W}(1) = 1$ because $W$ is normalized, and $\tilde{W}(\hat{A}) = 1$ by setting
$\hat{B} =\hat{A}$ in \eqref{cz2}. Therefore
\begin{equation}\label{cz7}
\tilde{W}(\hat{A}^{\perp}) = 0.
\end{equation}

\noindent Now, the linearity of $\tilde{W}$  implies that there exists a countable set of (unnormalized) vectors $|\phi_i\rangle$ such that for all operators $\hat{J}$
\begin{equation}\label{cz8}
\tilde{W}(\hat{J}) = \sum_{i} \langle\phi_i|\hat{J}|\phi_i\rangle.
\end{equation}

\noindent Setting $\hat{J}=\hat{A}^{\perp}$ in \eqref{cz8} and using \eqref{cz7}, we have
\begin{equation}\label{cz9}
0 = \sum_{i} \langle\phi_i|\hat{A}^{\perp}|\phi_i\rangle = \sum_{i} \langle\phi_i|\hat{A}^{\perp}\hat{A}^{\perp}|\phi_i\rangle,
\end{equation}

\noindent where each term is real $\geq 0$.  It follows that
\begin{equation}\label{cz10}
\hat{A}^{\perp}|\phi_i\rangle =0,
\end{equation}

\noindent for every $i$. To complete the proof of the lemma, we note that $\hat{B} = (\hat{A}+\hat{A}^{\perp})\hat{B}(\hat{A}+\hat{A}^{\perp})$ and so
\begin{equation}\label{cz11}
\tilde{W}(\hat{B}) = \tilde{W}(\hat{A}\hat{B}\hat{A}) + \tilde{W}(\hat{A}\hat{B}\hat{A}^{\perp}) + \tilde{W}(\hat{A}^{\perp}\hat{B}\hat{A}) + \tilde{W}(\hat{A}^{\perp}\hat{B}\hat{A}^{\perp}).
\end{equation}

\noindent Letting $\hat{J}$ in \eqref{cz8} be $\hat{A}\hat{B}\hat{A}^{\perp}$, $\hat{A}^{\perp}\hat{B}\hat{A}$, $\hat{A}^{\perp}\hat{B}\hat{A}^{\perp}$, in turn, we see from \eqref{cz10} that the last three terms in \eqref{cz11} vanish, and the lemma is proved since $W(\hat{B}) = \tilde{W}(\hat{B})$.

We now introduce the condition \eqref{cz4}.  To each eigenvalue $\lambda_i$ of $\hat{A}\hat{B}\hat{A}$ there corresponds an eigenspace whose projection operator may be called $\hat{P}_i$, and we have
\begin{equation}\label{cz12}
\hat{A}\hat{B}\hat{A} = \sum_{i} \lambda_i \hat{P}_i,
\end{equation}

\noindent as well as
\begin{equation}\label{cz13}
\hat{A}\hat{B}\hat{A} \hat{P}_i = \hat{P}_i \hat{A}\hat{B}\hat{A} = \lambda_i \hat{P}_i,
\end{equation}

\noindent for each $i$. Consider any $i$ for which $\lambda_i\neq 0$.  Using \eqref{cz13} and $\hat{A}^2 = \hat{A}$, we can write
\begin{eqnarray}\label{cz14}
\hat{P}_i  \hat{A} & = & \hat{P}_i (\hat{A}\hat{B}\hat{A}/\lambda_i)\hat{A} = \hat{P}_i (\hat{A}\hat{B}\hat{A}/\lambda_i)  =  \hat{P}_i,
\end{eqnarray}

\noindent and likewise
\begin{equation}\label{cz15}
\hat{A} \hat{P}_i = \hat{P}_i.
\end{equation}

\noindent Thus \eqref{cz1} is satisfied by replacing $\hat{B}$ with $\hat{P}_i$, and hence from \eqref{cz2}
\begin{equation}\label{cz16}
\tilde{W}(\hat{P}_i) = W(\hat{P}_i) = \frac{V(\hat{P}_i)}{V(\hat{A})} = \frac{\tilde{V}(\hat{P}_i)}{V(\hat{A})},
\end{equation}

\noindent provided that $\lambda_i\neq 0$. We now substitute \eqref{cz12} into \eqref{cz5}, obtaining
\begin{equation}\label{cz17}
W(\hat{B}) = \sum_{i} \lambda_i \tilde{W}(\hat{P}_i).
\end{equation}

\noindent If there is a $\lambda_i=0$, the corresponding term contributes zero to \eqref{cz17}, which
can therefore be written
\begin{equation}\label{cz18}
W(\hat{B}) = \sum_{i}^{\lambda_i\neq 0} \lambda_i \tilde{W}(\hat{P}_i).
\end{equation}

\noindent Equation \eqref{cz16} can now be applied to each term of \eqref{cz18}, yielding
\begin{equation}\label{cz19}
W(\hat{B}) = \frac{\sum_{i}^{\lambda_i\neq 0} \lambda_i \tilde{V}(\hat{P}_i)}{V(\hat{A})}.
\end{equation}

\noindent But from \eqref{cz12}, treating $\tilde{V}$ as we did $\tilde{W}$, we also obtain
\begin{equation}\label{cz20}
\tilde{V}(\hat{A}\hat{B}\hat{A}) = \sum_{i}^{\lambda_i\neq 0} \lambda_i \tilde{V}(\hat{P}_i).
\end{equation}

\noindent Comparing \eqref{cz20} to \eqref{cz19}, we obtain \eqref{cz3}, Q.E.D. If \eqref{cz4} is not assumed, the reasoning from \eqref{cz12} to \eqref{cz20} must be replaced by a more difficult argument given by \textcite{CZ} using the spectral theorem.

\vspace{10mm}

\vspace{10mm}

\section {Glossary}

\begin{itemize}

\item assertions: the set of correct statements a realistic theory makes about the objects in the ontology,
including statements of probability.

\item atom, atomic: $A$ is an atomic  member of a lattice if there exists no nonzero $B$ such that $B<A$.

\item beable: entity contained in the ontology. What a realistic theory is \lq about'.

\item Boolean lattice: a lattice in which the distributive law (as well as the lattice axioms) holds.

\item Born rule: a formula for evaluating the probability of a property or a history, given the state. In the present work it is sometimes referred to as the \lq microscopic' Born rule, in contrast to the \lq macroscopic' Born rule giving probabilities for the outcomes of measurement.

\item branch dependence: a feature of a family in which the sample space chosen at each time $t_n$ may depend on the prehistory selected up to $t_{n-1}$.

\item c-assertions (c-probability functions, c-truth): the set of correct but \emph{contextual} statements that can be made about quantum properties or histories within a single framework.

\item classical history: a history, all of whose properties are classical. Classical histories are discretized trajectories in phase space.

\item compatible frameworks: two (dynamic) frameworks are mutually compatible if their states are mutually compatible and if in addition, each history in one framework is compatible with all the histories in the other. Individual frameworks are internally compatible by definition, since all the histories in a family are mutually compatible.

\item compatible histories: two histories having the same projection times $t_1,...,t_N$ are (mutually) compatible if for each $n$, the $n$th property of one history is compatible with the $n$th property of the other.

\item compatible properties: two quantum properties (subspaces) whose projectors commute.

\item Compatible Quantum Theory (CQT):  the present realistic formulation of quantum mechanics that comprises a microscopic part (MIQM) and a macroscopic part (MAQM).

\item compatible sublattice: a lattice of mutually compatible properties.

\item compound property: a nonatomic member of the event space of a static framework.

\item compound history: a homogeneous event that is nonatomic in the event space of a family.

\item consistency conditions: conditions of orthogonality (decoherence) between elementary histories in a family. These conditions involve the state.

\item Consistent Histories theory (CH): the original formulation of the histories approach due to Griffiths and its later elaborations.

\item contextual: framework dependent.

\item decoherence: the physical mechanism by which properties of a quantum system lose correlations with other properties through interactions with an environment.

\item decoherence functional: a bilinear (hermitian) functional on histories in terms of which the decoherence or consistency conditions for frameworks are expressed, and whose
diagonal part gives the quadratic Born rule.

\item Decoherent Histories theory (DH): the version of the histories approach due to Gell-Mann and Hartle that had the formulation of quantum cosmology as its primary motivation. The approach considers an infinite system from the start, so that the distinction between the microscopic and macroscopic theories does not appear natural from this point of view. The formulation emphasizes the role of coarse graining and decoherence (hence its name) in the emergence of classical properties and histories in the macroscopic domain and in selecting physical frameworks.

\item dynamic case: the description of a quantum system at a sequence of $N$ times.

\item dynamic event: a member of the many-time event space in the dynamic case.

\item elementary history of a family: a member of the sample space, i.e. a history formed by selecting, at each of the projection times chosen for that family, one member of the static framework chosen at that time.

\item elementary property: a member of the one-time sample space, or equivalently an atom of the lattice or framework.

\item event: in probability theory, the argument of a probability function. In CQT, in the static case, a property that belongs to a static framework, determined by a subset of the sample space;  in the dynamic case, a member of a family, determined by a subset of the sample space.

\item event algebra/event space: the set of all events that form the domain of a probability function, i.e. a static framework of properties in the one-time case or a family of events in the many-time case.

\item family: a collection of many-time (dynamic) events generated algebraically by its elementary histories. The events may be homogenous or inhomogeneous.

\item framework: a static framework is a Boolean sublattice of  the lattice of properties in $\mathcal{H}$ whose elementary members (atoms) span $\mathcal{H}$. A dynamic framework is a pair, consisting of a state and a family, that satisfies consistency conditions.

\item history: a generalization of a property to multiple times, obtained by considering
one property at each of a sequence of times.  When considered as a member of a family, a history is the same as a homogeneous event.

\item history complex: a member of a family/event space that does not qualify as a history; an inhomogeneous event.

\item homogeneous event: a member of a family consisting of a time-ordered sequence of properties. Some homogeneous events are elementary (the properties are atomic), others are compound.  The distinction elementary/compound applies also to histories considered as members of a family.

\item inhomogeneous event: a member of a family that is not homogeneous; a history complex.

\item lattice: a partially ordered set satisfying additional properties spelled out in Appendix A. Quantum properties in Hilbert space form a non-Boolean lattice.

\item lattice measure: a function from quantum properties or histories to $[0,\infty]$ that satisfies a linearity condition like that of Kolmogorov, but with \lq exclusive' interpreted as \lq orthogonal' rather than \lq nonintersecting'.

\item macroscopic: a macroscopic system is a large (ideally infinite) system.

\item macroscopic quantum mechanics (MAQM): the part of CQT, pertaining to macroscopic systems which provides a mechanism for selecting a physical framework from the multiplicity of incompatible frameworks appearing in MIQM.

\item many-time case: equivalent to dynamic case.

\item  measure (classical): a function from subsets of some \lq universal' set $\Omega$ to $[0,\infty]$, satisfying
Kolmogorov's additivity condition.

\item measurement: a physical interaction between a system \textbf{S} and a classical measuring device whose purpose is to determine the truth values of selected properties of \textbf{S}. Measurements are defined only in MAQM.

\item measurement problem: this problem, which arises in the orthodox formulation of Subsec. V.B, concerns the explanation of wave function collapse.

\item medium decoherence: a relation, involving the decoherence functional, which is required for the consistency of a framework.

\item microscopic: a theory is microscopic if it applies to arbitrary closed systems, regardless of size.

\item noncontextual: framework independent.

\item normalized lattice measure: a lattice measure that maps the whole Hilbert space into 1. Its range is therefore the unit interval.

\item ontology: the set of all the beables in a theory.

\item operationalism (operationalist): a physical theory is said to be operationalist if in describing a system \textbf{S} it requires entities external to \textbf{S}.

\item physical framework selection: a mechanism for identifying, among all the mutually incompatible frameworks of a closed quantum system, a particular framework that incorporates physical truth and falsehood. The selection mechanism can be either \lq external', via a classical measurement, or \lq internal', constructed from a larger quantum system of which the system under study is a subsystem.

\item physical framework: a framework composed of physical histories.

\item physical history: a history, whose initial properties are quantum mechanical and whose final properties are classical.

\item physical truth: the truth associated with the physical assertions, pertaining to the selected physical framework.

\item prehistory: a history consisting of the first $n-1$ properties of a given subhistory $C_n$.

\item probability function: a function $p$ from a \lq universal' set $\Omega$ (known as the sample space) to the interval $[0,1]$, which maps $\Omega$ itself to 1.  Such a function naturally induces a function $\mathcal{P}$  from a field of subsets of $\Omega$ to $[0,1]$.

\item projection operator/projector: a self-adjoint operator $\hat{J}$ that satisfies $\hat{J}^2 = \hat{J}$.  Any such
operator is uniquely associated with some subspace $A$ according to the relation $\hat{J} = [A]$ defined in the paper. Then we say that $\hat{J}$ is the projector of $A$.

\item property: a classical property is a subset of phase space. A quantum property is a subspace of Hilbert space. Quantum properties can be represented by their projectors.

\item quantum state: same as state.

\item real (reality): the term \lq real' applied to a property or a state is ambiguous and is avoided in our formulation of quantum mechanics. We prefer to speak of \lq true' or \lq false' properties.

\item realism (realistic): the formulation of a physical theory is realistic if in describing a closed system \textbf{S} it uses only entities and concepts pertaining to $\mathcal{S}$.

\item sample space: in general, a set whose subsets may serve as arguments for a probability function. In CQT, a set of orthogonal properties (subspaces) that span the whole Hilbert space $\mathcal{H}$, or a set of orthogonal events that span $\mathcal{H}^N$. The sample space generates all the events in the event space.

\item state: the mathematical representation of the system under study. For classical mechanics it is a point in phase space in the deterministic case and a probability distribution over points in the stochastic case. For quantum mechanics it is a ray of vectors in Hilbert space for pure states, and a density matrix for mixed states. The state is the source of truth and of probability for properties and histories, which without the state have neither truth nor probability values.

\item static case: the description of a quantum system that does not take time dependence into account. It is equivalent to the one-time ($N=1$) dynamic case.

\item subhistory: a history consisting of the first $n$ properties of a history $C_N$, with $n<N$.

\item subspace: a subspace of a Hilbert space is a subset that is itself a Hilbert space.

\item true/false (truth/falsehood): defined by a truth function.

\item truth function: a function assigning a value 0 or 1 to each property or history in its domain.

\item weak decoherence: a necessary condition for the consistency of a framework, which involves the decoherence functional.

\end{itemize}
\bibliography{QR-ref}

\begin{thebibliography}{64}%
\makeatletter
\providecommand \@ifxundefined [1]{%
 \@ifx{#1\undefined}
}%
\providecommand \@ifnum [1]{%
 \ifnum #1\expandafter \@firstoftwo
 \else \expandafter \@secondoftwo
 \fi
}%
\providecommand \@ifx [1]{%
 \ifx #1\expandafter \@firstoftwo
 \else \expandafter \@secondoftwo
 \fi
}%
\providecommand \natexlab [1]{#1}%
\providecommand \enquote  [1]{``#1''}%
\providecommand \bibnamefont  [1]{#1}%
\providecommand \bibfnamefont [1]{#1}%
\providecommand \citenamefont [1]{#1}%
\providecommand \href@noop [0]{\@secondoftwo}%
\providecommand \href [0]{\begingroup \@sanitize@url \@href}%
\providecommand \@href[1]{\@@startlink{#1}\@@href}%
\providecommand \@@href[1]{\endgroup#1\@@endlink}%
\providecommand \@sanitize@url [0]{\catcode `\\12\catcode `\$12\catcode
  `\&12\catcode `\#12\catcode `\^12\catcode `\_12\catcode `\%12\relax}%
\providecommand \@@startlink[1]{}%
\providecommand \@@endlink[0]{}%
\providecommand \url  [0]{\begingroup\@sanitize@url \@url }%
\providecommand \@url [1]{\endgroup\@href {#1}{\urlprefix }}%
\providecommand \urlprefix  [0]{URL }%
\providecommand \Eprint [0]{\href }%
\providecommand \doibase [0]{http://dx.doi.org/}%
\providecommand \selectlanguage [0]{\@gobble}%
\providecommand \bibinfo  [0]{\@secondoftwo}%
\providecommand \bibfield  [0]{\@secondoftwo}%
\providecommand \translation [1]{[#1]}%
\providecommand \BibitemOpen [0]{}%
\providecommand \bibitemStop [0]{}%
\providecommand \bibitemNoStop [0]{.\EOS\space}%
\providecommand \EOS [0]{\spacefactor3000\relax}%
\providecommand \BibitemShut  [1]{\csname bibitem#1\endcsname}%
\let\auto@bib@innerbib\@empty
\bibitem [{\citenamefont {Adler}\ and\ \citenamefont {Bassi}(2009)}]{ab5}%
  \BibitemOpen
  \bibfield  {author} {\bibinfo {author} {\bibnamefont {Adler}, \bibfnamefont
  {Stephen~L}}, \ and\ \bibinfo {author} {\bibfnamefont {Angelo}\ \bibnamefont
  {Bassi}}} (\bibinfo {year} {2009}),\ \bibfield  {title} {\enquote {\bibinfo
  {title} {Is quantum theory exact?}}\ }\href {\doibase
  10.1126/science.1176858} {\bibfield  {journal} {\bibinfo  {journal}
  {Science}\ }\textbf {\bibinfo {volume} {325}}~(\bibinfo {number} {5938}),\
  \bibinfo {pages} {275--276}}\BibitemShut {NoStop}%
\bibitem [{\citenamefont {Allahverdyan}\ \emph {et~al.}(2013)\citenamefont
  {Allahverdyan}, \citenamefont {Balian},\ and\ \citenamefont
  {Nieuwenhuizen}}]{bal}%
  \BibitemOpen
  \bibfield  {author} {\bibinfo {author} {\bibnamefont {Allahverdyan},
  \bibfnamefont {Armen~E}}, \bibinfo {author} {\bibfnamefont {Roger}\
  \bibnamefont {Balian}}, \ and\ \bibinfo {author} {\bibfnamefont {Theo~M}\
  \bibnamefont {Nieuwenhuizen}}} (\bibinfo {year} {2013}),\ \bibfield  {title}
  {\enquote {\bibinfo {title} {Understanding quantum measurement from the
  solution of dynamical models},}\ }\href@noop {} {\bibfield  {journal}
  {\bibinfo  {journal} {Physics Reports}\ }\textbf {\bibinfo {volume}
  {525}}~(\bibinfo {number} {1}),\ \bibinfo {pages} {1--166}}\BibitemShut
  {NoStop}%
\bibitem [{\citenamefont {Bacciagaluppi}(2009)}]{bacc}%
  \BibitemOpen
  \bibfield  {author} {\bibinfo {author} {\bibnamefont {Bacciagaluppi},
  \bibfnamefont {Guido}}} (\bibinfo {year} {2009}),\ \href
  {http://philsci-archive.pitt.edu/3380/} {\enquote {\bibinfo {title} {Is logic
  empirical?}}\ }\bibinfo {note} {In: D. Gabbay, D. Lehmann and K. Engesser
  (eds), Handbook of Quantum Logic (Elsevier Science
  Publications).}\BibitemShut {Stop}%
\bibitem [{\citenamefont {Bell}(1964)}]{bell14}%
  \BibitemOpen
  \bibfield  {author} {\bibinfo {author} {\bibnamefont {Bell}, \bibfnamefont
  {J~S}}} (\bibinfo {year} {1964}),\ \bibfield  {title} {\enquote {\bibinfo
  {title} {{On the Einstein-Podolsky-Rosen paradox}},}\ }\href@noop {}
  {\bibfield  {journal} {\bibinfo  {journal} {Physics}\ }\textbf {\bibinfo
  {volume} {1}},\ \bibinfo {pages} {195}}\BibitemShut {NoStop}%
\bibitem [{\citenamefont {Bell}(2004)}]{b8}%
  \BibitemOpen
  \bibfield  {author} {\bibinfo {author} {\bibnamefont {Bell}, \bibfnamefont
  {J~S}}} (\bibinfo {year} {2004}),\ \href@noop {} {\emph {\bibinfo {title}
  {{Speakable and Unspeakable in Quantum Mechanics}}}}\ (\bibinfo  {publisher}
  {Cambridge University Press})\BibitemShut {NoStop}%
\bibitem [{\citenamefont {Bell}(1966)}]{bell1966}%
  \BibitemOpen
  \bibfield  {author} {\bibinfo {author} {\bibnamefont {Bell}, \bibfnamefont
  {John~S}}} (\bibinfo {year} {1966}),\ \bibfield  {title} {\enquote {\bibinfo
  {title} {On the problem of hidden variables in quantum mechanics},}\
  }\href@noop {} {\bibfield  {journal} {\bibinfo  {journal} {Reviews of Modern
  Physics}\ }\textbf {\bibinfo {volume} {38}},\ \bibinfo {pages}
  {447--452}}\BibitemShut {NoStop}%
\bibitem [{\citenamefont {Ben-Menahem}(2005)}]{hilary}%
  \BibitemOpen
  \bibfield  {author} {\bibinfo {author} {\bibnamefont {Ben-Menahem},
  \bibfnamefont {Yemima}}} (\bibinfo {year} {2005}),\ \href@noop {} {\emph
  {\bibinfo {title} {Hilary Putnam}}}\ (\bibinfo  {publisher} {Cambridge
  University Press})\BibitemShut {NoStop}%
\bibitem [{\citenamefont {Birkhoff}\ and\ \citenamefont {von
  Neumann}(1936)}]{bvn}%
  \BibitemOpen
  \bibfield  {author} {\bibinfo {author} {\bibnamefont {Birkhoff},
  \bibfnamefont {G}}, \ and\ \bibinfo {author} {\bibfnamefont {J}~\bibnamefont
  {von Neumann}}} (\bibinfo {year} {1936}),\ \bibfield  {title} {\enquote
  {\bibinfo {title} {{The logic of quantum mechanics}},}\ }\href@noop {}
  {\bibfield  {journal} {\bibinfo  {journal} {Annals of Mathematics}\ }\textbf
  {\bibinfo {volume} {37}},\ \bibinfo {pages} {823--843}}\BibitemShut {NoStop}%
\bibitem [{\citenamefont {Bohm}(1952{\natexlab{a}})}]{bohm15}%
  \BibitemOpen
  \bibfield  {author} {\bibinfo {author} {\bibnamefont {Bohm}, \bibfnamefont
  {D}}} (\bibinfo {year} {1952}{\natexlab{a}}),\ \bibfield  {title} {\enquote
  {\bibinfo {title} {{A suggested interpretation of the quantum theory in terms
  of "hidden" variables. I}},}\ }\href@noop {} {\bibfield  {journal} {\bibinfo
  {journal} {Phys. Rev.}\ }\textbf {\bibinfo {volume} {85}},\ \bibinfo {pages}
  {166}}\BibitemShut {NoStop}%
\bibitem [{\citenamefont {Bohm}(1952{\natexlab{b}})}]{bohm2}%
  \BibitemOpen
  \bibfield  {author} {\bibinfo {author} {\bibnamefont {Bohm}, \bibfnamefont
  {D}}} (\bibinfo {year} {1952}{\natexlab{b}}),\ \bibfield  {title} {\enquote
  {\bibinfo {title} {{A suggested interpretation of the quantum theory in terms
  of "hidden" variables: II}},}\ }\href@noop {} {\bibfield  {journal} {\bibinfo
   {journal} {Phys.Rev.}\ }\textbf {\bibinfo {volume} {85}},\ \bibinfo {pages}
  {180--193}}\BibitemShut {NoStop}%
\bibitem [{\citenamefont {de~Broglie}(1926)}]{debroglie}%
  \BibitemOpen
  \bibfield  {author} {\bibinfo {author} {\bibnamefont {de~Broglie},
  \bibfnamefont {L}}} (\bibinfo {year} {1926}),\ \bibfield  {title} {\enquote
  {\bibinfo {title} {The principles of the new undulatory mechanics},}\ }\href
  {\doibase 10.1051/jphysrad:01926007011032100} {\bibfield  {journal} {\bibinfo
   {journal} {Journal de Physique et le Radium}\ }\textbf {\bibinfo {volume}
  {{7}}},\ \bibinfo {pages} {321--337}}\BibitemShut {NoStop}%
\bibitem [{\citenamefont {Brukner}\ and\ \citenamefont
  {Zeilinger}(1999)}]{bruk-zeil}%
  \BibitemOpen
  \bibfield  {author} {\bibinfo {author} {\bibnamefont {Brukner}, \bibfnamefont
  {{\v{C}}aslav}}, \ and\ \bibinfo {author} {\bibfnamefont {Anton}\
  \bibnamefont {Zeilinger}}} (\bibinfo {year} {1999}),\ \bibfield  {title}
  {\enquote {\bibinfo {title} {Operationally invariant information in quantum
  measurements},}\ }\href@noop {} {\bibfield  {journal} {\bibinfo  {journal}
  {Physical Review Letters}\ }\textbf {\bibinfo {volume} {83}}~(\bibinfo
  {number} {17}),\ \bibinfo {pages} {3354}}\BibitemShut {NoStop}%
\bibitem [{\citenamefont {Bub}(1999)}]{bub}%
  \BibitemOpen
  \bibfield  {author} {\bibinfo {author} {\bibnamefont {Bub}, \bibfnamefont
  {Jeffrey}}} (\bibinfo {year} {1999}),\ \href@noop {} {\emph {\bibinfo {title}
  {Interpreting the quantum world}}}\ (\bibinfo  {publisher} {Cambridge
  University Press})\BibitemShut {NoStop}%
\bibitem [{\citenamefont {Byrne}(2007)}]{byrne}%
  \BibitemOpen
  \bibfield  {author} {\bibinfo {author} {\bibnamefont {Byrne}, \bibfnamefont
  {Peter}}} (\bibinfo {year} {2007}),\ \bibfield  {title} {\enquote {\bibinfo
  {title} {{The many worlds of Hugh Everett}},}\ }\href@noop {} {\bibfield
  {journal} {\bibinfo  {journal} {Scientific American Magazine}\ }\textbf
  {\bibinfo {volume} {297}}~(\bibinfo {number} {6}),\ \bibinfo {pages}
  {98--105}}\BibitemShut {NoStop}%
\bibitem [{\citenamefont {Cassinelli}\ and\ \citenamefont {Zanghi}(1983)}]{CZ}%
  \BibitemOpen
  \bibfield  {author} {\bibinfo {author} {\bibnamefont {Cassinelli},
  \bibfnamefont {Gianni}}, \ and\ \bibinfo {author} {\bibfnamefont
  {N}~\bibnamefont {Zanghi}}} (\bibinfo {year} {1983}),\ \bibfield  {title}
  {\enquote {\bibinfo {title} {{Conditional probabilities in quantum mechanics.
  I.—Conditioning with respect to a single event}},}\ }\href@noop {} {\bibfield
   {journal} {\bibinfo  {journal} {Il Nuovo Cimento B Series 11}\ }\textbf
  {\bibinfo {volume} {73}}~(\bibinfo {number} {2}),\ \bibinfo {pages}
  {237--245}}\BibitemShut {NoStop}%
\bibitem [{\citenamefont {Cassinello}\ and\ \citenamefont
  {S{\'a}nchez-G{\'o}mez}(1996)}]{cassinello}%
  \BibitemOpen
  \bibfield  {author} {\bibinfo {author} {\bibnamefont {Cassinello},
  \bibfnamefont {Andr{\'e}s}}, \ and\ \bibinfo {author} {\bibfnamefont
  {Jos{\'e}~Luis}\ \bibnamefont {S{\'a}nchez-G{\'o}mez}}} (\bibinfo {year}
  {1996}),\ \bibfield  {title} {\enquote {\bibinfo {title} {On the
  probabilistic postulate of quantum mechanics},}\ }\href@noop {} {\bibfield
  {journal} {\bibinfo  {journal} {Foundations of Physics}\ }\textbf {\bibinfo
  {volume} {26}}~(\bibinfo {number} {10}),\ \bibinfo {pages}
  {1357--1374}}\BibitemShut {NoStop}%
\bibitem [{\citenamefont {Caves}\ \emph {et~al.}(2002)\citenamefont {Caves},
  \citenamefont {Fuchs},\ and\ \citenamefont {Schack}}]{bayes}%
  \BibitemOpen
  \bibfield  {author} {\bibinfo {author} {\bibnamefont {Caves}, \bibfnamefont
  {Carlton~M}}, \bibinfo {author} {\bibfnamefont {Christopher~A}\ \bibnamefont
  {Fuchs}}, \ and\ \bibinfo {author} {\bibfnamefont {R{\"u}diger}\ \bibnamefont
  {Schack}}} (\bibinfo {year} {2002}),\ \bibfield  {title} {\enquote {\bibinfo
  {title} {{Quantum probabilities as Bayesian probabilities}},}\ }\href@noop {}
  {\bibfield  {journal} {\bibinfo  {journal} {Physical review A}\ }\textbf
  {\bibinfo {volume} {65}}~(\bibinfo {number} {2}),\ \bibinfo {pages}
  {022305}}\BibitemShut {NoStop}%
\bibitem [{\citenamefont {Chisolm}\ \emph {et~al.}(1996)\citenamefont
  {Chisolm}, \citenamefont {Sudarshan},\ and\ \citenamefont
  {Jordan}}]{chisolm1996}%
  \BibitemOpen
  \bibfield  {author} {\bibinfo {author} {\bibnamefont {Chisolm}, \bibfnamefont
  {Eric}}, \bibinfo {author} {\bibfnamefont {ECG}\ \bibnamefont {Sudarshan}}, \
  and\ \bibinfo {author} {\bibfnamefont {Thomas~F}\ \bibnamefont {Jordan}}}
  (\bibinfo {year} {1996}),\ \bibfield  {title} {\enquote {\bibinfo {title}
  {Weak decoherence and quantum trajectory graphs},}\ }\href@noop {} {\bibfield
   {journal} {\bibinfo  {journal} {International Journal of Theoretical
  Physics}\ }\textbf {\bibinfo {volume} {35}}~(\bibinfo {number} {3}),\
  \bibinfo {pages} {485--493}}\BibitemShut {NoStop}%
\bibitem [{\citenamefont {David}(2012)}]{david2012}%
  \BibitemOpen
  \bibfield  {author} {\bibinfo {author} {\bibnamefont {David}, \bibfnamefont
  {Fran{\c{c}}ois}}} (\bibinfo {year} {2012}),\ \bibfield  {title} {\enquote
  {\bibinfo {title} {A short introduction to the quantum formalism [s]},}\
  }\href@noop {} {\bibinfo  {journal} {arXiv preprint arXiv:1211.5627}\
  }\BibitemShut {NoStop}%
\bibitem [{\citenamefont {DeWitt}(1970)}]{dew}%
  \BibitemOpen
\bibfield  {journal} {  }\bibfield  {author} {\bibinfo {author} {\bibnamefont
  {DeWitt}, \bibfnamefont {BS}}} (\bibinfo {year} {1970}),\ \bibfield  {title}
  {\enquote {\bibinfo {title} {Quantum mechanics and reality},}\ }\href@noop {}
  {\bibfield  {journal} {\bibinfo  {journal} {Physics Today}\ }\textbf
  {\bibinfo {volume} {23}},\ \bibinfo {pages} {30--\&}}\BibitemShut {NoStop}%
\bibitem [{\citenamefont {Diosi}(2004)}]{diosi2004}%
  \BibitemOpen
  \bibfield  {author} {\bibinfo {author} {\bibnamefont {Diosi}, \bibfnamefont
  {Lajos}}} (\bibinfo {year} {2004}),\ \bibfield  {title} {\enquote {\bibinfo
  {title} {Anomalies of weakened decoherence criteria for quantum histories},}\
  }\href@noop {} {\bibfield  {journal} {\bibinfo  {journal} {Physical review
  letters}\ }\textbf {\bibinfo {volume} {92}}~(\bibinfo {number} {17}),\
  \bibinfo {pages} {170401}}\BibitemShut {NoStop}%
\bibitem [{\citenamefont {Driver}(2003)}]{Driver}%
  \BibitemOpen
  \bibfield  {author} {\bibinfo {author} {\bibnamefont {Driver}, \bibfnamefont
  {B~K}}} (\bibinfo {year} {2003}),\ \bibfield  {title} {\enquote {\bibinfo
  {title} {Analysis tools with applications, pp.222-226},}\ }\href@noop {} {\
  }\bibinfo {note} {Available online at
  \url{http://www.math.ucsd.edu/~bdriver/240-01-02/Lecture_Notes/anal2p.pdf}}\BibitemShut
  {NoStop}%
\bibitem [{\citenamefont {D{\"u}rr}\ \emph {et~al.}(2013)\citenamefont
  {D{\"u}rr}, \citenamefont {Goldstein},\ and\ \citenamefont
  {Zangh{\`\i}}}]{durr2013quantum}%
  \BibitemOpen
  \bibfield  {author} {\bibinfo {author} {\bibnamefont {D{\"u}rr},
  \bibfnamefont {Detlef}}, \bibinfo {author} {\bibfnamefont {Sheldon}\
  \bibnamefont {Goldstein}}, \ and\ \bibinfo {author} {\bibfnamefont {Nino}\
  \bibnamefont {Zangh{\`\i}}}} (\bibinfo {year} {2013}),\ \href@noop {} {\emph
  {\bibinfo {title} {Quantum Physics without Quantum Philosophy}}}\ (\bibinfo
  {publisher} {Springer})\BibitemShut {NoStop}%
\bibitem [{\citenamefont {Everett}(1957)}]{ev1}%
  \BibitemOpen
  \bibfield  {author} {\bibinfo {author} {\bibnamefont {Everett}, \bibfnamefont
  {H~III}}} (\bibinfo {year} {1957}),\ \bibfield  {title} {\enquote {\bibinfo
  {title} {{Relative State Formulation of Quantum Mechanics}},}\ }\href@noop {}
  {\bibfield  {journal} {\bibinfo  {journal} {Rev. Mod. Phys}\ }\textbf
  {\bibinfo {volume} {29}},\ \bibinfo {pages} {454}}\BibitemShut {NoStop}%
\bibitem [{\citenamefont {Fuchs}(2010)}]{qbism}%
  \BibitemOpen
  \bibfield  {author} {\bibinfo {author} {\bibnamefont {Fuchs}, \bibfnamefont
  {Christopher~A}}} (\bibinfo {year} {2010}),\ \bibfield  {title} {\enquote
  {\bibinfo {title} {{QBism, the perimeter of quantum Bayesianism}},}\
  }\href@noop {} {\bibinfo  {journal} {arXiv:1003.5209}\ }\BibitemShut
  {NoStop}%
\bibitem [{\citenamefont {Fuchs}\ and\ \citenamefont
  {Peres}(2000)}]{fuchs-peres}%
  \BibitemOpen
\bibfield  {journal} {  }\bibfield  {author} {\bibinfo {author} {\bibnamefont
  {Fuchs}, \bibfnamefont {Christopher~A}}, \ and\ \bibinfo {author}
  {\bibfnamefont {Asher}\ \bibnamefont {Peres}}} (\bibinfo {year} {2000}),\
  \bibfield  {title} {\enquote {\bibinfo {title} {Quantum theory needs no
  'interpretation'},}\ }\href@noop {} {\bibfield  {journal} {\bibinfo
  {journal} {Physics Today}\ }\textbf {\bibinfo {volume} {53}}~(\bibinfo
  {number} {3}),\ \bibinfo {pages} {70--71}}\BibitemShut {NoStop}%
\bibitem [{\citenamefont {Garson}(2009)}]{modal}%
  \BibitemOpen
  \bibfield  {author} {\bibinfo {author} {\bibnamefont {Garson}, \bibfnamefont
  {James}}} (\bibinfo {year} {2009}),\ \bibfield  {title} {\enquote {\bibinfo
  {title} {Modal logic},}\ }\href@noop {} {\bibinfo  {journal} {The Stanford
  Encyclopedia of Philosophy. Winter}\ }\BibitemShut {NoStop}%
\bibitem [{\citenamefont {Gell-Mann}\ and\ \citenamefont {Hartle}(1993)}]{gh3}%
  \BibitemOpen
\bibfield  {journal} {  }\bibfield  {author} {\bibinfo {author} {\bibnamefont
  {Gell-Mann}, \bibfnamefont {M}}, \ and\ \bibinfo {author} {\bibfnamefont
  {J.~B.}\ \bibnamefont {Hartle}}} (\bibinfo {year} {1993}),\ \bibfield
  {title} {\enquote {\bibinfo {title} {{Classical equations for quantum
  systems}},}\ }\href@noop {} {\bibfield  {journal} {\bibinfo  {journal} {Phys.
  Rev.}\ }\textbf {\bibinfo {volume} {D47}},\ \bibinfo {pages}
  {3345}}\BibitemShut {NoStop}%
\bibitem [{\citenamefont {Gell-Mann}\ and\ \citenamefont
  {Hartle}(2007)}]{gell2007quasiclassical}%
  \BibitemOpen
  \bibfield  {author} {\bibinfo {author} {\bibnamefont {Gell-Mann},
  \bibfnamefont {Murray}}, \ and\ \bibinfo {author} {\bibfnamefont {James~B}\
  \bibnamefont {Hartle}}} (\bibinfo {year} {2007}),\ \bibfield  {title}
  {\enquote {\bibinfo {title} {Quasiclassical coarse graining and thermodynamic
  entropy},}\ }\href@noop {} {\bibfield  {journal} {\bibinfo  {journal}
  {Physical Review A}\ }\textbf {\bibinfo {volume} {76}}~(\bibinfo {number}
  {2}),\ \bibinfo {pages} {022104}}\BibitemShut {NoStop}%
\bibitem [{\citenamefont {Gleason}(1957)}]{glea}%
  \BibitemOpen
  \bibfield  {author} {\bibinfo {author} {\bibnamefont {Gleason}, \bibfnamefont
  {A~M}}} (\bibinfo {year} {1957}),\ \bibfield  {title} {\enquote {\bibinfo
  {title} {{Measures on the closed subspaces of a Hilbert space}},}\
  }\href@noop {} {\bibfield  {journal} {\bibinfo  {journal} {J. Math. and
  Mechanics}\ }\textbf {\bibinfo {volume} {6}},\ \bibinfo {pages}
  {885}}\BibitemShut {NoStop}%
\bibitem [{\citenamefont {Goldstein}\ \emph {et~al.}(2011)\citenamefont
  {Goldstein}, \citenamefont {Norsen}, \citenamefont {Tausk},\ and\
  \citenamefont {Zanghi}}]{goldstein2011bell}%
  \BibitemOpen
  \bibfield  {author} {\bibinfo {author} {\bibnamefont {Goldstein},
  \bibfnamefont {Sheldon}}, \bibinfo {author} {\bibfnamefont {Travis}\
  \bibnamefont {Norsen}}, \bibinfo {author} {\bibfnamefont {Daniel~Victor}\
  \bibnamefont {Tausk}}, \ and\ \bibinfo {author} {\bibfnamefont {Nino}\
  \bibnamefont {Zanghi}}} (\bibinfo {year} {2011}),\ \bibfield  {title}
  {\enquote {\bibinfo {title} {Bell's theorem},}\ }\href@noop {} {\bibfield
  {journal} {\bibinfo  {journal} {Scholarpedia}\ }\textbf {\bibinfo {volume}
  {6}}~(\bibinfo {number} {10}),\ \bibinfo {pages} {8378}}\BibitemShut
  {NoStop}%
\bibitem [{\citenamefont {Goldstein}\ and\ \citenamefont {Page}(1995)}]{GP}%
  \BibitemOpen
  \bibfield  {author} {\bibinfo {author} {\bibnamefont {Goldstein},
  \bibfnamefont {Sheldon}}, \ and\ \bibinfo {author} {\bibfnamefont {Don~N}\
  \bibnamefont {Page}}} (\bibinfo {year} {1995}),\ \bibfield  {title} {\enquote
  {\bibinfo {title} {Linearly positive histories: probabilities for a robust
  family of sequences of quantum events},}\ }\href@noop {} {\bibfield
  {journal} {\bibinfo  {journal} {Physical review letters}\ }\textbf {\bibinfo
  {volume} {74}}~(\bibinfo {number} {19}),\ \bibinfo {pages}
  {3715--3719}}\BibitemShut {NoStop}%
\bibitem [{\citenamefont {Griffiths}(1984)}]{grif2}%
  \BibitemOpen
  \bibfield  {author} {\bibinfo {author} {\bibnamefont {Griffiths},
  \bibfnamefont {R~B}}} (\bibinfo {year} {1984}),\ \bibfield  {title} {\enquote
  {\bibinfo {title} {{Consistent histories and the interpretation of quantum
  mechanics}},}\ }\href@noop {} {\bibfield  {journal} {\bibinfo  {journal} {J.
  Stat. Phys.}\ }\textbf {\bibinfo {volume} {36}},\ \bibinfo {pages}
  {219}}\BibitemShut {NoStop}%
\bibitem [{\citenamefont {Griffiths}(2002)}]{grif1}%
  \BibitemOpen
  \bibfield  {author} {\bibinfo {author} {\bibnamefont {Griffiths},
  \bibfnamefont {R~B}}} (\bibinfo {year} {2002}),\ \href@noop {} {\emph
  {\bibinfo {title} {{Consistent Quantum Theory}}}}\ (\bibinfo  {publisher}
  {Cambridge University Press})\BibitemShut {NoStop}%
\bibitem [{\citenamefont {Griffiths}(2011{\natexlab{a}})}]{grif19}%
  \BibitemOpen
  \bibfield  {author} {\bibinfo {author} {\bibnamefont {Griffiths},
  \bibfnamefont {R~B}}} (\bibinfo {year} {2011}{\natexlab{a}}),\ \bibfield
  {title} {\enquote {\bibinfo {title} {{Consistent Histories: Questions and
  Answers}},}\ }\href@noop {} {\ }\Eprint
  {http://arxiv.org/abs/http://quantum.phys.cmu.edu/CHS/quest.html}
  {http://quantum.phys.cmu.edu/CHS/quest.html} \BibitemShut {NoStop}%
\bibitem [{\citenamefont {Griffiths}(2011{\natexlab{b}})}]{grifconsistent}%
  \BibitemOpen
  \bibfield  {author} {\bibinfo {author} {\bibnamefont {Griffiths},
  \bibfnamefont {Robert~B}}} (\bibinfo {year} {2011}{\natexlab{b}}),\ \bibfield
   {title} {\enquote {\bibinfo {title} {{A consistent quantum ontology}},}\
  }\href@noop {} {\bibinfo  {journal} {arXiv:1105.3932}\ }\BibitemShut
  {NoStop}%
\bibitem [{\citenamefont
  {Griffiths}(2011{\natexlab{c}})}]{griffiths2011quantum}%
  \BibitemOpen
\bibfield  {journal} {  }\bibfield  {author} {\bibinfo {author} {\bibnamefont
  {Griffiths}, \bibfnamefont {Robert~B}}} (\bibinfo {year}
  {2011}{\natexlab{c}}),\ \bibfield  {title} {\enquote {\bibinfo {title}
  {Quantum locality},}\ }\href@noop {} {\bibfield  {journal} {\bibinfo
  {journal} {Foundations of Physics}\ }\textbf {\bibinfo {volume}
  {41}}~(\bibinfo {number} {4}),\ \bibinfo {pages} {705--733}}\BibitemShut
  {NoStop}%
\bibitem [{\citenamefont {Griffiths}(2012)}]{grif3}%
  \BibitemOpen
  \bibfield  {author} {\bibinfo {author} {\bibnamefont {Griffiths},
  \bibfnamefont {Robert~B}}} (\bibinfo {year} {2012}),\ \bibfield  {title}
  {\enquote {\bibinfo {title} {{Measured responses to quantum Bayesianism}},}\
  }\href@noop {} {\bibfield  {journal} {\bibinfo  {journal} {Physics Today}\
  }\textbf {\bibinfo {volume} {65}},\ \bibinfo {pages} {8}}\BibitemShut
  {NoStop}%
\bibitem [{\citenamefont {Griffiths}(2013)}]{GrQL}%
  \BibitemOpen
  \bibfield  {author} {\bibinfo {author} {\bibnamefont {Griffiths},
  \bibfnamefont {Robert~B}}} (\bibinfo {year} {2013}),\ \bibfield  {title}
  {\enquote {\bibinfo {title} {The new quantum logic},}\ }\href@noop {}
  {\bibinfo  {journal} {arXiv preprint arXiv:1311.2619}\ }\BibitemShut
  {NoStop}%
\bibitem [{\citenamefont {Hartle}(1968)}]{hartle68}%
  \BibitemOpen
\bibfield  {journal} {  }\bibfield  {author} {\bibinfo {author} {\bibnamefont
  {Hartle}, \bibfnamefont {James~B}}} (\bibinfo {year} {1968}),\ \bibfield
  {title} {\enquote {\bibinfo {title} {Quantum mechanics of individual
  systems},}\ }\href@noop {} {\bibfield  {journal} {\bibinfo  {journal}
  {American Journal of Physics}\ }\textbf {\bibinfo {volume} {36}},\ \bibinfo
  {pages} {704}}\BibitemShut {NoStop}%
\bibitem [{\citenamefont {Hartle}(2005)}]{hartle2005}%
  \BibitemOpen
  \bibfield  {author} {\bibinfo {author} {\bibnamefont {Hartle}, \bibfnamefont
  {James~B}}} (\bibinfo {year} {2005}),\ \bibfield  {title} {\enquote {\bibinfo
  {title} {What connects different interpretations of quantum mechanics?}}\
  }\href@noop {} {\bibinfo  {journal} {Quo Vadis Quantum Mechanics?}\ ,\
  \bibinfo {pages} {73--82}}\BibitemShut {NoStop}%
\bibitem [{\citenamefont {Hartle}(2011)}]{jh1}%
  \BibitemOpen
\bibfield  {journal} {  }\bibfield  {author} {\bibinfo {author} {\bibnamefont
  {Hartle}, \bibfnamefont {James~B}}} (\bibinfo {year} {2011}),\ \bibfield
  {title} {\enquote {\bibinfo {title} {The quasiclassical realms of this
  quantum universe},}\ }\href@noop {} {\bibfield  {journal} {\bibinfo
  {journal} {Foundations of Physics}\ }\textbf {\bibinfo {volume}
  {41}}~(\bibinfo {number} {6}),\ \bibinfo {pages} {982--1006}}\BibitemShut
  {NoStop}%
\bibitem [{\citenamefont {Hohenberg}(2010)}]{pch}%
  \BibitemOpen
  \bibfield  {author} {\bibinfo {author} {\bibnamefont {Hohenberg},
  \bibfnamefont {Pierre~C}}} (\bibinfo {year} {2010}),\ \bibfield  {title}
  {\enquote {\bibinfo {title} {{Colloquium: An introduction to consistent
  quantum theory}},}\ }\href@noop {} {\bibfield  {journal} {\bibinfo  {journal}
  {Rev. Mod. Phys.}\ }\textbf {\bibinfo {volume} {82}},\ \bibinfo {pages}
  {2835}}\BibitemShut {NoStop}%
\bibitem [{\citenamefont {Isham}\ \emph {et~al.}(1994)\citenamefont {Isham},
  \citenamefont {Linden},\ and\ \citenamefont {Schreckenberg}}]{isham1994}%
  \BibitemOpen
  \bibfield  {author} {\bibinfo {author} {\bibnamefont {Isham}, \bibfnamefont
  {Chris~J}}, \bibinfo {author} {\bibfnamefont {Noah}\ \bibnamefont {Linden}},
  \ and\ \bibinfo {author} {\bibfnamefont {Stefan}\ \bibnamefont
  {Schreckenberg}}} (\bibinfo {year} {1994}),\ \bibfield  {title} {\enquote
  {\bibinfo {title} {The classification of decoherence functionals: an analog
  of gleason's theorem},}\ }\href@noop {} {\bibfield  {journal} {\bibinfo
  {journal} {Journal of Mathematical Physics}\ }\textbf {\bibinfo {volume}
  {35}}~(\bibinfo {number} {12}),\ \bibinfo {pages} {6360--6370}}\BibitemShut
  {NoStop}%
\bibitem [{\citenamefont {Kent}(1998)}]{kent}%
  \BibitemOpen
  \bibfield  {author} {\bibinfo {author} {\bibnamefont {Kent}, \bibfnamefont
  {A}}} (\bibinfo {year} {1998}),\ \bibfield  {title} {\enquote {\bibinfo
  {title} {Quantum histories},}\ }\href@noop {} {\bibfield  {journal} {\bibinfo
   {journal} {Physica Sripta}\ }\textbf {\bibinfo {volume} {T76}},\ \bibinfo
  {pages} {78--84}},\ \bibinfo {note} {nobel Symposium 104 on Modern Studies of
  Basic Quantum Concepts and Phenomena}\BibitemShut {NoStop}%
\bibitem [{\citenamefont {Kochen}\ and\ \citenamefont {Specker}(1967)}]{koc}%
  \BibitemOpen
  \bibfield  {author} {\bibinfo {author} {\bibnamefont {Kochen}, \bibfnamefont
  {S}}, \ and\ \bibinfo {author} {\bibfnamefont {E.}~\bibnamefont {Specker}}}
  (\bibinfo {year} {1967}),\ \bibfield  {title} {\enquote {\bibinfo {title}
  {The problem of hidden variables in quantum mechanics},}\ }\href@noop {}
  {\bibfield  {journal} {\bibinfo  {journal} {J. Math. Mech.}\ }\textbf
  {\bibinfo {volume} {17}},\ \bibinfo {pages} {59--87}}\BibitemShut {NoStop}%
\bibitem [{\citenamefont {Landau}\ and\ \citenamefont {Lifshitz}(1965)}]{ll6}%
  \BibitemOpen
  \bibfield  {author} {\bibinfo {author} {\bibnamefont {Landau}, \bibfnamefont
  {LD}}, \ and\ \bibinfo {author} {\bibfnamefont {EM}~\bibnamefont {Lifshitz}}}
  (\bibinfo {year} {1965}),\ \href@noop {} {\emph {\bibinfo {title} {{Course of
  Theoretical Physics: Vol.: 3: Quantum Mechanis: Non-Relativistic Theory}}}}\
  (\bibinfo  {publisher} {Pergamon Press})\BibitemShut {NoStop}%
\bibitem [{\citenamefont {Lebowitz}(1999)}]{leb}%
  \BibitemOpen
  \bibfield  {author} {\bibinfo {author} {\bibnamefont {Lebowitz},
  \bibfnamefont {JL}}} (\bibinfo {year} {1999}),\ \bibfield  {title} {\enquote
  {\bibinfo {title} {{Statistical mechanics: A selective review of two central
  issues}},}\ }\href@noop {} {\bibfield  {journal} {\bibinfo  {journal} {Rev.
  Mod. Phys.}\ }\textbf {\bibinfo {volume} {71}},\ \bibinfo {pages}
  {S346--S357}}\BibitemShut {NoStop}%
\bibitem [{\citenamefont {Lorentz}(1916)}]{lorentz}%
  \BibitemOpen
  \bibfield  {author} {\bibinfo {author} {\bibnamefont {Lorentz}, \bibfnamefont
  {Hendrik~Antoon}}} (\bibinfo {year} {1916}),\ \href@noop {} {\emph {\bibinfo
  {title} {The theory of electrons and its applications to the phenomena of
  light and radiant heat}}},\ Vol.~\bibinfo {volume} {29}\ (\bibinfo
  {publisher} {BG Teubner})\BibitemShut {NoStop}%
\bibitem [{\citenamefont {Mackey}(1957)}]{mackey1957quantum}%
  \BibitemOpen
  \bibfield  {author} {\bibinfo {author} {\bibnamefont {Mackey}, \bibfnamefont
  {George~W}}} (\bibinfo {year} {1957}),\ \bibfield  {title} {\enquote
  {\bibinfo {title} {{Quantum mechanics and Hilbert space}},}\ }\href@noop {}
  {\bibfield  {journal} {\bibinfo  {journal} {The American Mathematical
  Monthly}\ }\textbf {\bibinfo {volume} {64}}~(\bibinfo {number} {8}),\
  \bibinfo {pages} {45--57}}\BibitemShut {NoStop}%
\bibitem [{\citenamefont {Mermin}(1990)}]{mermin}%
  \BibitemOpen
  \bibfield  {author} {\bibinfo {author} {\bibnamefont {Mermin}, \bibfnamefont
  {N~David}}} (\bibinfo {year} {1990}),\ \bibfield  {title} {\enquote {\bibinfo
  {title} {Simple unified form for the major no-hidden-variables theorems},}\
  }\href {\doibase 10.1103/PhysRevLett.65.3373} {\bibfield  {journal} {\bibinfo
   {journal} {Phys. Rev. Lett.}\ }\textbf {\bibinfo {volume} {65}},\ \bibinfo
  {pages} {3373--3376}}\BibitemShut {NoStop}%
\bibitem [{\citenamefont {Mermin}(2013)}]{merint}%
  \BibitemOpen
  \bibfield  {author} {\bibinfo {author} {\bibnamefont {Mermin}, \bibfnamefont
  {N~David}}} (\bibinfo {year} {2013}),\ \bibfield  {title} {\enquote {\bibinfo
  {title} {{Annotated Interview with a QBist in the Making}},}\ }\href@noop {}
  {\bibinfo  {journal} {arXiv:1301.6551}\ }\BibitemShut {NoStop}%
\bibitem [{\citenamefont {von Neumann}(1932)}]{vN1932}%
  \BibitemOpen
\bibfield  {journal} {  }\bibfield  {author} {\bibinfo {author} {\bibnamefont
  {von Neumann}, \bibfnamefont {John}}} (\bibinfo {year} {1932}),\ \href@noop
  {} {\emph {\bibinfo {title} {{Mathematische Grundlagen der
  Quantenmechanik}}}}\ (\bibinfo  {publisher} {Springer})\BibitemShut {NoStop}%
\bibitem [{\citenamefont {von Neumann}(1996)}]{vn}%
  \BibitemOpen
  \bibfield  {author} {\bibinfo {author} {\bibnamefont {von Neumann},
  \bibfnamefont {John}}} (\bibinfo {year} {1996}),\ \href@noop {} {\emph
  {\bibinfo {title} {{Mathematical Foundations of Quantum Mechanics}}}},\
  Vol.~\bibinfo {volume} {2}\ (\bibinfo  {publisher} {Princeton University
  Press})\BibitemShut {NoStop}%
\bibitem [{\citenamefont {Nistic\`o}(1999)}]{nist}%
  \BibitemOpen
  \bibfield  {author} {\bibinfo {author} {\bibnamefont {Nistic\`o},
  \bibfnamefont {G}}} (\bibinfo {year} {1999}),\ \bibfield  {title} {\enquote
  {\bibinfo {title} {Consistency conditions for probabilities of quantum
  histories},}\ }\href@noop {} {\bibfield  {journal} {\bibinfo  {journal}
  {Found. Phys.}\ }\textbf {\bibinfo {volume} {29}},\ \bibinfo {pages}
  {221--229}}\BibitemShut {NoStop}%
\bibitem [{\citenamefont {Omn\`{e}s}(1999)}]{o4}%
  \BibitemOpen
  \bibfield  {author} {\bibinfo {author} {\bibnamefont {Omn\`{e}s},
  \bibfnamefont {R}}} (\bibinfo {year} {1999}),\ \href@noop {} {\emph {\bibinfo
  {title} {{Understanding Quantum Mechanics}}}}\ (\bibinfo  {publisher}
  {Princeton University Press})\BibitemShut {NoStop}%
\bibitem [{\citenamefont {Omn{\`e}s}(1992)}]{omnes1992}%
  \BibitemOpen
  \bibfield  {author} {\bibinfo {author} {\bibnamefont {Omn{\`e}s},
  \bibfnamefont {Roland}}} (\bibinfo {year} {1992}),\ \bibfield  {title}
  {\enquote {\bibinfo {title} {Consistent interpretations of quantum
  mechanics},}\ }\href@noop {} {\bibfield  {journal} {\bibinfo  {journal} {Rev.
  Mod. Phys.}\ }\textbf {\bibinfo {volume} {64}}~(\bibinfo {number} {2}),\
  \bibinfo {pages} {339}}\BibitemShut {NoStop}%
\bibitem [{\citenamefont {Peres}(1995)}]{peres}%
  \BibitemOpen
  \bibfield  {author} {\bibinfo {author} {\bibnamefont {Peres}, \bibfnamefont
  {Asher}}} (\bibinfo {year} {1995}),\ \href@noop {} {\emph {\bibinfo {title}
  {Quantum theory: concepts and methods}}},\ Vol.~\bibinfo {volume} {72}\
  (\bibinfo  {publisher} {Springer})\BibitemShut {NoStop}%
\bibitem [{\citenamefont {Putnam}(1969)}]{put69}%
  \BibitemOpen
  \bibfield  {author} {\bibinfo {author} {\bibnamefont {Putnam}, \bibfnamefont
  {Hilary}}} (\bibinfo {year} {1969}),\ \bibfield  {title} {\enquote {\bibinfo
  {title} {Is logic empirical},}\ }\href@noop {} {\bibfield  {journal}
  {\bibinfo  {journal} {Boston studies in the philosophy of science}\ }\textbf
  {\bibinfo {volume} {5}},\ \bibinfo {pages} {181--206}}\BibitemShut {NoStop}%
\bibitem [{\citenamefont {Putnam}(1981)}]{put81}%
  \BibitemOpen
  \bibfield  {author} {\bibinfo {author} {\bibnamefont {Putnam}, \bibfnamefont
  {Hilary}}} (\bibinfo {year} {1981}),\ \bibfield  {title} {\enquote {\bibinfo
  {title} {Quantum mechanics and the observer},}\ }\href@noop {} {\bibfield
  {journal} {\bibinfo  {journal} {Erkenntnis}\ }\textbf {\bibinfo {volume}
  {16}},\ \bibinfo {pages} {193--219}}\BibitemShut {NoStop}%
\bibitem [{\citenamefont {Sorkin}(1994)}]{sorkin1994}%
  \BibitemOpen
  \bibfield  {author} {\bibinfo {author} {\bibnamefont {Sorkin}, \bibfnamefont
  {Rafael~D}}} (\bibinfo {year} {1994}),\ \bibfield  {title} {\enquote
  {\bibinfo {title} {Quantum mechanics as quantum measure theory},}\
  }\href@noop {} {\bibfield  {journal} {\bibinfo  {journal} {Modern Physics
  Letters A}\ }\textbf {\bibinfo {volume} {9}}~(\bibinfo {number} {33}),\
  \bibinfo {pages} {3119--3127}}\BibitemShut {NoStop}%
\bibitem [{\citenamefont {Tegmark}(1997)}]{tegmark}%
  \BibitemOpen
  \bibfield  {author} {\bibinfo {author} {\bibnamefont {Tegmark}, \bibfnamefont
  {Max}}} (\bibinfo {year} {1997}),\ \bibfield  {title} {\enquote {\bibinfo
  {title} {{The interpretation of quantum mechanics: Many Worlds or many
  words?}}}\ }\href@noop {} {\bibinfo  {journal} {quant-ph/9709032}\
  }\BibitemShut {NoStop}%
\bibitem [{\citenamefont {Van~Kampen}(2008)}]{van2008}%
  \BibitemOpen
\bibfield  {journal} {  }\bibfield  {author} {\bibinfo {author} {\bibnamefont
  {Van~Kampen}, \bibfnamefont {NG}}} (\bibinfo {year} {2008}),\ \bibfield
  {title} {\enquote {\bibinfo {title} {The scandal of quantum mechanics},}\
  }\href@noop {} {\bibfield  {journal} {\bibinfo  {journal} {American Journal
  of Physics}\ }\textbf {\bibinfo {volume} {76}},\ \bibinfo {pages}
  {989}}\BibitemShut {NoStop}%
\bibitem [{\citenamefont {Weinberg}(2012)}]{weinbergqm}%
  \BibitemOpen
  \bibfield  {author} {\bibinfo {author} {\bibnamefont {Weinberg},
  \bibfnamefont {Steven}}} (\bibinfo {year} {2012}),\ \href@noop {} {\emph
  {\bibinfo {title} {Lectures on quantum mechanics}}}\ (\bibinfo  {publisher}
  {Cambridge University Press})\BibitemShut {NoStop}%
\end{thebibliography}%

\newpage

\end{document}